\documentclass[apj]{emulateapj}
\usepackage{apjfonts}
\usepackage{amsbsy}
\newcommand{\cmjj}{\mbox{${\rm cm^{-2}}$}}
\newcommand{\hI}{\mbox{${\rm H\,I}$}}
\newcommand{\civ}{\mbox{${\rm C\,IV}$}}
\newcommand{\lya}{\mbox{${\rm Ly}\alpha$}}
\newcommand{\ovi}{\mbox{${\rm O\,VI}$}}
\newcommand{\lyb}{\mbox{${\rm Ly}\beta$}}

\newcommand{\apg}{\gtrsim}
\newcommand{\apll}{\lesssim}
\newcommand{\etal}{\ensuremath{\mbox{et~al.}}}

\newcommand{\ibid}{\underline{\makebox[0.5in]{}}.}
\providecommand{\kms}{\,\ensuremath{\rm{km\,s}^{-1}}}

\shorttitle{Correlation of Galaxies and QSO Absorption-line Systems}
\shortauthors{Chen \& Mulchaey}

\begin{document}

\slugcomment{Accepted for Publication in the Astrophysical Journal}

\title{PROBING THE IGM-GALAXY CONNECTION AT z $<$ 0.5 I : \\ A GALAXY
SURVEY IN QSO FIELDS AND A GALAXY-ABSORBER CROSS-CORRELATION
STUDY$^{1,2}$}

\author{Hsiao-Wen Chen\altaffilmark{3} and John S.\ Mulchaey\altaffilmark{4}}

\altaffiltext{1}{Based in part on observations made with the NASA/ESA
Hubble Space Telescope, obtained at the Space Telescope Science
Institute, which is operated by the Association of Universities for
Research in Astronomy, Inc., under NASA contract NAS 5-26555. }

\altaffiltext{2}{Observations reported here were obtained in part at the
Magellan telescopes, a collaboration between the Observatories of the Carnegie
Institution of Washington, University of Arizona, Harvard University,
University of Michigan, and Massachusetts Institute of Technology.}

\altaffiltext{3}{Dept.\ of Astronomy \& Astrophysics and 
Kavli Institute for Cosmological Physics, 
University of Chicago, Chicago, IL, 60637, U.S.A. 
{\tt hchen@oddjob.uchicago.edu}} 

\altaffiltext{4}{The Observatories of the Carnegie Institution of Washington 
813 Santa Barbara Street, Pasadena, CA 91101, U.S.A. 
{\tt mulchaey@ociw.edu}}

\begin{abstract}

  We present an imaging and spectroscopic survey of galaxies in fields
around QSOs HE\,0226$-$4110, PKS\,0405$-$123, and PG\,1216$+$069.  The
fields are selected to have ultraviolet echelle spectra available,
which uncover 195 \lya\ absorbers and 13 \ovi\ absorbers along the
three sightlines.  We obtain robust redshifts for 1104 galaxies of
rest-frame absolute magnitude $M_R-5\log\,h\apll -16$ and at projected
physical distances $\rho\apll 4\ h^{-1}$ Mpc from the QSOs.  HST/WFPC2
images of the fields around PKS\,0405$-$123 and PG\,1216$+$069 are
available for studying the optical morphologies of absorbing galaxies.
Combining the absorber and galaxy data, we perform a cross-correlation
study to understand the physical origin of \lya\ and \ovi\ absorbers
and to constrain the properties of extended gas around galaxies.  The
results of our study are (1) both strong \lya\ absorbers of
$\log\,N(\hI)\ge 14$ and \ovi\ absorbers exhibit a comparable
clustering amplitude as emission-line dominated galaxies and a factor
of $\approx 6$ weaker amplitude than absorption-line dominated
galaxies on co-moving projected distance scales of $r_p<3\ h^{-1}$
Mpc; (2) weak \lya\ absorbers of $\log\,N(\hI)<13.5$ appear to cluster
very weakly around galaxies; (3) none of the absorption-line dominated
galaxies at $r_p\le 250\ h^{-1}$ kpc has a corresponding \ovi\
absorber to a sensitive upper limit of $W(1031)\apll 0.03$ \AA, while
the covering fraction of \ovi\ absorbing gas around emission-line
dominated galaxies is found to be $\kappa\approx 64$\%; and (4)
high-resolution images of five \ovi\ absorbing galaxies show that
these galaxies exhibit disk-like morphologies with mildly disturbed
features on the edge.  Together, the data indicate that \ovi\
absorbers arise preferentially in gas-rich galaxies.  In addition,
tidal debris in groups/galaxy pairs may be principally responsible for
the observed \ovi\ absorbers, particularly those of $W(1031)>70$ m\AA.

\end{abstract}


\section{INTRODUCTION}

  Traditional galaxy surveys trace large-scale structures visible
through stellar light or radio emission of cold neutral gas.  Stars
and known gaseous components account for roughly 12\% of all baryons
in the local universe and the rest are thought to reside in ionized
gaseous halos around galaxies or in intergalactic space (e.g.\
Fukugita 2004).  In principle, the diffuse halo gas and intergalactic
medium (IGM) can be probed by the forest of absorption line systems
observed in the spectra of background QSOs.  Studies of QSO
absorption-line systems are therefore expected to provide important
constraints for theoretical models that characterize the growth of
large scale structures.

  Indeed, the \lya\ forest is found to account for $\apg 95$\% of the
total baryons at redshift $z=2-3$ (e.g.\ Rauch \etal\ 1997).  In the
nearby universe, Penton, Stocke, \& Shull (2002, 2004) have argued
that \lya\ absorbers of neutral hydrogen column density $N(\hI) \le
10^{14.5}$ \cmjj\ may contain between $20-30$\% of the total baryons
based on a simple assumption for the cloud geometry.  Various
numerical simulations have further suggested that approximately 40\%
of the total baryons reside in diffuse intergalactic gas of
temperature $T < 10^{5-6}$ K (e.g.\ Dav\'e \etal\ 2001; Cen \&
Ostriker 2006).  This temperature range makes high-ionization
transitions, such as O\,VI\,$\lambda\lambda$1031,1037 and
Ne\,VIII\,$\lambda\lambda$770, 780, ideal tracers of this warm-hot
intergalactic medium (see e.g.\ Verner \etal\ 1994; Mulchaey \etal\
1996; Tripp \etal\ 2000).  However, recent studies have shown that
roughly 50\% of O\,VI absorbers are better explained by photo-ionized
gas of cooler temperatures (e.g.\ Tripp \etal\ 2008; Thom \& Chen
2008a,b).  Therefore, the nature of these high-ionization systems does
not appear to be unique.

  Whether or not QSO absorption-line systems trace the typical galaxy
population bears directly on the efforts to locate the missing baryons
in the present-day universe (Persic \& Salucci 1992; Fukugita, Hogan,
\& Peebles 1998), and to apply known statistical properties of the
absorbers for constraining statistical properties of faint galaxies.
This issue remains, however, unsettled (Lanzetta \etal\ 1995; Stocke
\etal\ 1995; Chen \etal\ 1998, 2001a; Penton \etal\ 2002; Churchill
\etal\ 2007).  While nearly all galaxies within a projected physical distance
of $\rho=100\ h^{-1}$ kpc from a background QSO have
associated absorbers produced by C\,IV\,$\lambda\lambda$1548,1550 and
Mg\,II\,$\lambda\lambda$2796,2803 (Chen \etal\ 2001b; Chen \& Tinker
2008) and nearly all galaxies within $\rho=180\ h^{-1}$ kpc have
associated \lya\ absorbers of $N(\hI) \ge 10^{14}$ \cmjj\ (Chen \etal\
1998, 2001a), not all \lya\ absorbers have a galaxy found within 1
$h^{-1}$ Mpc physical distance to a luminosity limit of $0.5\, L_*$
(Morris \etal\ 1993; Tripp \etal\ 1998; Stocke \etal\ 2006).  It is
not clear whether these absorbers are associated with fainter galaxies
or not related to galaxies at all.  In addition, the origin of O\,VI
absorbers in terms of their galactic environment is also poorly
quantified.  While Stocke \etal\ (2006) found 95\% of O\,VI absorbers
at $z<0.1$ are located at $\rho\apll 560\ h^{-1}$ kpc from an $L_*$
galaxy, only three of the six O\,VI absorbers studied in Prochaska
\etal\ (2006) have an $L_*$ galaxy found at $\rho<1\ h^{-1}$ Mpc.

To understand the origin of QSO absorption-line systems, Chen \etal\
(2005) presented a pilot study of the galaxy--\lya\ absorber
cross-correlation function based on the \lya\ absorbers and galaxies
identified along a single sightline toward PKS0405$-$123.  The
two-point cross-correlation analysis provides a quantitative measure
of the origin of the absorbers based on their clustering amplitude.
More massive systems arise in relatively higher overdensity
environments and are expected to exhibit stronger clustering amplitude
than low-mass objects arising in lower-overdensity environments.  A
primary result of Chen \etal\ (2005) was that the cross-correlation
function $\xi_{ga}$ including only emission-line dominated galaxies
and strong \lya\ absorbers of $\log\,N(\hI)\ge 14$ showed a comparable
strength to the galaxy auto-correlation function $\xi_{gg}$ on
co-moving, projection distance scales $r_p\le 1\,h^{-1}$ Mpc, while
there remained a lack of cross-correlation signal when using only
absorption-line dominated galaxies.  Absorption-line dominated
galaxies have red colers and are presumably evolved early-type
galaxies, while emission-line dominated galaxies are blue and
presumably younger star-forming systems.  Early-type galaxies are
found to cluster more strongly than younger, star-formig galaxies
(e.g.\ Madgwick \etal\ 2003; Zehavi \etal\ 2005), and are therefore
expected to reside in regions of higher matter overdensity.  The
comparable correlation amplitudes of emission-line galaxies and strong
\lya\ absorbers therefore suggested that strong absorbers of
$\log\,N(\hI)\ge 14$ reside in the same overdensity regions as
emission-line dominated galaxies.  It appeared that these absorbers do
not trace regions where more massive galaxies with dominant
absorption spectral features reside.

  This result of Chen \etal\ (2005) provided the first quantitative
constraint on the origin of \lya\ absorbers in terms of the
significance of the underlying matter density fluctuations around the
regions where they reside.  It also offered a physical explanation for
the on-average weaker clustering amplitude of \lya\ absorbers relative
to the clustering amplitude of the luminous galaxy population reported
earlier (e.g.\ Morris \etal\ 1993).  Wilman \etal\ (2007) performed a
similar analysis based on a galaxy and \lya\ absorber pair sample
established over 16 QSO sightlines.  While these authors concluded
that a spectral-type dependent galaxy and \lya\ cross-correlation
function was not confirmed by their analysis, the two-dimensional
cross-correlation function presented in Figure 4 of Wilman \etal\
clearly displays a strong signal between \lya\ absorbers and
emission-line galaxies that is absent in the cross-correlation
function measured using only absorption-line galaxies.

  To examine whether the initial results of Chen \etal\ (2005)
obtained based on a single field are representative of the general
\lya\ absorber population, we have been conducting a deep, wide-area
survey of galaxies in fields around QSOs at $z = 0.3 - 0.6$.  The QSO
fields are selected to have ultraviolet echelle spectra available from
the Far Ultraviolet Spectroscopic Explorer (FUSE) and the Space
Telescope Imaging Spectrograph (STIS) on board the Hubble Space
Telescope (HST).  The high-resolution UV spectra are necessary for
finding intervening hydrogen \lya\ and \ovi\ absorbers to form a
statistically representative sample (see e.g.\ Thom \& Chen 2008a;
Tripp \etal\ 2008), as well as identifying their associated metal-line
transitions for constraining the ionization state of the gas.  The
galaxy sample from our survey program therefore also allows us to
expand upon the initial \lya\ absorber study to understanding the
galactic environment of ionized gas traced by the O\,VI absorbers.
The primary objectives of our galaxy survey program are (i) to examine
the physical origin of \lya\ and \ovi\ absorbers based on their
respective clustering amplitudes, and (ii) to constrain the properties
of extended gas around galaxies.

  This paper is organized as follows.  In Section 2, we describe the
design of our galaxy survey program.  In Section 3, we describe our
observing program that includes both imaging observations, the
selection of candidate galaxies, and multi-slit spectroscopic
observations of three QSO fields.  In Section 4, we summarize the
results of the spectroscopic survey and examine the survey
completeness.  In Section 5, we summarize the properties of known
absorption-line systems along individual QSO sightlines and present a
new \lya\ absorber catalog for the sightline toward PG\,1216$+$069.
Descriptions of individual fields are presented in Section 6.  Results
of a galaxy--absorber cross-correlation study are presented in Section
7, and their implications are discussed in Section 8.  Finally, we
summary the main results of the paper in Section 9.  We adopt a
$\Lambda$CDM cosmology, $\Omega_{\rm M}=0.3$ and $\Omega_\Lambda =
0.7$, with a dimensionless Hubble constant $h = H_0/(100 \ {\rm km} \
{\rm s}^{-1}\ {\rm Mpc}^{-1})$ throughout the paper.

\section{GOALS AND DESIGN OF THE GALAXY SURVEY}

  The primary objectives of our galaxy survey program are (i) to
examine the physical origin of \lya\ and \ovi\ absorbers based on
their clustering amplitudes, and (ii) to constrain the properties of
extended gas around galaxies.  To reach the goals, we have been
conducting a wide-area survey of galaxies in fields around QSOs at $z
= 0.3 - 0.6$, for which echelle spectra at ultraviolet wavelengths are
available from FUSE and HST/STIS.  The high-resolution UV spectra are
necessary for finding intervening hydrogen \lya\ and \ovi\ absorbers
to form a statistically representative sample (see e.g.\ Thom \& Chen
2008a; Tripp \etal\ 2008), as well as identifying their associated
metal-line transitions for constraining the ionization state of the
gas.  The clustering amplitudes of \lya\ and \ovi\ absorbers are then
determined based on their cross-correlation signals with the galaxies
identified in the wide-field redshift survey.

  To characterize the cross-correlation function of galaxies and
\lya/\ovi\ absorbers, we aim to establish a statistically
representative sample of $L_*$ galaxies on scales of $\approx 1-4\
h^{-1}$ co-moving Mpc from the QSO lines of sight over a redshift
range of $z=0.1-0.5$.  These luminous galaxies are thought to reside
at the peaks of the underlying dark matter density fluctuations with a
small fraction ($\apll 20$\%) arising in group/cluster environments
(Zheng \etal\ 2007).  Therefore, they offer a unique tracer of the
large-scale matter overdensity in the distant universe.  Under the
$\Lambda$CDM paradigm, the relative clustering amplitudes of QSO
absorption-line systems with respect to these galaxies provide
quantitative measures of the overdensities of the regions where the
absorbers reside (Mo \& White 2002; Tinker \etal\ 2005).

  To examine the properties of extended gas around galaxies, we aim to
establish a complete sample of sub-$L_*$ galaxies at $z=0.1-0.5$
within a projected co-moving distance of $r_p\approx 250\ h^{-1}$ kpc
of the QSO lines of sight.  The galaxy and absorber pair sample allows
us to study the incidence and extent of ionized gas around galaxies of
a range of luminosity and stellar content.

  The short camera in the IMACS multi-object imaging spectrograph
(Dressler \etal\ 2006) on the Magellan Baade telescope contains eight
CCDs and has a plate scale of $0.2''$.  It observes a sky area of
$\approx 15\arcmin$ radius, corresponding to a co-moving distance of
$3.6\ h^{-1}$ Mpc at $z=0.3$.  In a single setup, IMACS provides the
necessary field coverage to efficiently carry out our galaxy survey
program.  To reach the scientific objectives outlined above, we have
selected three QSOs, HE\,0226$-$4110 ($z_{\rm QSO}=0.495$),
PKS\,0405$-$123 ($z_{\rm QSO}=0.573$), and PG\,1216$+$069 ($z_{\rm
QSO}=0.331$), all of which are accessible from the Las Campanas
Observatory.  Together, these QSO sightlines provide the largest
redshift pathlength available in the STIS echelle sample for the
absorber survey (see Figure 3 of Thom \& Chen 2008a), and allow us to
examine possible field to field variation in the properties of
absorbers and galaxies.

  We have targeted the spectroscopic survey to reach (1) $>80$\%
completeness for galaxies brighter than $R=22$ and at an angular
radius less than $\Delta\,\theta = 2'$ from the QSOs and (2) $\apg
50$\% completeness for galaxies brighter than $R=20$ and at
$\Delta\,\theta>2'$.  At $z=0.2$, $R=22$ corresponds to $\approx
0.04\,L_*$ galaxies and $\Delta\,\theta=2'$ corresponds to $\rho=280\
h^{-1}$ kpc.  At $z=0.5$, $R=22$ corresponds to $\approx 0.3\,L_*$
galaxies and $\Delta\,\theta=2'$ corresponds to $\rho=510\ h^{-1}$
kpc.  We note that the field around PKS\,0405$-$123 has been studied
by Chen \etal\ (2005) and Prochaska \etal\ (2006).  The previous
survey targeted galaxies brighter than $R=20$, which was only
sensitive to roughly $L_*$ galaxies at $z=0.5$.  Our program has been
designed to expand upon the earlier spectroscopic survey for measuring
redshifts of fainter galaxies along the QSO line of sight.

\begin{deluxetable*}{p{1.75 in}rcccl}[h]
\tabletypesize{\tiny}
\tablewidth{0pt}
\tablecaption{Journal of Imaging Observations} 
\tablehead{ & & & Exposure & FWHM &  \\
\multicolumn{1}{c}{Field} & \multicolumn{1}{c}{Instrument} & \multicolumn{1}{c}{Filter} & Time (s) & (arcsec) & \multicolumn{1}{c}{Date} \\
\multicolumn{1}{c}{(1)} & \multicolumn{1}{c}{(2)} & \multicolumn{1}{c}{(3)} &
\multicolumn{1}{c}{(4)} & \multicolumn{1}{c}{(5)} & \multicolumn{1}{c}{(6)} }
\startdata
HE\,0226$-$4110 ($z_{\rm QSO}=0.495$) \dotfill & IMACS & $B$ & 1200 & 0.8 & 2004-09-19 \nl
                                               & IMACS & $R$ & 1200 & 0.7 & 2004-09-19 \nl
                                               & IMACS & $I$ & 3000 & 0.7 & 2004-09-19 \nl
PKS\,0405$-$123 ($z_{\rm QSO}=0.573$) \dotfill & HST/WFPC2 & F702W & 2400 & 0.1 & 1996-01-08 \nl
                                               & HST/WFPC2 & F702W & 2100 & 0.1 & 1998-09-16 \nl
                                               & HST/WFPC2 & F702W & 2100 & 0.1 & 1998-09-23 \nl
PG\,1216$+$069 ($z_{\rm QSO}=0.331$) \dotfill    & IMACS & $B$ & 1200 & 0.9 & 2005-03-19 \nl
                                               & IMACS & $R$ & 1200 & 0.7 & 2005-03-19 \nl
                                               & IMACS & $I$ & 3000 & 0.8 & 2005-03-19 \nl
                                               & HST/WFPC2 & F702W & 2100 & 0.1 & 1998-03-30 \nl
                                               & HST/WFPC2 & F702W & 2100 & 0.1 & 1999-04-18 \nl
\enddata
\end{deluxetable*}

\section{OBSERVING PROGRAM}
 
The observing program of the galaxy survey consists of three phases:
(1) imaging observations of the QSO fields to identify candidate
galaxies, (2) object selection for follow-up faint galaxy
spectroscopy, and (3) multi-slit spectroscopic observations to measure
the redshifts of the selected galaxies.  In this section, we describe
each of the three phases.

\subsection{Imaging Observations and Data Reduction}

  Imaging observations of the fields around HE\,0226$-$4110 and
PG\,1216$+$069 were necessary in order to identify galaxies for
follow-up spectroscopy\footnote{We note that the field around
PG\,1216$+$069 is covered by the Sloan Digital Sky Survey (SDSS; York
\etal\ 2000), which has a targeted imaging depth of $r'=22.2$ (in the
AB magnitude system).  However, because of relatively large point
spread functions (PSFs) with a mean full-width-half-maximum
$\langle {\rm FWHM}\rangle=1.4''$, the available SDSS imaging data
become significantly incomplete for detecting distant faint galaxies
near the magnitude limit and for resolving objects near the QSO.
Additional imaging data are therefore necessary for the purpose of our
study.}.  Faint galaxies in the field around PKS\,0405$-$123 have been
published in Prochaska \etal\ (2006), which are complete to $R=22.5$.
We have therefore selected our spectroscopic targets of this field
from this galaxy catalog.

  Optical images of the field around HE\,0226$-$4110 were obtained on
2004 September 19, using the IMACS short camera and the $B$, $R$, and
$I$ filters on the Magellan Baade Telescope.  Optical images of the
field around PG\,1216$+$069 were obtained on 2005 March 19, using the
IMACS short camera and the $B$, $R$, and $I$ filters.  To increase the
efficiency in the spectroscopic follow-up, the multi-bandpass $B$,
$R$, and $I$ photometric measurements were necessary for selecting
candidate galaxies at $z<0.5$, excluding contaminating red stars and
possible luminous high-redshift galaxies.  The observations were
carried out in a series of four to six exposures of between 300 and
500 s each, and were dithered by between 20 and 30 arcsec in space to
cover the chip gaps between individual CCDs.  All the imaging data
were obtained under photometric conditions.  We also observed standard
star fields selected from Landolt (1992) through each of the $B$, $R$,
and $I$ filters on each night in order to calibrate the photometry in
our targeted fields.  Flat-field images were taken both at a flat
screen at the secondary mirror of the Magellan Baade telescope and at
a blank sky during evening twilight.  A journal of the IMACS imaging
observations is presented in Table 1, which lists in columns (1)
through (6) the object field, instrument, filter, total exposure time,
mean FWHM, and Date of observations, respectively.

\begin{figure*}
\begin{center}
\includegraphics[scale=0.5]{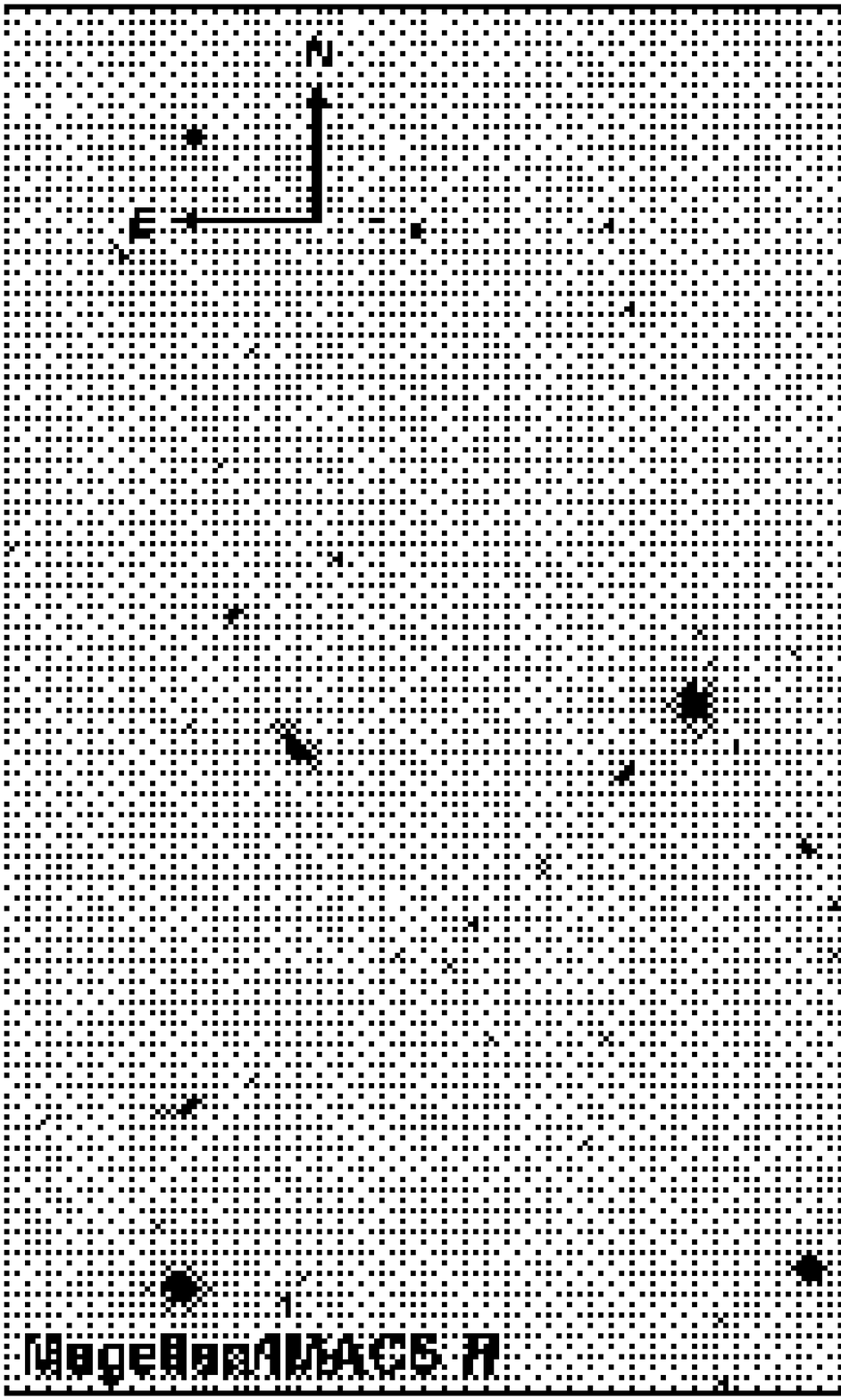}
\caption{A combined IMACS $R$-band image of the center $4'\times 4'$ field around
HE\,0226$-$4110 ($z_{\rm QSO}=0.495$). 
The QSO is at the center of the
panel.}
\end{center} 
\end{figure*}

  The reduction and processing of IMACS imaging data were complicated
because of the multi-CCD format and because of a substantial geometric
distortion across the field of the IMACS short camera.  Before
stacking individual exposures, we first formed a single,
geometrically-dewarped frame of every exposure according to the
following steps.  First, we subtracted the readout bias of individual
CCDs using the overscan regions of the chips.  Next, the bias
subtracted frames were flattened using a super sky flat formed from
median-filtering unregistered science exposures.  We found that in the
$B$ and $R$ images a super sky flat works more effectively than either
dome flats or twilight flats in removing the gain variations between
individual pixels.  For images obtained through the $I$ filter,
however, the fringes in individual science exposures made it impossible
to obtain an accurate flat-field image from the data.  We therefore
flattened the $I$-band exposures by first correcting the
pixel-to-pixel variation using a median dome flat.  The remaining
fringes were then removed by subtracting a median $I$-band sky image
formed from individual flattened science exposures.

  Next we registered individual CCDs of each exposure using known
stars found in the USNO A2.0 catalog.  We formed a geometrically
corrected mosaic frame of eight chips using the IRAF MSCRED package.
The resulting mosaic frame contained calibrated astrometry with a
typical r.m.s.\ scatter in stellar positions of $0.4''$.  Finally, we
combined individual mosaic frames to form a single stacked image of
the QSO field for each of the $BRI$ bandpasses.  The final stacked
images covered a contiguous sky area of $\approx 28\times 28$
arcmin$^2$ and have mean point spread functions (PSFs) in the range of
${\rm FHWM}=0.7''-0.9''$ in the center of the IMACS field.
Photometric calibrations were obtained using Landolt standard stars
observed on the same night.  The combined mosaic images reached 5
$\sigma$ limiting Vega magnitudes of $B=24.5$, $R=24.5$, and $I=23.5$.
False color images of the central $4'\times 4'$ regions around
HE\,0226$-$4110 and PG\,1216$+069$ are presented in Figures 1 and 2,
respectively.

\begin{figure*}
\begin{center}
\includegraphics[scale=0.5]{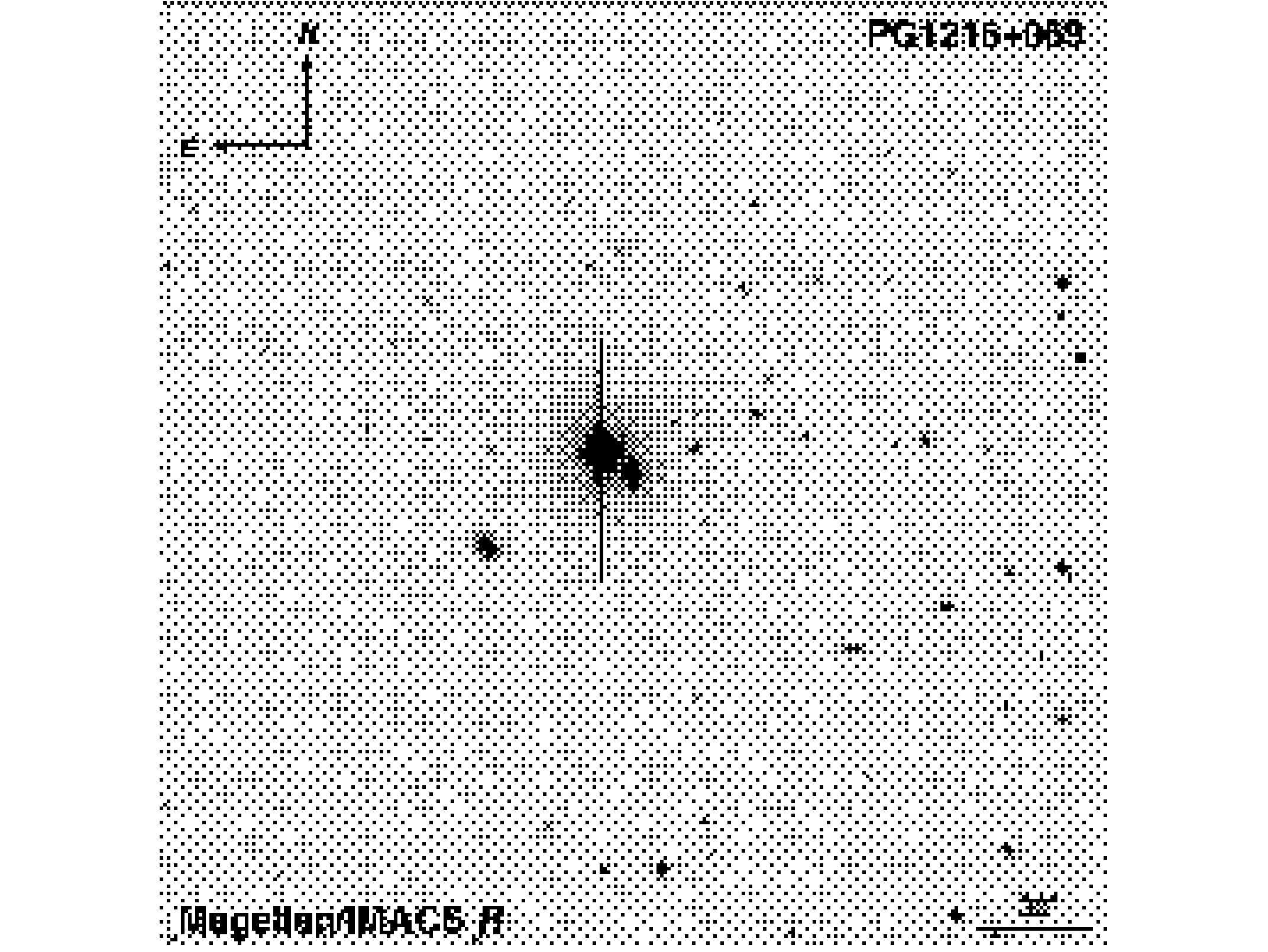}
\caption{A combined IMACS $R$-band image of the center $4'\times 4'$ field around
PG\,1216$+$069 ($z_{\rm QSO}=0.331$).  
The QSO is at the center of the
panel.  The bright star at roughly $10''$ northeast of the QSO makes
it challenging to accurately identify faint galaxies near the QSO
sightline.}
\end{center}
\end{figure*}

  Additional optical images of PKS\,0405$-$123 and PG1216$+$069
obtained with HST using the Wide Field and Planetary Camera2 (WFPC2)
with the F702W filter were accessed from the Hubble Space Telescope
(HST) data archive.  Individual exposures of each field were reduced
using standard pipeline techniques, registered to a common origin,
filtered for cosmic-ray events, and stacked to form a final combined
image.  These high spatial resolution images allowed us to resolve
faint galaxies close to the QSO line of sight and to study their
morphology in detail (see Figures 8 \& 9 below).  A journal of the
HST/WFPC2 observations is presented in Table 1 as well.

\subsection{Selection of Candidate Galaxies for Spectroscopic Follow-up}

  To select galaxies in the fields around HE\,0226$-$4110 and
PG\,1216$+$069 for spectroscopic follow-up, we first identified
objects in the stacked $R$-band mosaic image using the SExtractor
program (Bertin \& Arnout 1996).  Next, we measured object magnitudes
in the $B$, $R$, and $I$ images using the isophotal apertures of
individual objects determined by SExtractor.  This procedure yielded
948 and 1619 objects of $R=16-22$ within $\Delta\,\theta=11'$ angular
radius from HE\,0226$-$4110 and PG\,1216$+$069, respectively.  The
difference in the object surface density underscores the significance
of field to field variation.  The sightline toward PG\,1216$+$069 is
known to pass through the outskirts of the Virgo cluster at $z=0.004$
(and a projected physical distance $\rho\approx 1.4\ h^{-1}$ Mpc).  As
discussed below, the results of our spectroscopic survey have also
uncovered multiple large-scale overdensities along the sightline that
contribute to the high surface density of galaxies in this field.

  Separating stars and galaxies was challenging in these images,
because of non-negligible PSF distortions toward the edge of the IMACS
field.  To optimize the efficiency of the spectroscopic observations,
we applied the observed $B-R$ versus $R-I$ colors to exclude likely
contaminations from faint red stars and high-redshift galaxies.
Figure 3 shows the $B-R$ vs.\ $R-I$ color distribution of the objects
found in the field around HE\,0226$-$4110 (dots), together with
stellar colors calculated using the stellar library compiled by
Pickles (1998; star symbols) and simulated galaxy colors from $z=0$
(lower left) to $z=1$ (upper right) in steps of $\Delta\,z=0.1$ for
E/S0 galaxies (filled circles), Sab (filled triangles), Scd (filled
squares), and Irr (filled pentagon).  The galaxy templates are adopted
from Coleman \etal\ (1980).  The straight line indicates our color
selection, above which objects are likely to arise at $z\apll 0.7$ and
represent candidate galaxies for follow-up spectroscopic observations
with IMACS.  This procedure yielded 776 and 1551 objects of $R=16 -
22$ and $B-R+ 1.7 > 3.5 (R-I)$ within $\Delta\,\theta=11'$ angular
radius from HE\,0226$-$4110 and PG\,1216$+$069, respectively.

\begin{figure}[h]
\begin{center}
\includegraphics[scale=0.4]{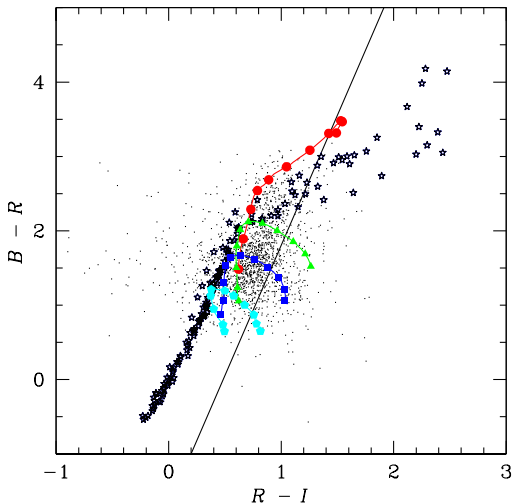}
\caption{The $B-R$ vs.\ $R-I$ color distribution of the objects found
in the field around HE\,0226$-$4110 (dots). Optical colors of stars
are shown in stellar symbols.  These are calculated using the stellar
library compiled by Pickles (1998).  Simulated galaxy colors for
different galaxy templates are also included for E/S0 (filled
circles), Sab (filled triangles), Scd (filled squares), and Irr
(filled pentagon) from $z=0$ (lower left) to $z=1$ (upper right) in
steps of $\Delta\,z=0.1$ .  The straight line indicates our color
selection, above which objects are likely to arise at $z\apll 0.7$ and
are included in the spectroscopic observations with IMACS.}
\end{center} 
\end{figure}

  The goals of our galaxy survery, as outlined in \S\ 2, are (1) to
establish a statistically representative sample of $L_*$ galaxies for
tracing the large-scale matter overdensities along the lines of sight
toward the QSOs and (2) to identify a highly complete sample of
sub-$L_*$ galaxies within projected co-moving distance $r_p\approx
250\ h^{-1}$ kpc from the QSO lines of sight for studying extended gas
around individual galaxies and galaxy groups.  In designing the
multi-slit masks, we therefore aimed to include nearly all galaxies
within $\Delta\,\theta=2'$ of the background QSO.  Because of spectral
collisions between closely located objects, this goal dictates the
number of masks required to reach a high survey completeness.  We were
able to reach the goal in five IMACS masks and included 693 and 750
objects in the multi-slit spectroscopic observations of
HE\,0226$-$4110 and PG\,1216$+$069, respectively\footnote{Some objects
were duplicated in different masks.}.

  For the field around PKS\,0405$-$123, we extracted 524 galaxies
within $\Delta\,\theta=11'$ of the QSO and brighter than $R=22$ from
the photometric catalog of Prochaska \etal\ (2006).  These galaxies
had not been observed in the previous survey.  We aimed to complete
the redshift survey of galaxies to $R\le 22$ in this QSO field with
IMACS observations.  We designed five IMACS masks to observe 450
objects in this field.

\subsection{Spectroscopic Observations and Data Reduction}

  Multi-slit spectroscopic observations of objects selected from the
imaging program described in \S\ 3.2 were carried out using IMACS and
the short camera during three different runs in November 2004, May
2006, and April 2007.  We were able to complete the observations of
five masks for the field around HE\,0226$-$4110, four masks in the
field around PKS\,0405$-$123, and five masks in the field around
PG\,1216$+$069.  Each mask contained between 103 to 218 slitlets of
$1.2''$ in width.  Typical slit lengths were five arcseconds for
compact sources.  For extended objects, we adopted the Kron radius
estimated in SExtractor and expanded by a factor of three to allow
accurate sky subtraction.  We used the 200 l/mm grism that offers a
spectral coverage of $\lambda=5000-9000$ \AA\ with $\approx 2$ \AA\
per pixel resolution.  Observations of faint galaxies were carried out
in a series of two to three exposures of between 1200 to 1800 s
each. Observations of He-Ne-Ar lamps and a quartz lamp were performed
every hour for wavelength and flat-field calibrations.  A journal of
the IMACS spectroscopic observations is presented in Table 2, which
lists in columns (1) through (7) the object field, instrument set-up,
spectral resolution (FWHM), mask number, number of objects observed
per mask, total exposure time, and Date of observations, respectively.

  Additional spectroscopic observations of 27 objects at
$\Delta\,\theta<2'$ from HE\,0226$-$4110 and eight objects from
PKS\,0405$-$123 were attempted in October 2007 and December 2008,
using the Low Dispersion Survey Spectrograph 3 (LDSS3) on the Magellan
Clay telescope.  We included galaxies fainter than $R=22$ and clearly
visible near the QSO sightlines in the LDSS3 observations, to further
improve the survey depth in the immediate vicinity of the QSOs.  LDSS3
observes a field of $8.3'$ diameter at a pixel scale of $0.189''$.  We
used the volume phase holographic (VPH) {\it blue} and VPH {\it all}
grisms that cover a spectral range of $\lambda=4000-6500$ \AA\ and
$\lambda=6000-9000$ \AA, respectively, and $1.0''\times 7''$ slitlets.
The additional LDSS3 multi-slit spectroscopic observations allow us to
reach 100\% survey completeness of faint galaxies near the QSOs.  A
journal of the LDSS3 observations is included in Table 2 as well.

\begin{deluxetable*}{p{1.75 in}lcrrcl}
\tabletypesize{\tiny}
\tablewidth{0pt}
\tablecaption{Journal of Multi-slit Spectroscopic Observations} 
\tablehead{ &  \multicolumn{1}{c}{Instrument/} & FHWM & & & Exposure &  \\
\multicolumn{1}{c}{Field} & \multicolumn{1}{c}{Disperser} & \multicolumn{1}{c}{\AA} & \multicolumn{1}{c}{Mask} & \multicolumn{1}{c}{No.\ of Slitlets} & Time (s) & \multicolumn{1}{c}{Date} \\
\multicolumn{1}{c}{(1)} & \multicolumn{1}{c}{(2)} & \multicolumn{1}{c}{(3)} &
\multicolumn{1}{c}{(4)} & \multicolumn{1}{c}{(5)} & \multicolumn{1}{c}{(6)} & \multicolumn{1}{c}{(7)} }
\startdata
HE\,0226$-$4110 ($z_{\rm QSO}=0.495$) \dotfill & IMACS/f2/200l & 12 & 1 & 187 & 3600 & 2004 November \nl
                                               & IMACS/f2/200l & 12 & 2 & 168 & 3600 & 2004 November \nl
                                               & IMACS/f2/200l & 12 & 3 & 119 & 3600 & 2004 November \nl
                                               & IMACS/f2/200l & 12 & 4 & 116 & 2400 & 2004 November \nl
                                               & IMACS/f2/200l & 12 & 5 & 103 & 3600 & 2004 November \nl
                                               & LDSS3/VPH Blue & 4.3 & ... & 27 & 3600 & 2007 October \nl
                                               & LDSS3/VPH All & 12 & ... & 27 & 3600 & 2008 December \nl
PKS\,0405$-$123 ($z_{\rm QSO}=0.573$) \dotfill & IMACS/f2/200l & 12 & 1 & 159 & 3600 & 2004 November \nl
                                               & IMACS/f2/200l & 12 & 2 & 141 & 4200 & 2004 November \nl
                                               & IMACS/f2/200l & 12 & 3 & 134 & 2400 & 2004 November \nl
                                               & IMACS/f2/200l & 12 & 5 &  48 & 2400 & 2004 November \nl
                                               & LDSS3/VPH All & 12 & ... & 8 & 3600 & 2008 December \nl
PG\,1216$+$069 ($z_{\rm QSO}=0.331$) \dotfill  & IMACS/f2/200l & 12 & 1 & 218 & 3600 & 2006 May \nl
                                               & IMACS/f2/200l & 12 & 2 & 199 & 7200 & 2007 April \nl
                                               & IMACS/f2/200l & 12 & 3 & 182 & 3600 & 2006 May \nl
                                               & IMACS/f2/200l & 12 & 4 & 142 & 5400 & 2007 April \nl
                                               & IMACS/f2/200l & 12 & 5 & 151 & 3600 & 2006 May \nl
\enddata
\end{deluxetable*}

  The multi-slit spectroscopic data were processed and reduced using
the Carnegie Observatories System for MultiObject Spectroscopy
(COSMOS) program\footnote{\scriptsize
http://www.lco.cl/lco/telescopes-information/magellan/magellan/instruments/imacs/cosmos/cosmos.html/view}
that has been developed and tested by A.\ Oemler, K.\ Clardy, D.\
Kelson, and G.\ Walth.  The program adopts the optical model of the
spectrograph and makes initial guesses both for the slit locations in
the 2D spectral frame and for the wavelength calibrations of
individual spectra.  These initial gueses are then further refined
using known spectral lines in the He-Ne-Ar frame obtained close in
time with the science data.  The rms scatter of the wavelength
solution is typically a fraction of a pixel, i.e.\ $\apll 1$ \AA.
Each raw frame of our spectral data was processed for bias subtraction
and corrected for flat-fielding variations, following the standard
procedure.  Spectral extraction was performed using optimal weights
determined according to the noise of relevant pixels.  An error
spectrum was generated from each raw frame, following noise counting
statistics and the spectrum extraction procedure, and propagated
through the pipeline.  Finally, the extracted 1D spectra and their
associated errors were flux-calibrated using a crude spectral
sensitivity function calculated from the spectra of the alignment
stars.

\section{RESULTS OF THE SPECTROSCOPIC SURVEY}

\subsection{Redshift Measurements and Spectral Classification}

\begin{figure}
\begin{center}
\includegraphics[scale=0.4]{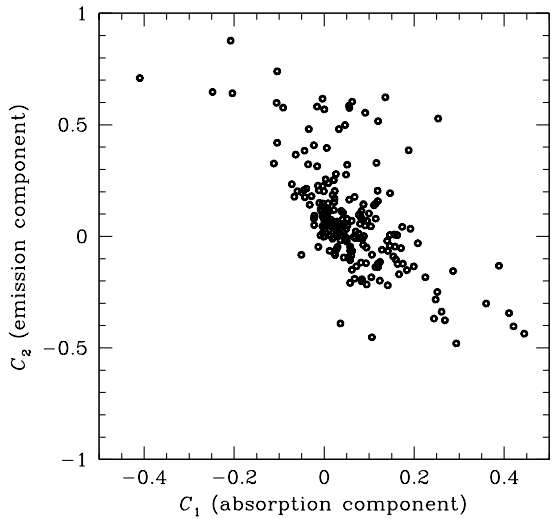}
\caption{Relative significances between absorption ($C_1$) and
emission ($C_2$) components for 220 galaxies in the field around
PKS\,0405$-$123. The absorption coefficient $C_1$ is determined by the
best-fit coefficient of the first eigen spectrum.  The emission
coefficient $C_2$ is determined by the best-fit coefficient of the
second eigen spectrum, corrected for the continuum slope by
subtracting off the best-fit coefficients of the remaining two eigen
spectra. }
\end{center} 
\end{figure}

  The redshifts of objects were determined independently by each of us
in two steps.  First, we performed separate automatic $\chi^2$ fitting
routines that determine a best-fit redshift based on cross-correlating
the flux-calibrated object spectrum with an input model template.  One
of the cross-correlation routines was written by D.\ Kelson.  This
routine adopts SDSS spectral templates\footnote{See
http://www.sdss.org/dr7/algorithms/spectemplates/index.html} for
early-type (SDSS template \# 24) and late-type galaxies (\#28) as
input models, and determines a best-fit redshift based on matching
absorption or emission line features.  The other cross-correlation
software was written by one of us (HWC), which employs four galaxy
eigen spectra kindly provided by S.\ Burles.  These eigen templates
were obtained from analyzing $>100,000$ SDSS galaxy spectra, using a
principal component analysis (see e.g.\ Yip \etal\ 2004).  The first
two eigen spectra are characterized by predominantly absorption
features and predominantly emission features, respectively.  The last
two eigen spectra offer additional modifications in the continuum
slope.  A linear combination of the four eigen function formed a model
template to be compared with a flux-calibrated object spectrum, and a
$\chi^2$ routine was performed to determine the best-fit linear
coefficients and redshift for each object.

  Next, the best-fit redshifts returned from the cross-correlation
routines were then visually inspected by each of us to determine a
final and robust redshift for every object.  In some cases, the
cross-correlation routines failed to identify a correct redshift due
to contaminating residuals of bright skylines or imperfect background
sky subtraction.  We were able to recover the redshifts during this
visual inspection process based on the presence of Ca\,II H\&K, or the
presence of H\,$\alpha$ and [N\,II].  A comparison between redshifts
determined independently by the two of us shows a typical scatter of
$|\Delta\,z|\approx 0.0003$, corresponding to 1-$\sigma$ uncertainty
of 70 \kms.

  The procedure described above led to robust redshift measurements
for 432 and 448 galaxies in the fields around
HE\,0226$-$4110 and PG\,1216$+$069, respectively.  In addition, we
measured 224 new redshifts for galaxies in the field around
PKS\,0405$-$123, in comparison to 91 galaxy redshifts in the same
area from Prochaska \etal\ (2006).  The remaining galaxies turned out
to have insufficient $S/N$ in their spectra for a robust redshift
measurement.

  Spectral classification was guided by the relative magnitudes
between the best-fit coefficients of the eigen spectra described
above.  These coefficients provide a quantitative evaluation of the
fractional contributions from different spectral components in a
galaxy, such as absorption spectra due to low-mass stars or emission
spectra due to H\,II regions.  The contribution due to absorption
components may be determined by the best-fit coefficient of the first
eigen spectrum ($C_1$).  The contribution due to emission components
may be determined by the best-fit coefficient of the second eigen
spectrum, corrected for the continuum slope by subtracting off the
best-fit coefficients of the remaining two eigen spectra ($C_2$).
Figure 4 shows that despite a large scatter, there exists a linear
distribution between emission $C_2$ and absorption $C_1$ components
for 224 galaxies found in the field around PKS\,0405$-$123.  We
classified galaxies with $C_1>0.09$ and $C_2<0.3$ as absorption-line
dominated early-type galaxies and the rest as emission-line dominated
late-type galaxies.

\subsection{Survey Completeness and Galaxy Redshift Distribution}

  The results of our spectroscopic survey program in three QSO fields
are summarized in Figure 5.  For the field around HE\,0226$-$4110,
the open circles indicate the sky positions of all objects brighter
than $R=23$ and the closed circles indicate the ones with available
spectroscopic redshifts.  The QSO is at ${\rm RA(J2000)}=37.0633$ deg and
${\rm Dec(J2000)}=-40.9544$ deg. For the fields around PKS\,0405$-$123 and
PG\,1216$+$069, the open circles indicate the sky positions of all
objects brighter than $R=22$ and the closed circles indicate the ones
with available spectroscopic redshifts.  The QSOs are at ${\rm
RA(J2000)}=61.9518$ deg and ${\rm Dec(J2000)}=-12.1935$ deg, and ${\rm
RA(J2000)}=184.8372$ deg and ${\rm Dec(J2000)}=6.6440$ deg, respectively.

\begin{figure*}
\begin{center}
\includegraphics[scale=0.55]{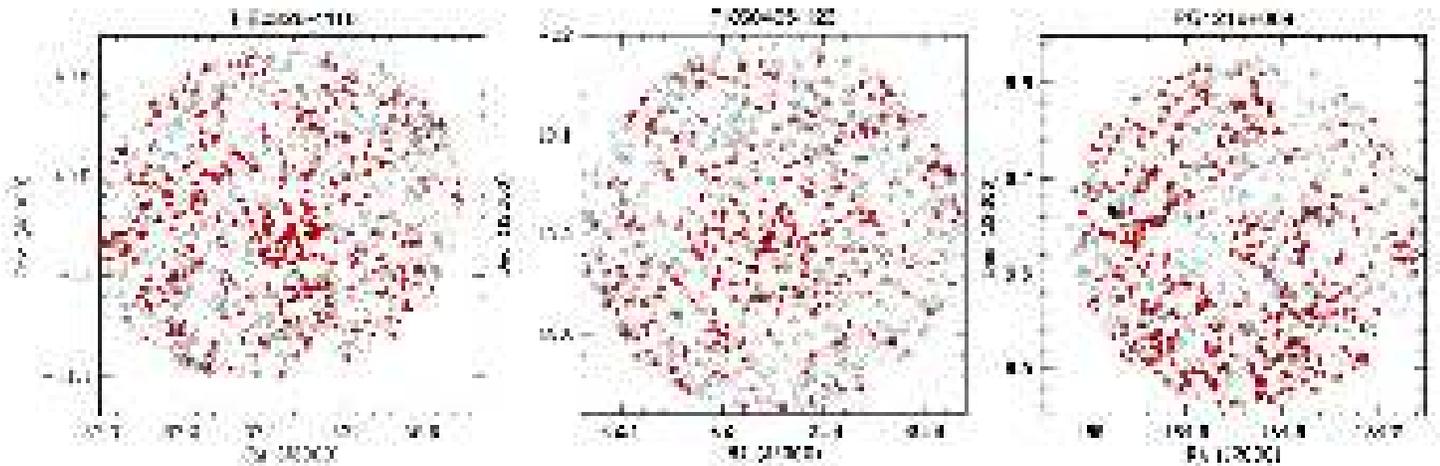}
\caption{Summary of the spectroscopic survey in the fields around
HE\,0226$-$4110 (left), PKS\,0405$-$123 (middle), and PG\,1216$+$069
(right).  Open circles represent all objects of $R\le 23$ in
HE\,0226$-$4110, and $R\le 22$ in PKS\,0405$-$123 and PG\,1216$+$069.
Closed circles represent those with available spectroscopic redshifts.
The QSOs are at ${\rm RA(J2000)}=37.0633$ deg and ${\rm
Dec(J2000)}=-40.9544$ deg, ${\rm RA(J2000)}=61.9518$ deg and ${\rm
Dec(J2000)}=-12.1935$ deg, and ${\rm RA(J2000)}=184.8372$ deg and ${\rm
Dec(J2000)}=6.6440$ deg, respectively.}
\end{center} 
\end{figure*}

  We estimate the completeness of our survey by calculating the
fraction of photometrically identified galaxies that have available
spectroscopic redshifts.  The calculations were performed for
different minimum brightness in the $R$ band and at different angular
distances $\Delta\theta$ to the QSO.  The results are presented in the
top panels of Figure 6.  We find that our survey is most complete in
the field around HE\,0226$-$4110, reaching 100\% completeness for
galaxies brighter than $R=23$ at angular distances $\Delta\theta\le
2'$.  For PKS\,0405$-$123 and PG\,1216$+$069, we have also reached
100\% completeness for galaxies brighter than $R=20$ (solid lines),
$>80$\% at $R=21$ (dotted lines) and $>60$\% at $R=22$ (dash-dotted
lines) in the inner $2'$ radius.  The survey completeness becomes
$\approx 50$\% at larger radii.

\begin{figure*}
\begin{center}
\includegraphics[scale=0.5]{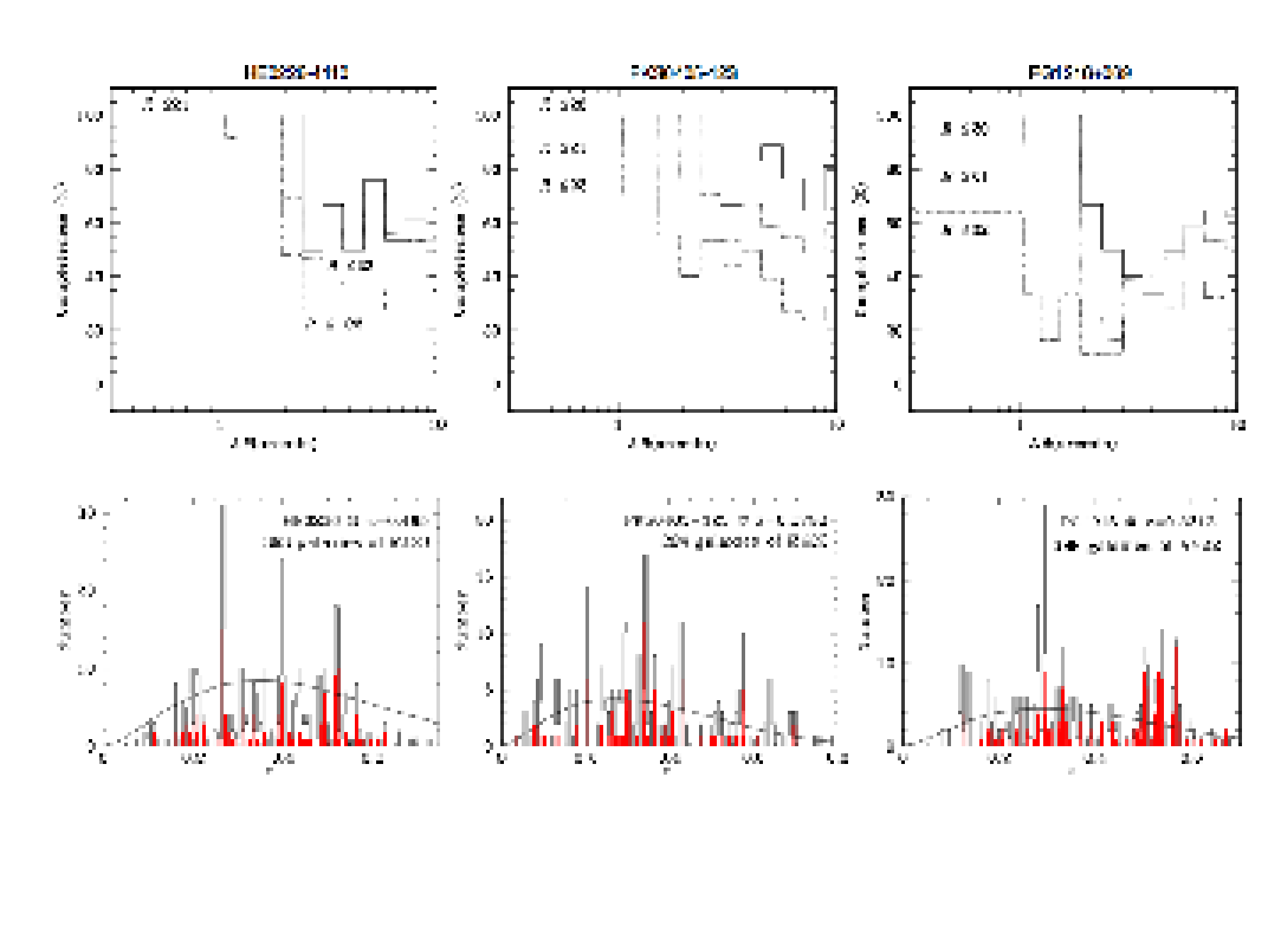}
\caption{Summary of the survey completeness (top) and redshift
distributions (bottom) of galaxies in the fields around
HE\,0226$-$4110 (left), PKS\,0405$-$123 (middle), and PG\,1216$+$069
(right).  The survey completeness was calculated for different
brightness limits, from $R\le 20$ (solid lines), to $R\le 21$ (dotted
lines) and to $R\le 22$ (dash-dotted lines).  The bottom panels show
the redshift distributions of spectroscopically identified galaxies in
the three fields, in comparison to model expectations (solid curves)
based on a non-evolving rest-frame $R$-band galaxy luminosity function
of Blanton \etal\ (2003) and the respective completeness functions
displayed in the top panels.  The redshift distribution of
absorption-line dominated galaxies in each field is shown in solid
histograms.  Large-scale galaxy overdensities along the QSO lines of
sight are evident through the presence of redshift spikes.}
\end{center} 
\end{figure*}

  The bottom panels of Figure 6 show the redshift distributions of the
spectroscopically identified galaxies in the three fields.  The solid
curve in each panel represents the model expectation from convolving a
non-evolving rest-frame $R$-band galaxy luminosity function of Blanton
\etal\ (2003), which is characterized by $M_{R_*}-5\,\log\,h=-20.4$,
$\alpha=-1.1$, and $\phi_*=0.015\,h^{3}\,{\rm Mpc}^{-3}$, with the
completeness functions displayed in the top panels.  Redshift spikes
indicate the presence of large-scale galaxy overdensities along the
QSO lines of sight.  The comparison between observations and model
expectations in HE\,0226$-$4110 and PG\,1216$+$069 further
demonstrates that the color selection criterion described in \S\ 3.2
and Figure 3 has effectively excluded most galaxies at $z>0.6$.


\section{CATALOGS OF QSO ABSORPTION-LINE SYSTEMS}

  A necessary component of the galaxy--absorber cross-correlation
analysis is an absorber catalog.  A catalog of 57 \lya\ absorbers
along the sightline toward HE\,0226$-$4110 has been published by
Lehner \etal\ (2006).  These \lya\ absorbers exhibit a range of \hI\
column density from $\log\,N(\hI)=12.5$ to $\log\,N(\hI)=15.1$ at
$z_{\lya}=0.017-0.4$.  We adopt their catalog in our
cross-correlation analysis in this field.  To summarize, the left
panels of Figure 7 display the redshift distribution (top) and the
cumulative $N(\hI)$ distribution function (bottom) of these absorbers.

  The sightline toward PKS\,0405$-$123 has been studied by Williger
\etal\ (2006) and Lehner \etal\ (2007).  We adopt the revised catalog
of 76 \lya\ absorbers from Lehner \etal\ (2007) in our analysis. These
\lya\ absorbers exhibit a range of \hI\ column density from
$\log\,N(\hI)=12.8$ to $\log\,N(\hI)=16.3$ at $z_{\lya}=0.012-0.410$.
The middle panels of Figure 7 display their redshift distribution
(top) and cumulative $N(\hI)$ distribution.

\begin{figure*}
\begin{center}
\includegraphics[scale=0.5]{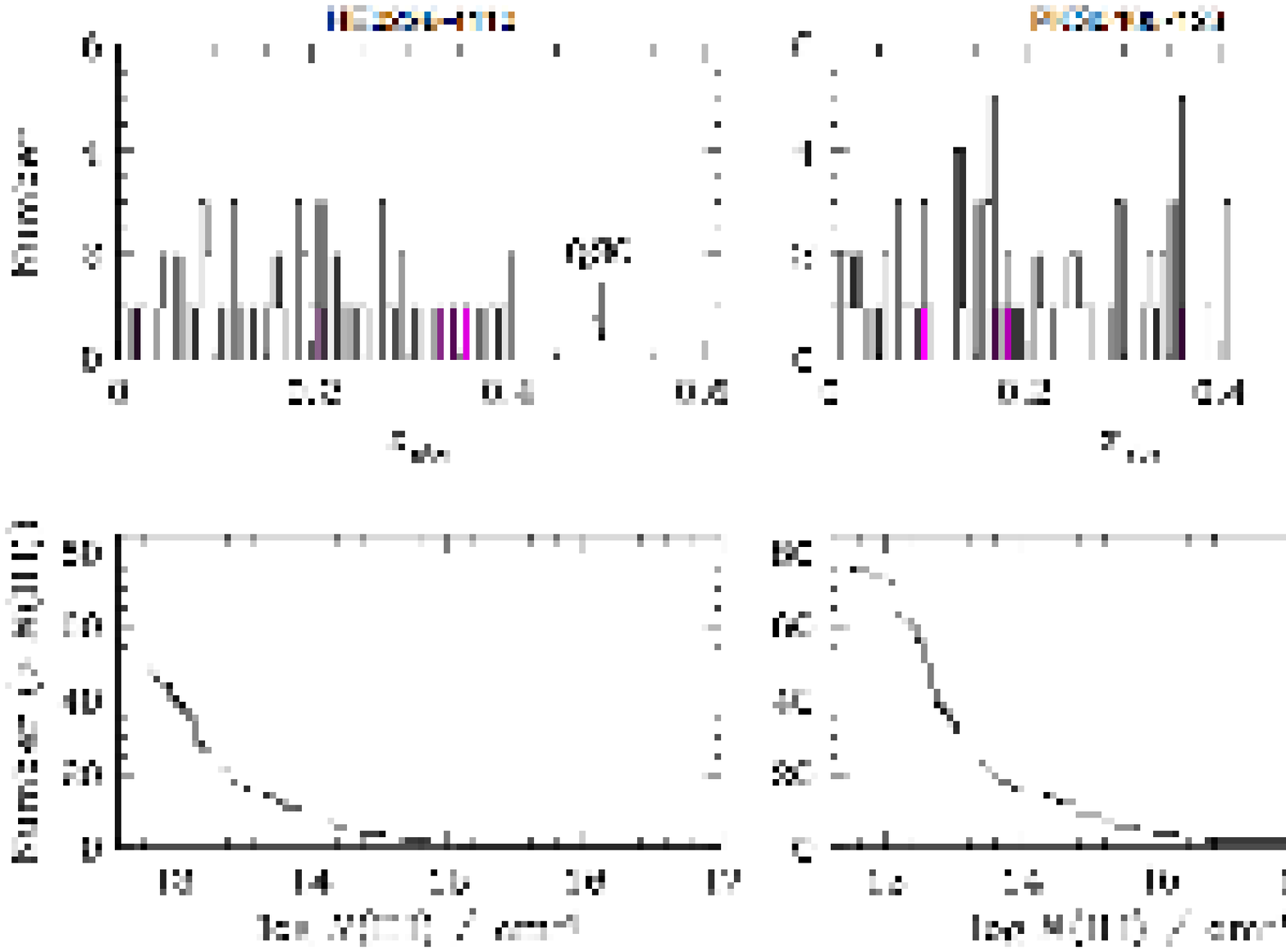}
\caption{The redshift distribution (top) and cumulative $N(\hI)$
distribution (bottom) of \lya\ absorbers identified along the
sightline toward HE\,0226$-$4110 (left), PKS\,0405$-$123 (middle), and
PG\,1216$+$069 (right).  The shaded histograms in the top panels show
the redshift distributions of \ovi\ absorbers along the respective
sightlines.}
\end{center} 
\end{figure*}

  A systematic survey of \lya\ absorbers for the sightline toward
PG\,1216$+$069 has only been conducted for strong systems by Jannuzi
\etal\ (1998) in low-resolution (${\rm FWHM}\,\approx\,250$ \kms) spectra
obtained using the Faint Object Spectrograph.  These authors
identified nine strong \lya\ absorbers with $\log\,N(\hI)>14$.  To
complete the study of low-column density gas ($\log\,N(\hI)<14$) along
the sightline, we have conducted our own search of intervening \lya\
lines in available echelle spectra of the QSO obtained by STIS.

  Details regarding the reduction and processing of the STIS echelle
spectra can be found in Thom \& Chen (2008a).  The final combined
spectrum of PG\,1216$+$069 covers a spectral range of $\lambda=1160 -
1700$ \AA\ with a spectral resolution of $\approx 6.8$ \kms\ and has
SNR of $\approx 7$ per resolution element in most of the spectral
region.  To establish a complete catalog of \lya\ absorbers, we first
identify absorption features that are at $>5$-$\sigma$ level of
significance.  Next, we identify known features due to either
interstellar absorption of the Milky Way, or metal absorption lines
and higher order Lyman series associated with known strong \lya\
absorbers.  We consider the remaining unidentified absorption lines as
intervening \lya\ absorbers.  Finally, we perform a Voigt profile
analysis using the VPFIT package\footnote{see
http://www.ast.cam.ac.uk/\char'176rfc/vpfit.html.} for estimating the
underlying $N(\hI)$ and the Doppler parameter $b$ (e.g.\ Carswell
\etal\ 1991).  This procedure has yielded 66 \lya\ absorbers of \hI\
column density $\log\,N(\hI)=12.6-19.3$ at $z_{\lya}=0.003-0.321$.
The results of our search are presented in Table 3, which for each
absorber lists the absorber redshift, $N(\hI)$ and the associated
error, and $b$.  The redshift distribution and cumulative $N(\hI)$
distribution of these \lya\ absorbers are presented in the right
panels of Figure 7.


\begin{deluxetable*}{lrrclrr}
\tabletypesize{\tiny}
\tablewidth{0pt}
\tablecaption{IGM \lya\ Absorbers Identified Along the Sightline toward PG\,1216$+$069} 
\tablehead{\multicolumn{1}{c}{$z_{\displaystyle\lya}$} & \multicolumn{1}{c}{$\log\,N(\hI)$} & \multicolumn{1}{c}{$b$ (\kms)} & & 
\multicolumn{1}{c}{$z_{\displaystyle\lya}$} & \multicolumn{1}{c}{$\log\,N(\hI)$} & \multicolumn{1}{c}{$b$ (\kms)} \\
\multicolumn{1}{c}{(1)} & \multicolumn{1}{c}{(2)} & \multicolumn{1}{c}{(3)} & & 
\multicolumn{1}{c}{(4)} & \multicolumn{1}{c}{(5)} & \multicolumn{1}{c}{(6)} }
\startdata
0.00362 & $13.32\pm0.17$ & $49\pm24$ & | & 0.13503$^b$ & $14.41\pm0.04$ & $35\pm 2$ \nl 
0.00630$^{a,b}$ & $19.32\pm0.03$ & ...   & | & 0.15475 & $13.07\pm0.09$ & $29\pm 7$ \nl
0.01264 & $13.94\pm0.03$ & $40\pm 3$ & | & 0.15512 & $12.75\pm0.16$ & $24\pm 11$ \nl
0.01493 & $13.07\pm0.10$ & $34\pm 9$ & | & 0.15576 & $12.77\pm0.13$ & $18\pm 7$ \nl
0.01682 & $13.00\pm0.14$ & $59\pm 22$ & | & 0.15601 & $13.10\pm0.07$ & $14\pm 3$ \nl
0.01752 & $12.62\pm0.18$ & $20\pm 10$ & | & 0.16076 & $13.14\pm0.10$ & $67\pm 19$ \nl
0.02393 & $13.58\pm0.03$ & $26\pm 2$  & | & 0.17993$^b$ & $13.97\pm0.03$ & $34\pm 2$ \nl
0.05725 & $12.87\pm0.12$ & $35\pm 12$ & | & 0.18026 & $13.23\pm0.10$ & $34\pm 9$ \nl
0.06508 & $12.89\pm0.11$ & $25\pm 7$  & | & 0.18071 & $13.48\pm0.04$ & $33\pm 3$ \nl
0.07381 & $12.83\pm0.12$ & $23\pm 8$  & | & 0.18698 & $13.26\pm0.06$ & $40\pm 6$ \nl
0.07421 & $13.61\pm0.03$ & $35\pm 3$  & | & 0.19007 & $12.96\pm0.11$ & $23\pm 7$ \nl
0.07606 & $13.02\pm0.08$ & $31\pm 7$  & | & 0.20040$^b$ & $13.98\pm0.03$ & $93\pm 7$ \nl
0.08035 & $13.79\pm0.24$ & $45\pm 11$ & | & 0.20114 & $13.17\pm0.09$ & $23\pm 5$ \nl
0.08053$^b$ & $13.99\pm0.15$ & $32\pm 4$  & | & 0.20442 & $12.94\pm0.12$ & $25\pm 8$ \nl
0.08253 & $12.61\pm0.19$ & $16\pm 8$  & | & 0.22189 & $13.34\pm0.06$ & $42\pm 7$ \nl
0.08278 & $12.83\pm0.16$ & $35\pm 16$ & | & 0.22578 & $12.99\pm0.13$ & $22\pm 9$ \nl
0.08313 & $12.77\pm0.12$ & $16\pm 5$  & | & 0.22728 & $13.09\pm0.10$ & $32\pm 9$ \nl
0.08348 & $13.22\pm0.07$ & $37\pm 7$  & | & 0.23035 & $12.67\pm0.18$ & $17\pm 9$ \nl
0.08400 & $12.86\pm0.18$ & $64\pm 32$ & | & 0.23641 & $13.45\pm0.12$ & $43\pm 15$ \nl
0.08813 & $13.22\pm0.07$ & $54\pm 10$ & | & 0.25035 & $12.93\pm0.14$ & $17\pm 6$ \nl
0.09137 & $12.98\pm0.08$ & $30\pm 6$  & | & 0.26267 & $13.12\pm0.12$ & $26\pm 9$ \nl
0.09415 & $13.07\pm0.10$ & $40\pm 12$ & | & 0.26624 & $12.93\pm0.14$ & $28\pm 10$ \nl
0.09578 & $13.61\pm0.03$ & $39\pm 3$  & | & 0.26763$^b$ & $14.01\pm0.04$ & $32\pm 2$ \nl
0.09727 & $12.72\pm0.20$ & $44\pm 26$ & | & 0.27154 & $12.81\pm0.12$ & $12\pm 4$ \nl
0.09999 & $13.12\pm0.05$ & $20\pm 3$  & | & 0.27352 & $14.43\pm0.07$ & $22\pm 3$ \nl
0.10234 & $13.00\pm0.15$ & $78\pm 32$ & | & 0.27865 & $12.98\pm0.15$ & $42\pm 17$ \nl
0.10309 & $12.67\pm0.15$ & $15\pm 7$  & | & 0.28108 & $12.72\pm0.18$ & $16\pm 8$ \nl
0.10440 & $13.09\pm0.08$ & $32\pm 6$  & | & 0.28226$^b$ & $16.34\pm0.10$ & $21\pm 2$ \nl
0.12357 & $14.36\pm0.08$ & $29\pm 2$  & | & 0.28298 & $12.88\pm0.19$ & $21\pm 10$ \nl
0.12388$^b$ & $14.53\pm0.14$ & $28\pm 3$  & | & 0.30394 & $13.40\pm0.09$ & $78\pm 20$ \nl
0.12461$^b$ & $14.30\pm0.02$ & $53\pm 3$  & | & 0.31377 & $12.97\pm0.13$ & $47\pm 17$ \nl
0.12493 & $14.06\pm0.07$ & $20\pm 2$  & | & 0.31857 & $13.23\pm0.11$ & $79\pm 24$ \nl
0.12700 & $12.82\pm0.09$ & $20\pm 5$  & | & 0.32102 & $12.83\pm0.12$ & $21\pm 7$ \nl
\enddata
\tablenotetext{a}{The measurement is adopted from Tripp \etal\ (2005).} 
\tablenotetext{b}{These lines have also been identified by Jannuzi \etal\ (1998) in low-resolution data obtained using the Faint Object Spectrograph.}
\end{deluxetable*}

For \ovi\ absorbers along the sightlines toward the three QSOs in our
survey, we adopt the catalog of Thom \& Chen (2008a,b) and include
absorbers at $z<0.14$ from Tripp \etal\ (2008).  There are five \ovi\
absorbers known along the sightline toward HE\,0226$-$4110, six \ovi\
absorbers known along the sightline toward PKS\,0405$-$123 (see also
Prochaska \etal\ 2004), and two \ovi\ absorbers known along the
sightline toward PG\,1216$+$069.  The \ovi\ absorber at $z=0.1829$
along the sightline toward PKS\,0405$-$123 shows two distinct
components separated by $\approx 87$ \kms\ (see Thom \& Chen 2008b;
Prochaska \etal\ 2004).  The \ovi\ absorber at $z=0.1236$ along the
sightline toward PG\,1216$+$069 also shows two dominant components well
separated by $\approx 350$ \kms\ (see \S\ 6.3 and Figure 12 below).
We consider these components separate objects in the
cross-correlation study presented in \S\ 7.  The redshifts of these
\ovi\ absorbers range from $z_{\ovi}=0.017$ to $z_{\ovi}=0.495$.  The
column densities of these absorbers range from $\log\,N(\ovi)=13.4$ to
$\log\,N(\ovi)=14.7$.  The redshift distribution of these \ovi\
absorbers is shown in shaded histograms in the top panels of Figure 7.

\begin{figure*}
\begin{center}
\includegraphics[scale=0.5]{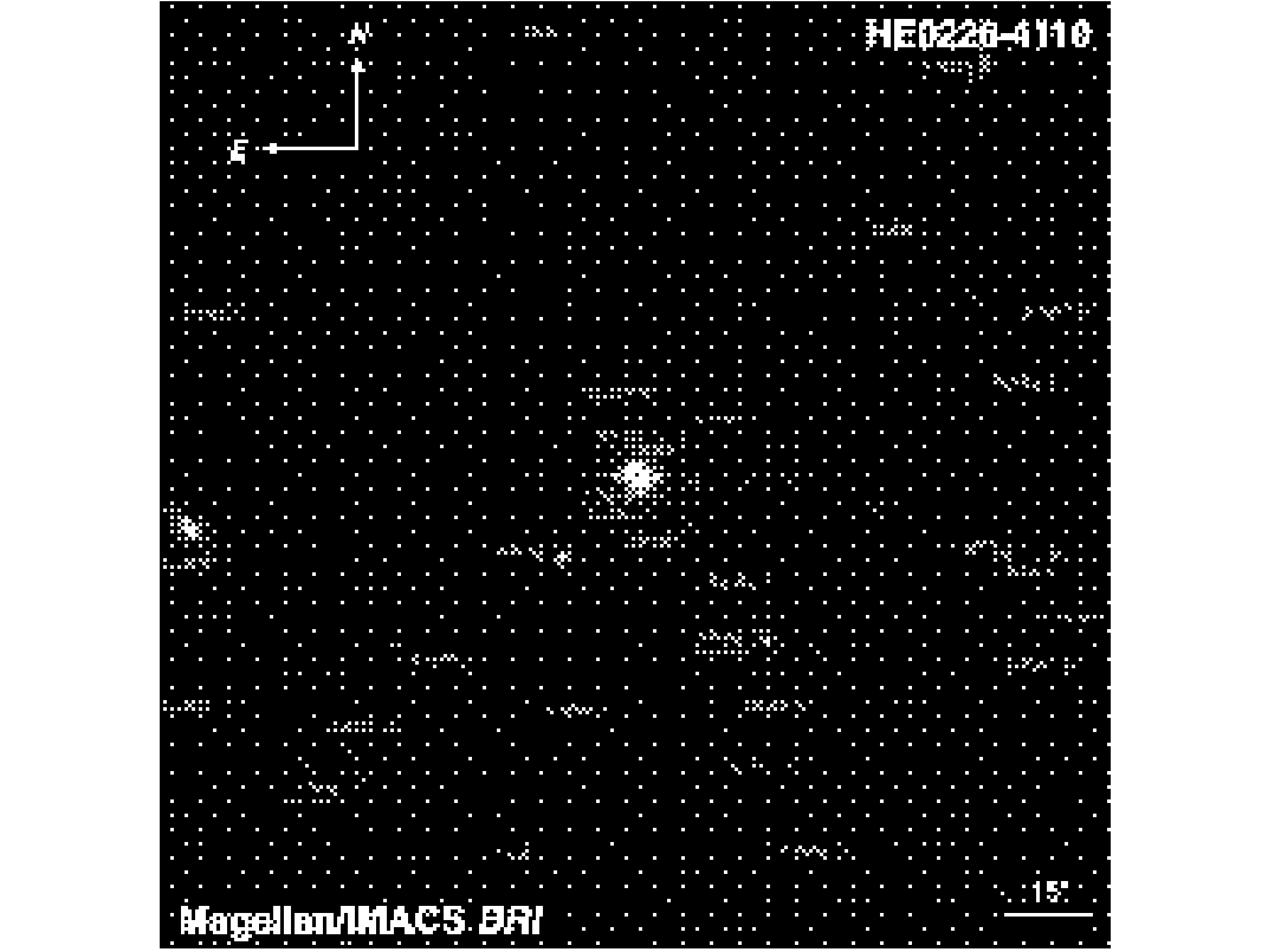}
\caption{A summary of the spectroscopic survey of faint galaxies in
the center $2.7'\times 2.7'$ field around HE\,0226$-$4110 ($z_{\rm
QSO}=0.495$).  The false color image was produced using IMACS $B$,
$R$, and $I$ images from our preimaging data.  We have reached 100\%
completeness for galaxies brighter than $R=23$ in this area.  Our best
estimated redshifts are shown to the left of individual sources.
Galaxies with redshift coincident with known O\,VI absorbers
($|\Delta\,v|<300$ \kms) are marked by a rectangular box.  Most
interestingly, at the location of the Ne\,VIII absorbers, $z=0.20701$
(Savage \etal\ 2005), we found three galaxies at projected physical
distances $\rho<200\ h^{-1} kpc$.}
\end{center} 
\end{figure*}

\section{DESCRIPTION OF INDIVIDUAL FIELDS}

In this section, we review the galaxy and absorber properties in each
individual field.

\subsection{The Field toward HE0226$-$4110 at $z_{\rm QSO}=0.495$}

The sightline toward HE\,0226$-$4110 exhibits 57 \lya\ absorbers
(Lehner \etal\ 2006) and five \ovi\ absorbers (Tripp \etal\ 2008; Thom
\& Chen 2008a,b) over the redshift range from $z_{\lya}=0.017$ to
$z_{\lya}=0.4$.  The column density of these \lya\ absorbers span a
range from $\log\,N(\hI)=12.5$ to $\log\,N(\hI)=15.1$, and the column
density of the \ovi\ absorbers span a range from $\log\,N(\ovi)=13.6$
to $\log\,N(\ovi)=14.4$.  This field is particularly interesting
because it exhibits a Ne\,VIII\,$\lambda\lambda$770,780 absorber at
$z=0.20701$ (Savage \etal\ 2005), which is so far the only known
detection of such high-ionization species in the IGM.  The high
ionization potential required to produce Ne$^{7+}$ (207.28 e.V.) and
the relatively large cosmic abundance of Ne make the Ne\,VIII doublet
transitions a sensitive probe of warm-hot gas of $T=(0.5-1)\times
10^6$ K.  Savage \etal\ (2005) performed a detailed analysis of the
ionization state of the gas, taking into account the relative
abudances of additional highly ionized species such as S\,VI and O\,VI,
and concluded that the observations are best explained by a
collisional ionized gas of $T=5.4\times 10^5$ K.  While the detection
of Ne\,VIII at $z=0.207$ provides exciting support for the presence
of warm-hot gas, this system remains unique and continuing efforts to
search for more Ne\,VIII absorbers have uncovered no new systems
(e.g.\ Lehner \etal\ 2006).

  Our galaxy survey is most complete in the field around
HE\,0226$-$4110, reaching 100\% completeness for galaxies brighter
than $R=23$ at angular distances $\Delta\theta\le 2'$ (upper-left
panel of Figure 6).  The redshift distribution shown in the
bottom-left panel of Figure 6 displays clear galaxy overdensities at
$z=0.27$ and $z=0.4$ in front of the QSO.  While there are \lya\
absorbers present at these redshifts, no \ovi\ absorbers are found to
coincide with these large-scale galaxy overdensities to the limit of
$\log\,N(\ovi)=13.5$.  

Figure 8 shows the redshifts of galaxies in the inner $2.7'\times
2.7'$ field around the QSO.  We have been able to obtain high $S/N$
spectra of galaxies at angular distance as close as $\Delta\theta\apll
3''$ to the background bright QSO.  Galaxies with redshift coincide
with known O\,VI absorbers ($|\Delta\,v|<300$ \kms) are marked by a
rectangular box.  A complete photometric and spectroscopic catalog of
galaxies with $R\le 23$ and at angular distance $\Delta\theta\le 11'$
of the QSO is available electronically at
http://lambda.uchicago.edu/public/cat/cat\_0226.html.  An example of
the first 30 targets in the catalog is presented in Table 4, which
lists from columns (1) through (13) the object ID, the right ascension
(RA) and declination (Dec), the position offsets in RA
($\Delta\alpha$) and Dec ($\Delta\delta$) of the galaxy from the QSO,
the angular distance to the QSO ($\Delta\theta$), the projected
distance in physical $h^{-1}$ kpc, the $BRI$ magnitudes and
uncertainties, the spectroscopic redshift $z_{\rm spec}$ ($-1$
indicates an absence of spectroscopic redshift measurement), spectral
type, and rest-frame $R$-band absolute magnitude\footnote{The
  rest-frame $R$-band magnitude of each spectroscopically identified
  object was estimated based on the observed $R$-band magnitude and
  spectral type.  For absorption-line dominated galaxies, we evaluate
  the $k$-corretion using the early-type E/S0 and Sab galaxy templates
  from Coleman \etal\ (1980).  For emission-line dominated galaxies,
  we evaluate the $k$-correction using the Irr templates.}.

Of the five O\,VI absorbers found along the sightline toward
HE\,0226$-$4110, we have identified coincident galaxies at velocity
offsets $|\Delta\,v|<300$ \kms\ and projected physical distances
$\rho\le 250\ h^{-1}$ kpc for two absorbers, including the Ne\,VIII
absorber at $z=0.207$.  No galaxies are found near the \ovi\ absorber
at $z=0.01746$ (Lehner \etal\ 2006; Tripp \etal\ 2008), although our
spectroscopic survey only covers $\rho<160\ h^{-1}$ physical kpc at
this low redshift.

Three galaxies of $R=20.3-21.9$ are found within $|\Delta\,v|<300$
\kms\ and $\rho<200\ h^{-1}$ physical kpc of the \ovi\ and Ne\,VIII
absorber at $z=0.207$, all of which are sub-$L_*$ galaxies with
rest-frame $R$-band absolute magnitudes spanning a range from
$M_R-5\log\,h=-17.2$ to $M_R-5\log\,h=-18.9$.  The survey completness
rules out the presence of additional galaxies that are more luminous
than $M_R-5\log\,h=-16.1$ within $\rho=285\ h^{-1}$ kpc.  A detailed
analysis of the galactic environment of the Ne\,VIII absorber is
presented in Mulchaey \& Chen (2009).

The nearest galaxy to the \ovi\ absorber at $z=0.32629$ (Thom \& Chen
2008a,b) has $R=23.0$ and $\rho=381\ h^{-1}$ physical kpc, with a
corresponding rest-frame $R$-band absolute magnitude of
$M_R-5\log\,h=-17.7$.  The galaxy spectrum is dominated by absorption
features.  Three more galaxies are found at $|\Delta\,v|<300$ \kms\
from the absorber redshift, but they are over $\rho=840\ h^{-1}$ kpc
to $\rho=2.6\ h^{-1}$ Mpc physical distances away.  The survey
completness rules out the presence of additional star-forming galaxies
that are more luminous than $M_R-5\log\,h=-17.3$ within $\rho=396\
h^{-1}$ kpc.

The nearest galaxy to the \ovi\ absorber at $z=0.34034$ (Lehner \etal\
2006; Tripp \etal\ 2008; Thom \& Chen 2008a,b) has $R=22.1$ and
$\rho=213\ h^{-1}$ physical kpc, with a corresponding rest-frame
$R$-band absolute magnitude of $M_R-5\log\,h=-18.3$.  The galaxy
spectrum is dominated by emission features.  Two more galaxies are
found at $|\Delta\,v|<300$ \kms\ from the absorber redshift, and have
$\rho=432\ h^{-1}$ kpc and $\rho=643\ h^{-1}$ kpc physical distances,
respectively.  The survey completness rules out the presence of
additional galaxies that are more luminous than $M_R-5\log\,h=-17.4$
within $\rho=407\ h^{-1}$ kpc.

The nearest galaxy to the \ovi\ absorber at $z=0.35529$ (Lehner \etal\
2006; Tripp \etal\ 2008; Thom \& Chen 2008a,b) has $R=22.0$ and
$\rho=306\ h^{-1}$ physical kpc, with a corresponding rest-frame
$R$-band absolute magnitude of $M_R-5\log\,h=-18.5$.  The galaxy
spectrum is dominated by emission features.  Ten more galaxies are
found at $|\Delta\,v|<300$ \kms\ from the absorber redshift over a
physical distance range of $\rho=387\ h^{-1}$ kpc to $\rho=1.8\
h^{-1}$ Mpc.  The survey completness rules out the presence of
additional galaxies that are more luminous than $M_R-5\log\,h=-17.5$
within $\rho=419\ h^{-1}$ kpc.

\begin{deluxetable*}{lccrrrrcccrcr}
\tabletypesize{\tiny}
\tablewidth{0pt}
\tablecaption{An Example of the Photometric and Spectroscopic Catalog of Objects in the Field around HE0226$-$4110$^a$}
\tablehead{ &  &  & \multicolumn{1}{c}{$\Delta\,\alpha$} & \multicolumn{1}{c}{$\Delta\,\delta$} & \multicolumn{1}{c}{$\Delta\,\theta$} & 
\multicolumn{1}{c}{$\rho$} & & & & & & \multicolumn{1}{c}{$M_R$} \\
\multicolumn{1}{c}{ID} & \multicolumn{1}{c}{RA(J2000)} & \multicolumn{1}{c}{Dec(J2000)} & \multicolumn{1}{c}{($''$)} & 
\multicolumn{1}{c}{($''$)} & \multicolumn{1}{c}{($''$)} & \multicolumn{1}{c}{($h^{-1}$ kpc)} & 
\multicolumn{1}{c}{$B$} & \multicolumn{1}{c}{$R$} & \multicolumn{1}{c}{$I$} & \multicolumn{1}{c}{$z_{\rm spec}$} & \multicolumn{1}{c}{Type$^b$} & \multicolumn{1}{c}{$- 5\,\log\,h$} \\
\multicolumn{1}{c}{(1)} & \multicolumn{1}{c}{(2)} & \multicolumn{1}{c}{(3)} & \multicolumn{1}{c}{(4)} & 
\multicolumn{1}{c}{(5)} & \multicolumn{1}{c}{(6)} & \multicolumn{1}{c}{(7)} & 
\multicolumn{1}{c}{(8)} & \multicolumn{1}{c}{(9)} & \multicolumn{1}{c}{(10)} & \multicolumn{1}{c}{(11)} & \multicolumn{1}{c}{(12)} & \multicolumn{1}{c}{(13)}}
\startdata
00443 & 02:29:21.173 & $-40$:54:57.72 & $   743.2$ & $   132.2$ & $   754.9$ & $       -1.00$ & $    22.740\pm  0.034$ & $    20.199\pm  0.006$ & $    18.865\pm  0.002$ & $ -1.0000$ &     0 & $    0.00$ \nl 
00542 & 02:29:19.348 & $-40$:53:34.44 & $   723.2$ & $   215.3$ & $   754.5$ & $       -1.00$ & $    23.831\pm  0.066$ & $    22.208\pm  0.024$ & $    21.471\pm  0.021$ & $ -1.0000$ &     0 & $    0.00$ \nl 
00551 & 02:29:19.107 & $-40$:52:55.99 & $   720.7$ & $   253.5$ & $   764.0$ & $       -1.00$ & $    23.595\pm  0.048$ & $    22.316\pm  0.022$ & $    21.853\pm  0.025$ & $ -1.0000$ &     0 & $    0.00$ \nl 
00562 & 02:29:19.090 & $-40$:53:31.52 & $   720.3$ & $   218.2$ & $   752.6$ & $       -1.00$ & $    25.019\pm  0.200$ & $    22.435\pm  0.031$ & $    21.368\pm  0.020$ & $ -1.0000$ &     0 & $    0.00$ \nl 
00578 & 02:29:19.162 & $-40$:56:50.16 & $   719.9$ & $    20.5$ & $   720.2$ & $       -1.00$ & $    22.480\pm  0.024$ & $    20.453\pm  0.006$ & $    19.700\pm  0.005$ & $ -1.0000$ &     0 & $    0.00$ \nl 
00595 & 02:29:19.137 & $-40$:57:35.04 & $   719.4$ & $   -24.1$ & $   719.8$ & $       -1.00$ & $    23.495\pm  0.066$ & $    21.802\pm  0.023$ & $    20.779\pm  0.015$ & $ -1.0000$ &     0 & $    0.00$ \nl 
00604 & 02:29:18.789 & $-40$:57:29.05 & $   715.5$ & $   -18.1$ & $   715.7$ & $       -1.00$ & $    24.723\pm  0.147$ & $    22.970\pm  0.048$ & $    22.043\pm  0.036$ & $ -1.0000$ &     0 & $    0.00$ \nl 
00676 & 02:29:18.104 & $-40$:54:59.59 & $   708.6$ & $   130.6$ & $   720.6$ & $       -1.00$ & $    22.600\pm  0.035$ & $    19.784\pm  0.004$ & $    18.270\pm  0.002$ & $ -1.0000$ &     0 & $    0.00$ \nl 
00685 & 02:29:17.937 & $-40$:54:45.17 & $   706.8$ & $   145.0$ & $   721.6$ & $     2422.79$ & $    22.052\pm  0.021$ & $    20.544\pm  0.009$ & $    19.855\pm  0.008$ & $  0.3347$ &     2 & $  -19.81$ \nl 
00686 & 02:29:18.299 & $-40$:55:16.06 & $   710.7$ & $   114.2$ & $   719.9$ & $     2838.03$ & $    23.014\pm  0.047$ & $    21.567\pm  0.021$ & $    21.286\pm  0.021$ & $  0.4329$ &     2 & $  -19.45$ \nl 
00693 & 02:29:17.587 & $-40$:57:16.51 & $   702.0$ & $    -5.6$ & $   702.0$ & $       -1.00$ & $    25.241\pm  0.267$ & $    22.480\pm  0.035$ & $    21.038\pm  0.016$ & $ -1.0000$ &     0 & $    0.00$ \nl 
00694 & 02:29:17.554 & $-40$:58:39.99 & $   701.1$ & $   -88.6$ & $   706.7$ & $     2640.43$ & $    23.287\pm  0.051$ & $    21.406\pm  0.014$ & $    20.639\pm  0.012$ & $  0.3957$ &     1 & $  -19.87$ \nl 
00699 & 02:29:17.643 & $-40$:55:33.34 & $   703.3$ & $    97.1$ & $   709.9$ & $       -1.00$ & $    24.376\pm  0.160$ & $    22.318\pm  0.038$ & $    22.169\pm  0.047$ & $ -1.0000$ &     0 & $    0.00$ \nl 
00731 & 02:29:17.061 & $-40$:56:26.82 & $   696.4$ & $    43.9$ & $   697.8$ & $       -1.00$ & $    24.646\pm  0.137$ & $    22.940\pm  0.045$ & $    22.294\pm  0.036$ & $ -1.0000$ &     0 & $    0.00$ \nl 
00733 & 02:29:17.169 & $-40$:57:12.21 & $   697.3$ & $    -1.2$ & $   697.3$ & $       -1.00$ & $    22.885\pm  0.041$ & $    19.942\pm  0.005$ & $    18.311\pm  0.002$ & $ -1.0000$ &     0 & $    0.00$ \nl 
00734 & 02:29:16.844 & $-40$:55:31.43 & $   694.3$ & $    99.1$ & $   701.3$ & $       -1.00$ & $    24.860\pm  0.220$ & $    22.249\pm  0.031$ & $    21.780\pm  0.028$ & $ -1.0000$ &     0 & $    0.00$ \nl 
00766 & 02:29:16.521 & $-40$:52:29.86 & $   691.7$ & $   279.7$ & $   746.1$ & $     1806.70$ & $    21.877\pm  0.016$ & $    19.504\pm  0.003$ & $    18.689\pm  0.002$ & $  0.2125$ &     1 & $  -20.00$ \nl 
00768 & 02:29:16.668 & $-40$:55:35.82 & $   692.2$ & $    94.7$ & $   698.7$ & $       -1.00$ & $    24.080\pm  0.137$ & $    21.605\pm  0.021$ & $    20.676\pm  0.012$ & $ -1.0000$ &     0 & $    0.00$ \nl 
00808 & 02:29:15.823 & $-40$:58:33.06 & $   681.7$ & $   -81.6$ & $   686.6$ & $       -1.00$ & $    24.985\pm  0.197$ & $    21.960\pm  0.020$ & $    20.525\pm  0.009$ & $ -1.0000$ &     0 & $    0.00$ \nl 
00811 & 02:29:15.782 & $-40$:54:39.60 & $   682.6$ & $   150.7$ & $   699.0$ & $       -1.00$ & $    23.763\pm  0.060$ & $    22.377\pm  0.029$ & $    21.780\pm  0.028$ & $ -1.0000$ &     0 & $    0.00$ \nl 
00820 & 02:29:15.782 & $-40$:59:06.36 & $   681.0$ & $  -114.7$ & $   690.6$ & $     2935.54$ & $    23.482\pm  0.060$ & $    21.809\pm  0.020$ & $    20.971\pm  0.016$ & $  0.4951$ &     2 & $  -19.58$ \nl 
00830 & 02:29:15.386 & $-40$:52:03.47 & $   679.0$ & $   306.1$ & $   744.8$ & $       -1.00$ & $    24.118\pm  0.095$ & $    22.542\pm  0.036$ & $    21.686\pm  0.026$ & $ -1.0000$ &     0 & $    0.00$ \nl 
00855 & 02:29:15.283 & $-40$:59:43.43 & $   675.2$ & $  -151.6$ & $   692.0$ & $       -1.00$ & $    24.084\pm  0.095$ & $    22.935\pm  0.049$ & $    21.812\pm  0.029$ & $ -1.0000$ &     0 & $    0.00$ \nl 
00861 & 02:29:14.925 & $-40$:57:10.49 & $   672.1$ & $     0.6$ & $   672.1$ & $       -1.00$ & $    24.847\pm  0.172$ & $    22.933\pm  0.046$ & $    21.701\pm  0.025$ & $ -1.0000$ &     0 & $    0.00$ \nl 
00862 & 02:29:14.974 & $-40$:57:27.68 & $   672.5$ & $   -16.5$ & $   672.7$ & $       -1.00$ & $    25.048\pm  0.182$ & $    22.919\pm  0.042$ & $    21.489\pm  0.018$ & $ -1.0000$ &     0 & $    0.00$ \nl 
00863 & 02:29:14.971 & $-40$:52:05.68 & $   674.3$ & $   303.9$ & $   739.7$ & $       -1.00$ & $    24.107\pm  0.099$ & $    22.633\pm  0.041$ & $    22.033\pm  0.036$ & $ -1.0000$ &     0 & $    0.00$ \nl 
00871 & 02:29:15.132 & $-40$:57:02.04 & $   674.4$ & $     9.0$ & $   674.5$ & $       -1.00$ & $    24.085\pm  0.081$ & $    22.600\pm  0.034$ & $    21.911\pm  0.029$ & $ -1.0000$ &     0 & $    0.00$ \nl 
00875 & 02:29:15.616 & $-40$:59:51.01 & $   678.9$ & $  -159.1$ & $   697.3$ & $       -1.00$ & $    22.372\pm  0.037$ & $    20.852\pm  0.015$ & $    20.062\pm  0.012$ & $ -1.0000$ &     0 & $    0.00$ \nl 
00879 & 02:29:15.087 & $-40$:55:48.33 & $   674.4$ & $    82.4$ & $   679.4$ & $       -1.00$ & $    23.696\pm  0.076$ & $    22.361\pm  0.035$ & $    21.305\pm  0.022$ & $ -1.0000$ &     0 & $    0.00$ \nl 
00885 & 02:29:14.503 & $-40$:52:28.29 & $   668.9$ & $   281.4$ & $   725.7$ & $       -1.00$ & $    25.539\pm  0.282$ & $    22.831\pm  0.038$ & $    21.314\pm  0.014$ & $ -1.0000$ &     0 & $    0.00$ \nl 
\enddata \tablenotetext{a}{A complete catalog is available
electronically at
http://lambda.uchicago.edu/public/cat/cat\_0226.html.}
\tablenotetext{b}{Spectral type of the galaxy: 1 $\rightarrow$ absorption-line dominated and 2 $\rightarrow$ emission-line dominated.}
\end{deluxetable*}

\subsection{The Field toward PKS0405$-$123 at $z_{\rm QSO}=0.573$}

The sightline toward PKS\,0405$-$123 exhibits 76 \lya\ absorbers
(Lehner \etal\ 2006) and six \ovi\ absorbers (Prochaska \etal\ 2004;
Tripp \etal\ 2008; Thom \& Chen 2008a,b) over the redshift range from
$z_{\lya}=0.012$ to $z_{\lya}=0.538$.  The column density of these
\lya\ absorbers span a range from $\log\,N(\hI)=12.5$ to
$\log\,N(\hI)=16.3$, and the column density of the \ovi\ absorbers
span a range from $\log\,N(\ovi)=13.5$ to $\log\,N(\ovi)=14.7$.  

The field around PKS\,0405$-$123 has been surveyed by Prochaska \etal\
(2006).  This previous survey has yielded robust redshifts for 95\% of
all galaxies brighter than $R=20$, corresponding to roughly $L_*$
galaxies at $z=0.5$.  Our program expands upon the earlier
spectroscopic survey for uncovering fainter galaxies along the QSO
line of sight, reaching 100\% completeness at $R=20$ and $> 70$\% at
$R=22$ within $\Delta\theta\le 2'$ (upper-middle panel of Figure 6).
The redshift distribution shown in the bottom-middle panel of Figure 6
displays numerous galaxy overdensities in front of the QSO, but only
one \ovi\ absorber is found to coincide with the large-scale galaxy
overdensity at $z=0.0966$.

Figure 9 displays a combined image of the field around
PKS\,0405$-$123, covering roughly a $2.7'\times 2.7'$ area.  Galaxies
with known spectroscopic redshifts are indicated by their redshifts to
the left.  Blue values represent redshift measurements obtained by
previous authors (Spinrad \etal\ 1993; Prochaska \etal\ 2006).  Red
values represent new redshifts obtained in our survey.  Galaxies with
redshift coincident with known O\,VI absorbers ($|\Delta\,v|<300$ \kms)
are marked by a rectangular box.  While this field contains a
relatively high surface density of galaxies, the majority of the
galaxies turn out to reside in the QSO host environment.  A complete
photometric and spectroscopic catalog of galaxies with $R\le 22$ and
at angular distance $\Delta\theta\le 11'$ of the QSO is available
electronically at
http://lambda.uchicago.edu/public/cat/cat\_0405.html.  An example of
the first 30 targets in the catalog is presented in Table 5.

\begin{figure*}
\begin{center}
\includegraphics[scale=0.5]{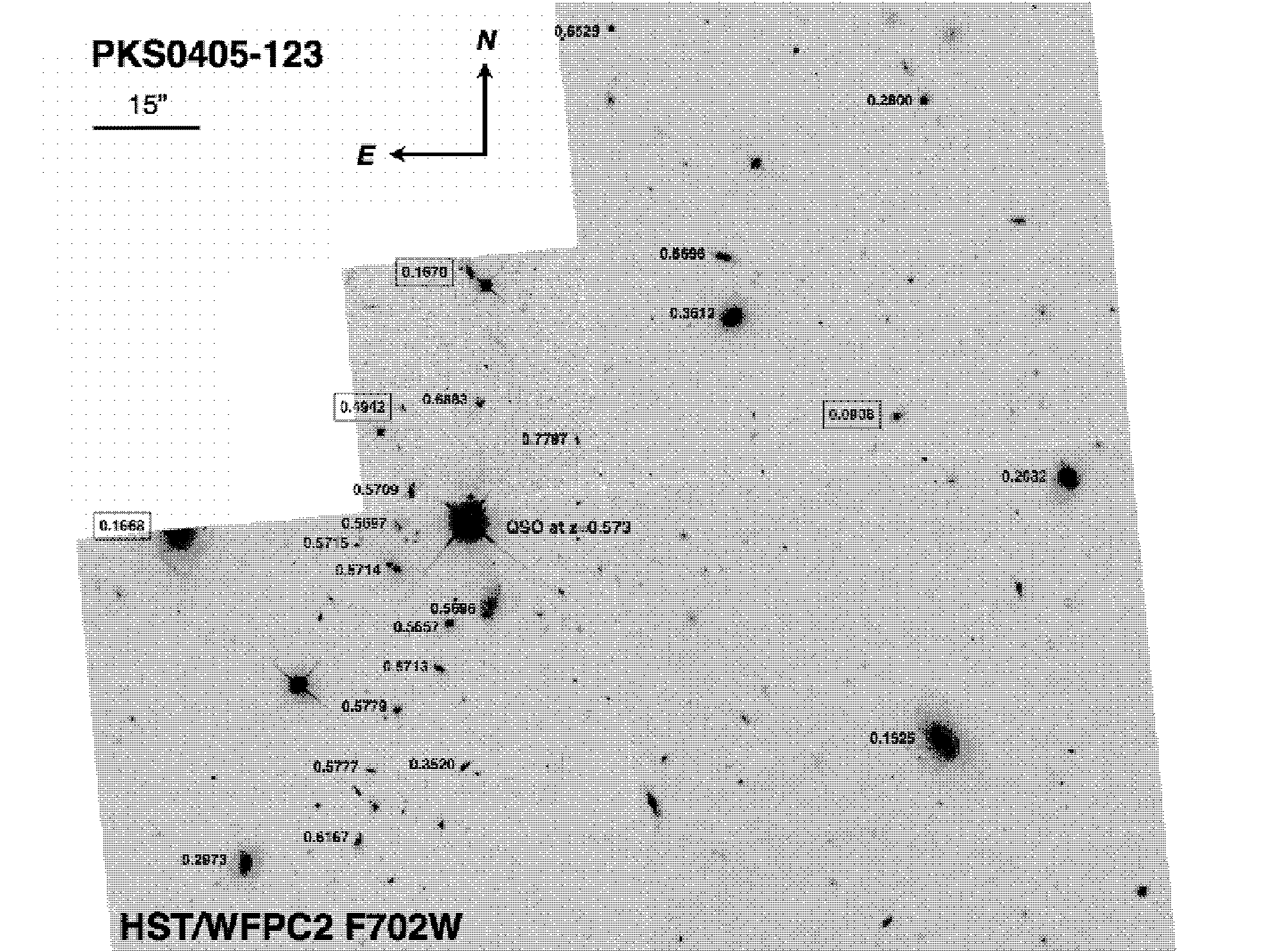}
\caption{A combined HST image obtained using WFPC2 and the F702W
filter.  The image is roughly $2.7'$ on a side.  Galaxies with known
spectroscopic redshifts are indicated by their redshifts to the left.
Blue values represent redshift measurements obtained by previous
authors (Spinrad \etal\ 1993; Prochaska \etal\ 2006).  Red values
represent new redshifts obtained in our survey.  Similar to Figure 8,
galaxies with redshift coincident with known O\,VI absorbers
($|\Delta\,v|<300$ \kms) are marked by a rectangular box.}
\end{center} 
\end{figure*}

Of the six O\,VI absorbers found along the sightline toward
PKS\,0405$-$123, four are identified with coincident galaxies at
$|\Delta\,v|<300$ \kms\ and $\rho\le 250\ h^{-1}$ kpc.  For the \ovi\
absorber at $z=0.0918$ (Prochaska \etal\ 2004), the closest galaxy
found in previous surveys was at $\rho=306\ h^{-1}$ kpc with $R=19.7$
and $\Delta\,v=-341$ \kms\ (Prochaska \etal\ 2006).  We have
identified three new galaxies of $R=21.3-21.9$ at $|\Delta\,v|=27-275$
\kms\ and $\rho=74-209 \ h^{-1}$ kpc from the absorber, including one
that is observed in the combined HST/WFPC2 image (left panel of Figure
10).  All three galaxies are faint dwarfs with emission-line dominated
spectral features (see the top three panels of Figure 15 below).  The
corresponding rest-frame $R$-band absolute magnitudes span a range
from $M_R-5\log\,h^{-1}=-15.4$ to $M_R-5\log\,h^{-1}=-16.1$.  The HST
image presented in Figure 10 shows that the dwarf galaxy at
$\rho=73.5\ h^{-1}$ kpc exhibits a compact core with low-surface
brightness emission extended to the north of the galaxy.

Only one galaxy is found within $|\Delta\,v|<300$ \kms\ and $\rho<200\
h^{-1}$ physical kpc of the \ovi\ absorber at $z=0.09658$ (Prochaska
\etal\ 2004).  The galaxy has $R=19.0$, corresponding to
$M_R-5\log\,h=-18.2$ (Prochaska \etal\ 2006).  Eight additional
galaxies are found at $|\Delta\,v|<300$ \kms\ from the absorber
redshift and between $\rho=257\ h^{-1}$ kpc to $\rho=945\ h^{-1}$ kpc
physical distances away.

The strong \ovi\ absorber of $\log\,N(\ovi)=14.7$ at $z=0.1671$ has
been studied extensively by Chen \& Prochaska (2000) and Prochaska
\etal\ (2004).  Two galaxies of $R=17.43$ and $R=21.0$ are found
within $|\Delta\,v|<200$ \kms\ and $\rho<100\ h^{-1}$ physical kpc of
the \ovi\ absorber at $z=0.1670$ (see the discussions in Chen \&
Prochaska 2000 and Prochaska \etal\ 2006).  No additional galaxies are
found within $\rho=1\ h^{-1}$ Mpc of the absorber.  The luminous
galaxy at $\Delta\,\delta=40''$ (corresponding to $\rho=80.9\ h^{-1}$
kpc) from the QSO has $M_R-5\log\,h=-20.9$ and exhibits post
starburst spectral features (Prochaska \etal\ 2006).  The WFPC2 image
presented in Figure 9 covers only part of the galaxy.  A detailed
examination of the WFPC2 image shows regular spiral structures with
several compact H\,II regions along the spiral arms (center-left panel
of Figure 10).  The dwarf galaxy at $\rho=67.8\ h^{-1}$ kpc has
$M_R-5\log\,h=-18.1$.  The disk-like morphology appears to be mildly
distburbed by a faint companion at the northeast edge (center-right
panel of Figure 10).

\begin{figure*}
\begin{center}
\includegraphics[scale=0.5]{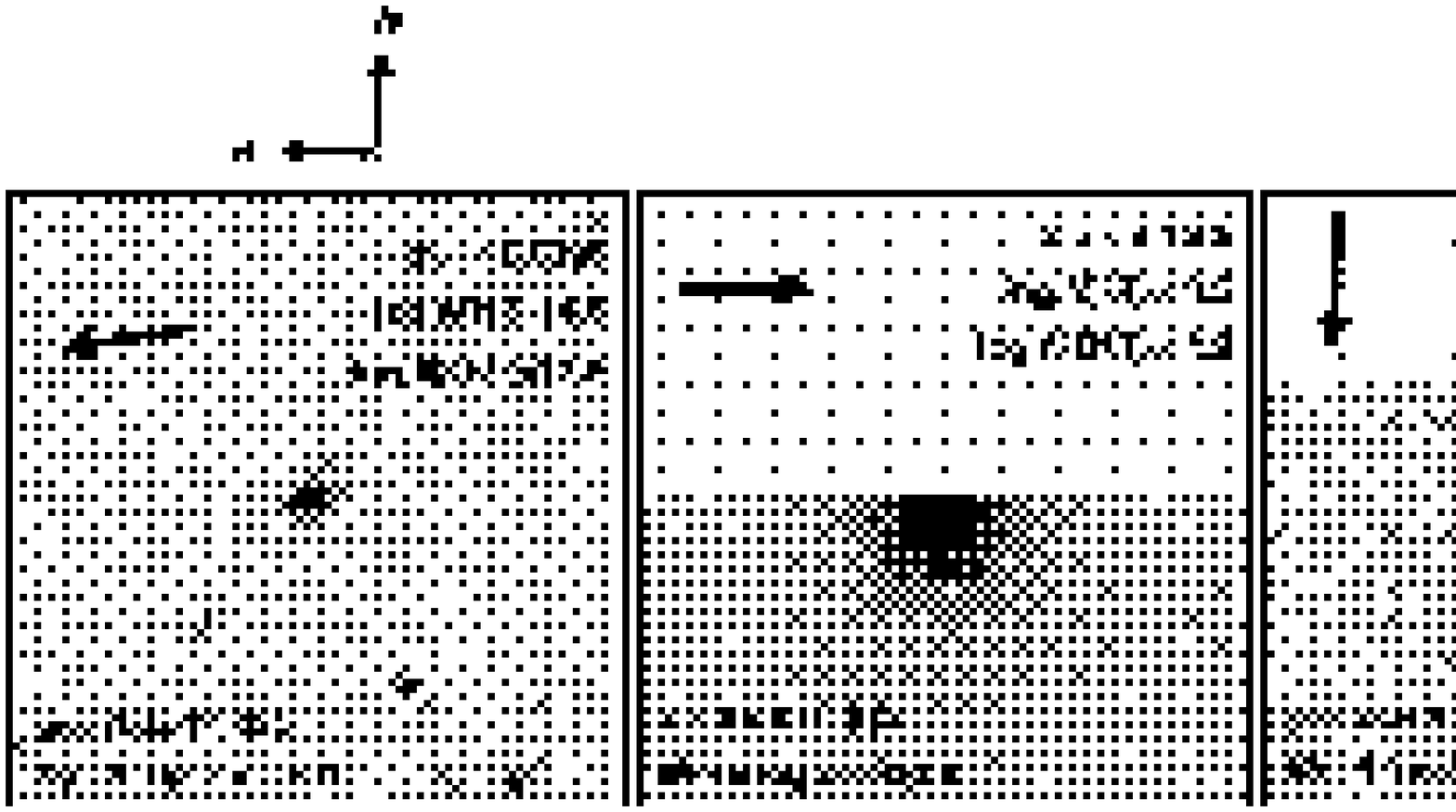}
\caption{Optical morphologies of galaxies that are associated with
known \ovi\ absorbers along the sightline toward PKS\,0405$-$123.
Each panel is a subset of the combined WFPC2/F702W image presented in
Figure 9.  The absorbing galaxy is located at the center of each
panel, which corresponds to roughly $25\ h^{-1}$ physical kpc on a
side at the redshift of the galaxy.  The arrow in each panel indicates
the direction toward the QSO sightline.}
\end{center} 
\end{figure*}

No galaxies are found near the \ovi\ absorber at $z=0.1829$ (Prochaska
\etal\ 2004; Tripp \etal\ 2008; Thom \& Chen 2008a,b).  The high
completness of our survey to $R=22$ within $\Delta\theta=2'$ suggests
that the galaxies associated with the \ovi\ absorber is likely fainter
than $M_R-5\log\,h=-16.6$.

No galaxies are found near the \ovi\ absorber at $z=0.3633$ (Prochaska
\etal\ 2004; Tripp \etal\ 2008; Thom \& Chen 2008a,b).  The high
completness of our survey to $R=22$ within $\Delta\theta=2'$ suggests
that the galaxies associated with the \ovi\ absorber is likely fainter
than $M_R-5\log\,h=-18.5$.

The strong \ovi\ absorber of $\log\,N(\ovi)=14.5$ at $z=0.4951$ has
been studied extensively by Howk \etal\ (2009).  The early search by
Prochaska \etal\ (2006) did not uncover any galaxies within $\rho=2.6\
h^{-1}$ Mpc.  Our spectroscopic survey has uncovered a galaxy of
$R=22.63$ at $\Delta\,v=-181$ \kms\ and $\rho=77\ h^{-1}$ physical kpc
from the absorber, with a corresponding rest-frame $R$-band absolute
magnitude of $M_R-5\log\,h=-18.8$.  The spectrum is dominated by
emission-line features due to [O\,II], H$\gamma$, H$\beta$, and
[O\,III] (see the 4th panel in Figure 15 below).  The morphology
appears to be extended in the HST image in Figure 10 (right panel).  No
additional galaxies are found at $|\Delta\,v|<300$ \kms\ and $\rho<1\
h^{-1}$ Mpc.

\begin{deluxetable*}{lccrrrrcrcr}
\tabletypesize{\tiny}
\tablewidth{0pt}
\tablecaption{An Example of the Photometric and Spectroscopic Catalog of Objects in the Field around PKS0405$-$123$^a$}
\tablehead{ &  &  & \multicolumn{1}{c}{$\Delta\,\alpha$} & \multicolumn{1}{c}{$\Delta\,\delta$} & \multicolumn{1}{c}{$\Delta\,\theta$} & 
\multicolumn{1}{c}{$\rho$} & & & & \multicolumn{1}{c}{$M_R$} \\
\multicolumn{1}{c}{ID} & \multicolumn{1}{c}{RA(J2000)} & \multicolumn{1}{c}{Dec(J2000)} & \multicolumn{1}{c}{($''$)} & 
\multicolumn{1}{c}{($''$)} & \multicolumn{1}{c}{($''$)} & \multicolumn{1}{c}{($h^{-1}$ kpc)} & 
\multicolumn{1}{c}{$R$} & \multicolumn{1}{c}{$z_{\rm spec}$} & \multicolumn{1}{c}{Type$^b$} & \multicolumn{1}{c}{$- 5\,\log\,h$} \\
\multicolumn{1}{c}{(1)} & \multicolumn{1}{c}{(2)} & \multicolumn{1}{c}{(3)} & \multicolumn{1}{c}{(4)} & 
\multicolumn{1}{c}{(5)} & \multicolumn{1}{c}{(6)} & \multicolumn{1}{c}{(7)} & 
\multicolumn{1}{c}{(8)} & \multicolumn{1}{c}{(9)} & \multicolumn{1}{c}{(10)} & \multicolumn{1}{c}{(11)}}
\startdata
00071 & 04:08:33.10 & $-12$:10:48.0 & $   655.0$ & $    48.7$ &    656.8 & $       -1.00$ & $    21.340\pm  0.110$ &  -1.0000 &     0 & $    0.00$ \nl 
00078$^c$ & 04:08:32.80 & $-12$:13: 9.0 & $   650.5$ & $   -92.3$ &    657.0 & $     2980.30$ & $    20.510\pm  0.090$ &   0.5639 &     1 & $  -22.03$ \nl 
00089 & 04:08:32.50 & $-12$:13:42.0 & $   646.0$ & $  -125.3$ &    658.1 & $       -1.00$ & $    20.470\pm  0.080$ &  -1.0000 &     0 & $    0.00$ \nl 
00107 & 04:08:32.10 & $-12$:14:17.0 & $   640.2$ & $  -160.3$ &    659.9 & $       -1.00$ & $    19.740\pm  0.070$ &  -1.0000 &     0 & $    0.00$ \nl 
00108 & 04:08:32.30 & $-12$:11:06.0 & $   643.2$ & $    30.7$ &    643.9 & $       -1.00$ & $    21.230\pm  0.100$ &  -1.0000 &     0 & $    0.00$ \nl 
00114 & 04:08:32.20 & $-12$:10:39.0 & $   641.8$ & $    57.7$ &    644.3 & $       -1.00$ & $    21.290\pm  0.090$ &  -1.0000 &     0 & $    0.00$ \nl 
00120 & 04:08:32.00 & $-12$:09:42.0 & $   638.9$ & $   114.7$ &    649.1 & $       -1.00$ & $    16.970\pm  0.060$ &  -1.0000 &     0 & $    0.00$ \nl 
00123 & 04:08:31.80 & $-12$:12:32.0 & $   635.8$ & $   -55.3$ &    638.2 & $       -1.00$ & $    18.090\pm  0.060$ &  -1.0000 &     0 & $    0.00$ \nl 
00129 & 04:08:31.70 & $-12$:14:01.0 & $   634.3$ & $  -144.3$ &    650.5 & $       -1.00$ & $    20.810\pm  0.090$ &  -1.0000 &     0 & $    0.00$ \nl 
00138 & 04:08:31.40 & $-12$:10:44.0 & $   630.0$ & $    52.7$ &    632.2 & $       -1.00$ & $    17.290\pm  0.060$ &  -1.0000 &     0 & $    0.00$ \nl 
00142$^c$ & 04:08:32.00 & $-12$:14:10.0 & $   638.7$ & $  -153.3$ &    656.8 & $     2981.03$ & $    20.890\pm  0.090$ &   0.5645 &     2 & $  -20.86$ \nl 
00144 & 04:08:31.20 & $-12$:13:59.0 & $   627.0$ & $  -142.3$ &    642.9 & $       -1.00$ & $    21.320\pm  0.090$ &  -1.0000 &     0 & $    0.00$ \nl 
00145 & 04:08:31.20 & $-12$:11:14.0 & $   627.1$ & $    22.7$ &    627.5 & $       -1.00$ & $    17.570\pm  0.060$ &  -1.0000 &     0 & $    0.00$ \nl 
00149 & 04:08:31.00 & $-12$:14:55.0 & $   624.0$ & $  -198.3$ &    654.8 & $       -1.00$ & $    21.580\pm  0.120$ &  -1.0000 &     0 & $    0.00$ \nl 
00153 & 04:08:31.00 & $-12$:13:02.0 & $   624.1$ & $   -85.3$ &    629.9 & $       -1.00$ & $    21.600\pm  0.110$ &  -1.0000 &     0 & $    0.00$ \nl 
00162 & 04:08:31.00 & $-12$:11:01.0 & $   624.2$ & $    35.7$ &    625.2 & $       -1.00$ & $    21.890\pm  0.130$ &  -1.0000 &     0 & $    0.00$ \nl 
00163 & 04:08:30.70 & $-12$:13:09.0 & $   619.7$ & $   -92.3$ &    626.5 & $       -1.00$ & $    21.370\pm  0.100$ &  -1.0000 &     0 & $    0.00$ \nl 
00165 & 04:08:30.80 & $-12$:11:30.0 & $   621.2$ & $     6.7$ &    621.2 & $       -1.00$ & $    17.600\pm  0.060$ &  -1.0000 &     0 & $    0.00$ \nl 
00169 & 04:08:30.30 & $-12$:15:00.0 & $   613.7$ & $  -203.3$ &    646.5 & $       -1.00$ & $    16.240\pm  0.060$ &  -1.0000 &     0 & $    0.00$ \nl 
00174 & 04:08:30.60 & $-12$:10:56.0 & $   618.3$ & $    40.7$ &    619.6 & $       -1.00$ & $    19.140\pm  0.070$ &  -1.0000 &     0 & $    0.00$ \nl 
00175 & 04:08:30.80 & $-12$:08:27.0 & $   621.3$ & $   189.7$ &    649.6 & $       -1.00$ & $    21.220\pm  0.090$ &  -1.0000 &     0 & $    0.00$ \nl 
00179 & 04:08:30.60 & $-12$:11:14.0 & $   618.3$ & $    22.7$ &    618.7 & $       -1.00$ & $    21.590\pm  0.110$ &  -1.0000 &     0 & $    0.00$ \nl 
00190 & 04:08:30.60 & $-12$:08:09.0 & $   618.4$ & $   207.7$ &    652.3 & $       -1.00$ & $    21.760\pm  0.120$ &  -1.0000 &     0 & $    0.00$ \nl 
00191 & 04:08:30.00 & $-12$:13:17.0 & $   609.4$ & $  -100.3$ &    617.6 & $       -1.00$ & $    21.980\pm  0.130$ &  -1.0000 &     0 & $    0.00$ \nl 
00203 & 04:08:29.60 & $-12$:13:26.0 & $   603.5$ & $  -109.3$ &    613.4 & $       -1.00$ & $    20.830\pm  0.080$ &  -1.0000 &     0 & $    0.00$ \nl 
00209 & 04:08:29.80 & $-12$:09:00.0 & $   606.6$ & $   156.7$ &    626.5 & $       -1.00$ & $    19.400\pm  0.070$ &  -1.0000 &     0 & $    0.00$ \nl 
00210 & 04:08:29.90 & $-12$:08:04.0 & $   608.1$ & $   212.7$ &    644.2 & $       -1.00$ & $    20.800\pm  0.090$ &  -1.0000 &     0 & $    0.00$ \nl 
00213 & 04:08:29.40 & $-12$:15:16.0 & $   600.5$ & $  -219.3$ &    639.3 & $       -1.00$ & $    21.170\pm  0.100$ &  -1.0000 &     0 & $    0.00$ \nl 
00214 & 04:08:29.50 & $-12$:11:19.0 & $   602.2$ & $    17.7$ &    602.4 & $       -1.00$ & $    21.560\pm  0.110$ &  -1.0000 &     0 & $    0.00$ \nl 
00221$^d$ & 04:08:29.00 & $-12$:13:28.0 & $   594.7$ & $  -111.3$ &    605.1 & $     2079.39$ & $    19.400\pm  0.070$ &   0.3466 &     1 & $  -20.41$ \nl 
\enddata
\tablenotetext{a}{A complete catalog is available electronically at http://lambda.uchicago.edu/public/cat/cat\_0405.html.  The Object ID, coordinates, and $R$-band photometry are adopted from Prochaska \etal\ (2006).}
\tablenotetext{b}{Spectral type of the galaxy: 1 $\rightarrow$ absorption-line dominated and 2 $\rightarrow$ emission-line dominated.}
\tablenotetext{c}{New Spectroscopic redshifts obtained in our survey.}
\tablenotetext{d}{Spectroscopic redshifts published in Prochaska \etal\ (2006).}
\end{deluxetable*}

\subsection{The Field toward PG1216$+$069 at $z_{\rm QSO}=0.3313$}

The search of \lya\ absorbers described in \S\ 5 has identified 66
\lya\ absorbers along the sightline toward PG\,1216$+$069 over the
redshift range from $z_{\lya}=0.0036$ to $z_{\lya}=0.321$ (Table 3).
The column density of these \lya\ absorbers span a range from
$\log\,N(\hI)=12.6$ to $\log\,N(\hI)=19.3$.  Two \ovi\ absorbers are
found at $z=0.1242$ and $z=0.2823$ with $\log\,N(\ovi)=14.7$ and
$\log\,N(\ovi)=13.4$, respectively (Tripp \etal\ 2008; Thom \& Chen
2008a,b)\footnote{We note that Tripp \etal\ (2008) reported one more
\ovi\ absorber at $z=0.26768$ along this sightline.  However, the
expected O\,VI 1037 transition falls at a wavelength that has been
contaminated by a saturated \lyb\ feature at $z=0.2819$.  No other
metal absorption lines are found for the absorber to confirm the
nature of the presumed O\,VI 1031 feature.  We therefore consider this
absorption feature as \lya\ at $z=0.0760$.}.

  Our galaxy survey in the field around PG\,1216$+$069 in the least
incomplete of all three, reaching $>80$\% completeness for galaxies
brighter than $R=21$ at angular distances $\Delta\theta\le 2'$
(upper-right panel of Figure 6).  The redshift distribution shown in
the bottom-right panel of Figure 6 displays a significant galaxy
overdensity at $z=0.26-0.28$ in front of the QSO.  One of the two
\ovi\ absorbers does coincide with this large-scale galaxy
overdensity.  Figure 11 shows the redshifts of galaxies in the inner
$2.7'\times 2.7'$ field around the QSO.  We have not been able to
obtain high quality spectra of galaxies at angular distance
$\Delta\theta\apll 5''$ due to the presence of a bright star and the
background QSO.  Only one foreground galaxy is found at $\rho<200\
h^{-1}$ kpc.  A complete photometric and spectroscopic catalog of
galaxies with $R\le 22$ and at angular distance $\Delta\theta\le 11'$
of the QSO is available electronically at
http://lambda.uchicago.edu/public/cat/cat\_1216.html.  An example of
the first 30 targets in the catalog is presented in Table 6.

\begin{figure*}
\begin{center}
\includegraphics[scale=0.5]{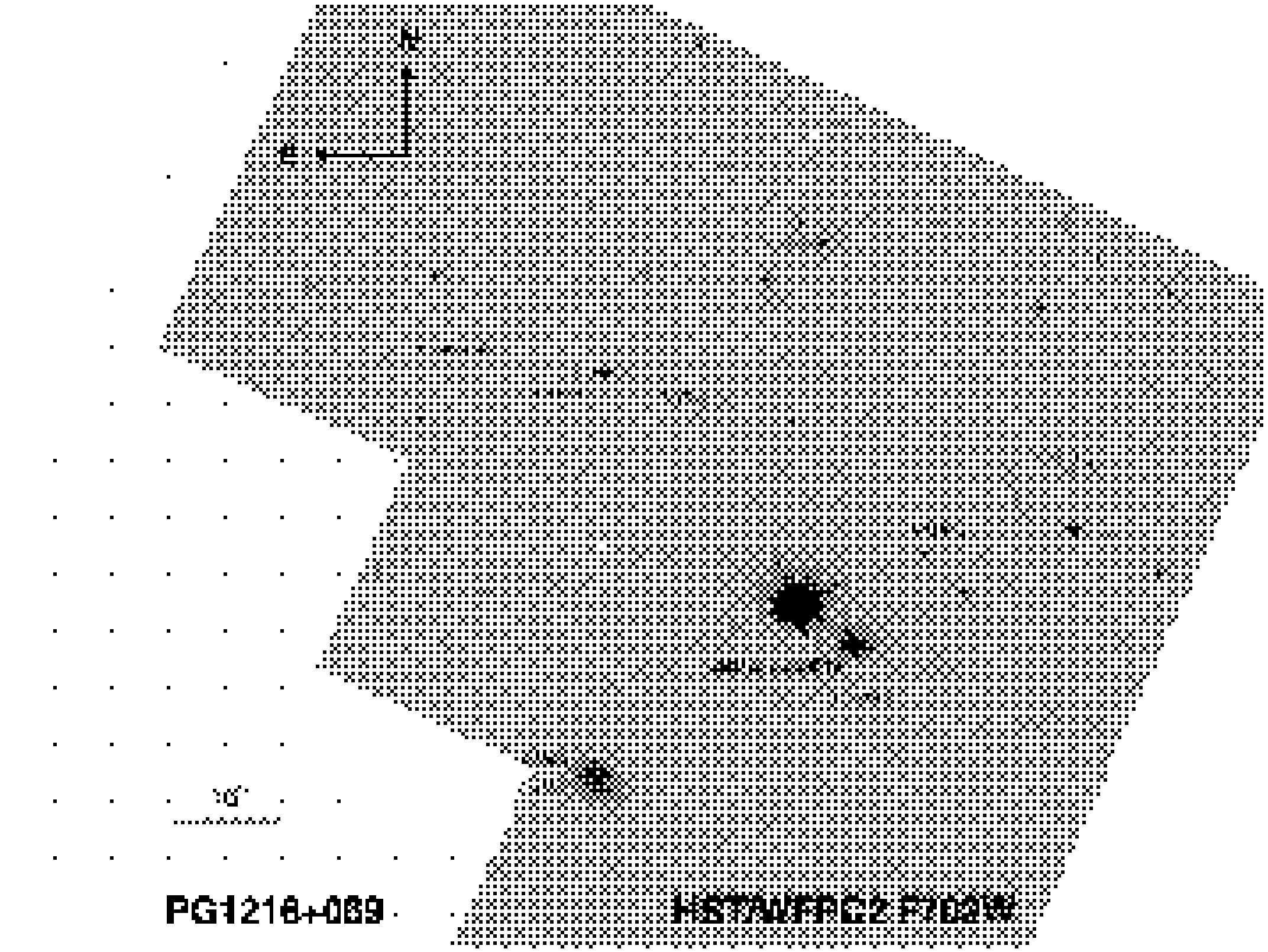}
\caption{A combined HST image obtained using WFPC2 and the F702W
filter.  The image is roughly $2.7'$ on a side.  Galaxies with known
spectroscopic redshifts are indicated by their redshifts to the left.
Known redshifts obtained prior to our survey are marked in blue (Chen
\etal\ 2001a).  Red values represent our own measurements.  The galaxy
with a redshift coincident with the O\,VI absorber at $z=0.1236$ is
marked by a rectangular box.  The compact source at NE of the galaxy
turns out to be a star.  No additional foreground galaxies have been
found at $\rho<200\ h^{-1}$ kpc.}
\end{center} 
\end{figure*}

\begin{deluxetable*}{lccrrrrcccrcr}
\tabletypesize{\tiny}
\tablewidth{0pt}
\tablecaption{An Example of the Photometric and Spectroscopic Catalog of Objects in the Field around PG1216$+$069$^a$}
\tablehead{ &  &  & \multicolumn{1}{c}{$\Delta\,\alpha$} & \multicolumn{1}{c}{$\Delta\,\delta$} & \multicolumn{1}{c}{$\Delta\,\theta$} & 
\multicolumn{1}{c}{$\rho$} & & & & & & \multicolumn{1}{c}{$M_R$} \\
\multicolumn{1}{c}{ID} & \multicolumn{1}{c}{RA(J2000)} & \multicolumn{1}{c}{Dec(J2000)} & \multicolumn{1}{c}{($''$)} & 
\multicolumn{1}{c}{($''$)} & \multicolumn{1}{c}{($''$)} & \multicolumn{1}{c}{($h^{-1}$ kpc)} & 
\multicolumn{1}{c}{$B$} & \multicolumn{1}{c}{$R$} & \multicolumn{1}{c}{$I$} & \multicolumn{1}{c}{$z_{\rm spec}$} & \multicolumn{1}{c}{Type$^b$} & \multicolumn{1}{c}{$- 5\,\log\,h$} \\
\multicolumn{1}{c}{(1)} & \multicolumn{1}{c}{(2)} & \multicolumn{1}{c}{(3)} & \multicolumn{1}{c}{(4)} & 
\multicolumn{1}{c}{(5)} & \multicolumn{1}{c}{(6)} & \multicolumn{1}{c}{(7)} & 
\multicolumn{1}{c}{(8)} & \multicolumn{1}{c}{(9)} & \multicolumn{1}{c}{(10)} & \multicolumn{1}{c}{(11)} & \multicolumn{1}{c}{(12)} & \multicolumn{1}{c}{(13)}}
\startdata
00539 & 12:18:48.159 & $+$06:32:41.11 & $  -484.7$ & $  -358.0$ & $   602.6$ & $       -1.00$ & $    18.723\pm  0.023$ & $    16.923\pm  0.005$ & $    16.654\pm  0.005$ &  $-1.0000$ &     0 & $    0.00$ \nl 
01375 & 12:19:20.821 & $+$06:27:37.63 & $     0.9$ & $  -657.5$ & $   657.5$ & $     2023.15$ & $    21.500\pm  0.014$ & $    20.237\pm  0.005$ & $    19.765\pm  0.006$ &   0.2943 &     2 & $  -19.79$ \nl 
01419 & 12:19:21.951 & $+$06:27:38.90 & $    17.7$ & $  -656.2$ & $   656.4$ & $       -1.00$ & $    21.148\pm  0.012$ & $    18.744\pm  0.002$ & $    17.645\pm  0.001$ &  $-1.0000$ &     0 & $    0.00$ \nl 
01454 & 12:19:21.826 & $+$06:27:48.01 & $    15.8$ & $  -647.1$ & $   647.3$ & $       -1.00$ & $    22.316\pm  0.019$ & $    21.763\pm  0.015$ & $    21.491\pm  0.021$ &  $-1.0000$ &     0 & $    0.00$ \nl 
01567 & 12:19:14.802 & $+$06:27:54.42 & $   -88.4$ & $  -641.2$ & $   647.3$ & $     2551.46$ & $    22.493\pm  0.026$ & $    21.123\pm  0.010$ & $    20.820\pm  0.012$ &   0.4328 &     2 & $  -19.89$ \nl 
01591 & 12:19:13.861 & $+$06:27:58.57 & $  -102.4$ & $  -637.2$ & $   645.3$ & $       -1.00$ & $    23.090\pm  0.046$ & $    20.541\pm  0.006$ & $    19.378\pm  0.004$ &  $-1.0000$ &     0 & $    0.00$ \nl 
01610 & 12:19:27.817 & $+$06:27:50.45 & $   104.6$ & $  -644.3$ & $   652.7$ & $     1357.74$ & $    20.959\pm  0.015$ & $    19.128\pm  0.004$ & $    18.510\pm  0.004$ &   0.1753 &     2 & $  -19.67$ \nl 
01619 & 12:19:10.432 & $+$06:27:56.42 & $  -153.2$ & $  -639.5$ & $   657.6$ & $       -1.00$ & $    22.830\pm  0.040$ & $    21.748\pm  0.019$ & $    21.337\pm  0.024$ &  $-1.0000$ &     0 & $    0.00$ \nl 
01621 & 12:19:24.357 & $+$06:27:49.33 & $    53.3$ & $  -645.6$ & $   647.8$ & $     2044.82$ & $    21.010\pm  0.011$ & $    19.615\pm  0.004$ & $    19.123\pm  0.004$ &   0.3054 &     2 & $  -20.50$ \nl 
01693 & 12:19:28.263 & $+$06:27:56.16 & $   111.2$ & $  -638.6$ & $   648.2$ & $       -1.00$ & $    22.168\pm  0.038$ & $    20.307\pm  0.009$ & $    19.929\pm  0.012$ &  $-1.0000$ &     0 & $    0.00$ \nl 
01740 & 12:19:27.316 & $+$06:27:58.82 & $    97.1$ & $  -636.0$ & $   643.4$ & $       -1.00$ & $    18.688\pm  0.002$ & $    17.676\pm  0.001$ & $    17.384\pm  0.001$ &  $-1.0000$ &     0 & $    0.00$ \nl 
01785 & 12:19:26.316 & $+$06:28:10.08 & $    82.3$ & $  -624.9$ & $   630.2$ & $       -1.00$ & $    22.555\pm  0.034$ & $    20.485\pm  0.006$ & $    19.921\pm  0.005$ &  $-1.0000$ &     0 & $    0.00$ \nl 
01794 & 12:19:12.163 & $+$06:28:13.40 & $  -127.6$ & $  -622.5$ & $   635.5$ & $       -1.00$ & $    22.901\pm  0.029$ & $    21.660\pm  0.013$ & $    21.428\pm  0.020$ &  $-1.0000$ &     0 & $    0.00$ \nl 
01814 & 12:19:23.459 & $+$06:27:58.51 & $    40.0$ & $  -636.6$ & $   637.8$ & $     1979.71$ & $    21.963\pm  0.028$ & $    20.423\pm  0.009$ & $    19.919\pm  0.010$ &   0.2981 &     2 & $  -19.64$ \nl 
01819 & 12:19:18.003 & $+$06:28:07.61 & $   -41.0$ & $  -627.9$ & $   629.2$ & $     1953.51$ & $    22.342\pm  0.034$ & $    20.752\pm  0.010$ & $    20.270\pm  0.012$ &   0.2981 &     2 & $  -19.31$ \nl 
01852 & 12:19:15.205 & $+$06:28:10.43 & $   -82.5$ & $  -625.3$ & $   630.7$ & $     2756.59$ & $    23.582\pm  0.111$ & $    20.399\pm  0.008$ & $    19.560\pm  0.006$ &   0.5220 &     1 & $  -21.84$ \nl 
01875 & 12:19:25.887 & $+$06:28:17.02 & $    75.9$ & $  -618.0$ & $   622.6$ & $       -1.00$ & $    22.975\pm  0.035$ & $    21.821\pm  0.017$ & $    21.236\pm  0.016$ &  $-1.0000$ &     0 & $    0.00$ \nl 
01897 & 12:19:32.437 & $+$06:28:15.65 & $   173.0$ & $  -618.9$ & $   642.6$ & $       -1.00$ & $    22.797\pm  0.046$ & $    21.590\pm  0.020$ & $    21.289\pm  0.024$ &  $-1.0000$ &     0 & $    0.00$ \nl 
01898 & 12:19:09.029 & $+$06:28:19.40 & $  -174.1$ & $  -616.8$ & $   640.9$ & $       -1.00$ & $    23.362\pm  0.065$ & $    21.496\pm  0.015$ & $    20.783\pm  0.015$ &  $-1.0000$ &     0 & $    0.00$ \nl 
01915 & 12:19:33.091 & $+$06:28:20.09 & $   182.7$ & $  -614.4$ & $   641.0$ & $       -1.00$ & $    23.848\pm  0.101$ & $    21.792\pm  0.019$ & $    21.147\pm  0.017$ &  $-1.0000$ &     0 & $    0.00$ \nl 
01921 & 12:19:13.717 & $+$06:28:24.51 & $  -104.6$ & $  -611.4$ & $   620.3$ & $       -1.00$ & $    22.682\pm  0.022$ & $    21.873\pm  0.015$ & $    21.701\pm  0.025$ &  $-1.0000$ &     0 & $    0.00$ \nl 
01972 & 12:19:17.545 & $+$06:28:20.58 & $   -47.8$ & $  -615.0$ & $   616.9$ & $     1746.19$ & $    22.692\pm  0.045$ & $    21.316\pm  0.016$ & $    20.763\pm  0.020$ &   0.2618 &     2 & $  -18.43$ \nl 
01996 & 12:19:35.051 & $+$06:28:26.57 & $   211.7$ & $  -607.8$ & $   643.6$ & $       -1.00$ & $    22.038\pm  0.018$ & $    21.173\pm  0.010$ & $    20.940\pm  0.014$ &  $-1.0000$ &     0 & $    0.00$ \nl 
02007 & 12:19:19.594 & $+$06:28:27.37 & $   -17.5$ & $  -608.1$ & $   608.4$ & $       -1.00$ & $    22.915\pm  0.042$ & $    19.955\pm  0.004$ & $    18.404\pm  0.002$ &  $-1.0000$ &     0 & $    0.00$ \nl 
02025 & 12:19:16.440 & $+$06:28:31.24 & $   -64.3$ & $  -604.5$ & $   607.9$ & $     2661.45$ & $    23.481\pm  0.064$ & $    21.396\pm  0.014$ & $    20.673\pm  0.012$ &   0.5237 &     2 & $  -20.14$ \nl 
02028 & 12:19:13.644 & $+$06:28:23.06 & $  -105.7$ & $  -612.8$ & $   621.9$ & $       -1.00$ & $    21.271\pm  0.016$ & $    20.161\pm  0.008$ & $    19.790\pm  0.010$ &  $-1.0000$ &     0 & $    0.00$ \nl 
02044 & 12:19:34.051 & $+$06:28:29.44 & $   196.9$ & $  -605.1$ & $   636.3$ & $       -1.00$ & $    23.178\pm  0.052$ & $    21.980\pm  0.022$ & $    21.772\pm  0.031$ &  $-1.0000$ &     0 & $    0.00$ \nl 
02048 & 12:19:17.171 & $+$06:28:25.84 & $   -53.4$ & $  -609.8$ & $   612.2$ & $       -1.00$ & $    23.015\pm  0.061$ & $    21.382\pm  0.018$ & $    20.722\pm  0.018$ &  $-1.0000$ &     0 & $    0.00$ \nl 
02108 & 12:19:37.598 & $+$06:28:37.06 & $   249.4$ & $  -597.2$ & $   647.2$ & $       -1.00$ & $    23.376\pm  0.058$ & $    21.848\pm  0.020$ & $    21.412\pm  0.021$ &  $-1.0000$ &     0 & $    0.00$ \nl 
02111 & 12:19:13.945 & $+$06:28:38.92 & $  -101.3$ & $  -597.0$ & $   605.6$ & $       -1.00$ & $    21.601\pm  0.011$ & $    20.708\pm  0.006$ & $    20.529\pm  0.010$ &  $-1.0000$ &     0 & $    0.00$ \nl 
\enddata
\tablenotetext{a}{A complete catalog is available electronically at http://lambda.uchicago.edu/public/cat/cat\_1216.html.}
\tablenotetext{b}{Spectral type of the galaxies: ``1'' indicates
absorption-line dominated galaxies, ``2'' indicates emission-line
dominated galaxies, and ``0'' indicates absence of spectra.}
\end{deluxetable*}

A galaxy is identified at $|\Delta\,v|<300$ \kms\ and $\rho\le 250\
h^{-1}$ kpc from the absorber at $z=0.1242$ (see Chen \etal\ 2001a).
The galaxy has $R=19.3$, corresponding to $M_R-5\log\,h=-20.0$, and
$\rho=64\ h^{-1}$ kpc.  The morphology of the galaxy is indicative of
a galaxy (at the southern edge) being tidally torn by a more massive
galaxy during the process of merging (top-left panel of Figure 12).
The spectrum of the system shown in the bottom left panel of Figure 12
exhibits prominent emission features, such as H$\alpha$, [N\,II], and
[S\,II], suggesting that the ISM is chemically enriched to solar
metallicity.  The large H$\alpha$ to H$\beta$ line ratio also suggests
the presence of substantial dust in the ISM.

The absorber at $z=0.1242$ exhibits a complex absorption profile.  We
show in the right panel of Figure 12 the absorption features
identified in the QSO spectra obtained using both HST/STIS and FUSE.
Reduction and processing of the STIS spectra are described in Thom \&
Chen (2008).  Reduction and processing of the FUSE spectra are
described in Scott \etal\ (2004).  The combined STIS spectrum has a
spectral resolution of $\delta\,v\approx 6.8$ \kms; the combined FUSE
spectrum has a spectral resolution of $\delta\,v\approx 30$ \kms.

At least four absorption components ($\Delta\,v=-158, -82, +120$ and
$+188$ \kms) are seen in the H\,I, C\,III, and O\,VI absorption
transitions (see also Tripp \etal\ 2008 for a brief discussion of this
system).  Narrow absorption due to the Si\,III$\lambda$1206 transition
is also detected, but only in three components.  All four components
of the \lya\ absorption are saturated.  We have performed a Voigt
profile analysis to determine the column densities of individual
components.  The results are presented in Table 7, which lists from
columns (2) through (9) the best-fit column density, Doppler parameter
($b$), and their associated uncertainties.  Column (10) of Table 7
gives the total estimated column densities of individual transitions.
Given the saturated profiles of the \lya\ transition and relatively
low resolution in the observation of \lyb, we place conservative
limits for the underlying $N(\hI)$.  The Si\,III components all appear
to be very narrow and unresolved in the STIS echelle data.  We
estimate the column densities based on a fixed $b$ value of $b=2.4$
\kms\ that corresponds to gas of $\approx 10^4$ K.  The best-fit Voigt
profiles are shown in the right panel of Figure 12 as smooth curves.

While the blue-shifted and redshifted components of \lya\ and \lyb\
show nearly symmetric kinematic signatures (and possibly in the C\,III
absorption as well) that are indicative of an origin in expanding
supperbubble shells (see e.g.\ Bond \etal\ 2001 and Simcoe \etal\
2002, although see also Kawata \& Rauch 2007 who showed that such
absorption features could also be produced by filamentary accretion
onto a central galaxy or galaxy group), there exists a strong
abundance gradient in Si\,III and O\,VI from $\Delta\,v=-158$ \kms\ to
$\Delta\,v=+188$ \kms\ that suggests a significant variation in the
underlying gas density (assuming a uniform background radiation
field).  Combining the optical morphology of the galaxy and the
differential relative abundances of different ionization species
indicates that the absorber is likely to originate in disrupted tidal
tails as a result of a merger.

\begin{figure*}
\begin{center}
\includegraphics[scale=0.5]{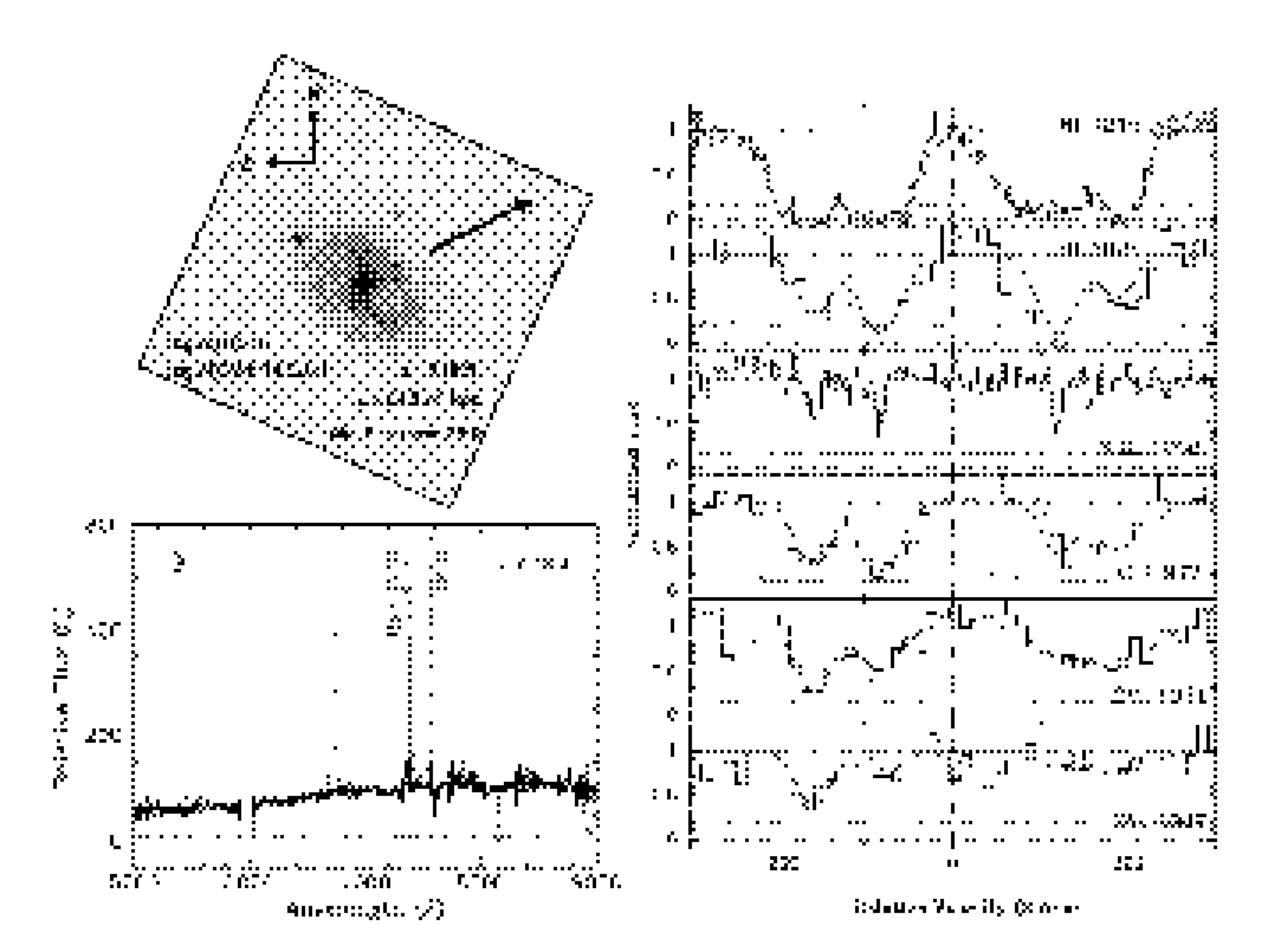}
\caption{{\it Top Left}: Optical morphology of the \ovi\ absorbing
galaxy at $z=0.1239$, as revealed in the WFPC2/F702W image.  The image
corresponds to roughly $25\ h^{-1}$ kpc on a side at the galaxy
redshift.  The arrow indicates the direction toward the QSO sightline.
{\it Bottom left}: The IMACS spectrum of the galaxy in the top panel,
showing prominent emission features of H$\alpha$, [N\,II], and [S\,II]
(marked by dotted lines) that indicate an ISM metallicity comparable
to solar.  The weak H$\beta$ line suggests the presence of substantial
dust.  The gap at 6000 \AA\ is due to a chip gap between IMACS CCDs.
{\it Right}: Absorption profiles of neutral and ionic species found at
$z=0.1242$ (corresponding to zero velocity in the absorption profile).
The spectra were obtained using HST/STIS with a spectral resolution of
$\delta\,v\approx 6.8$ \kms\ (Thom \& Chen 2008a) and FUSE with a
spectral resolution of $\delta\,v\approx 30$ \kms\ (kindly provided by
J.\ Scott; see also Scott \etal\ 2004).  The 1-$\sigma$ error array is
represented by the thin histograms above the zero level.  The
zero-flux level and normalized continuum level are marked by the
dash-dotted lines.  The At least four distint components are present
in the H\,I, C\,III, and O\,VI absorption transitions.  Narrow
absorption due to the Si\,III$\lambda$1206 transition is also
detected, but only in three components.  Smooth curves represent the
best-fit model from a Voigt profile analysis.}
\end{center} 
\end{figure*}

The nearest galaxy to the \ovi\ absorber at $z=0.28232$ (Tripp \etal\
2008; Thom \& Chen 2008a,b) has $R=19.9$ and $\rho=374\ h^{-1}$
physical kpc, with a corresponding rest-frame $R$-band absolute
magnitude of $M_R-5\log\,h=-20.3$.  The galaxy spectrum is dominated
by absorption features.  Three more galaxies are found at
$|\Delta\,v|<300$ \kms\ from the absorber redshift, but they are over
$\rho=835\ h^{-1}$ kpc to $\rho=2.0\ h^{-1}$ Mpc physical distances
away.  

\begin{deluxetable*}{p{0.75in}crrcrrcrrcrrrr}
\tabletypesize{\tiny} 
\tablewidth{0pt} 
\tablecaption{Column Density Measurements of The $\lya+\ovi$ Absorber at $z=0.1242$ Along the PG\,1216$+$069 Sightline} 
\tablehead{\multicolumn{1}{c}{} & &
\multicolumn{2}{c}{$\Delta\,v=-158$ \kms} & &
\multicolumn{2}{c}{$\Delta\,v=-82$ \kms} & &
\multicolumn{2}{c}{$\Delta\,v=+120$ \kms} & &
\multicolumn{2}{c}{$\Delta\,v=+188$ \kms} & & \\ \cline{3-4}
\cline{6-7} \cline{9-10} \cline{12-13} \\
\multicolumn{1}{c}{Transition} & & \multicolumn{1}{c}{$\log\,N$} &
\multicolumn{1}{c}{$b$} & & \multicolumn{1}{c}{$\log\,N$} &
\multicolumn{1}{c}{$b$} & & \multicolumn{1}{c}{$\log\,N$} &
\multicolumn{1}{c}{$b$} & & \multicolumn{1}{c}{$\log\,N$} &
\multicolumn{1}{c}{$b$} & & \multicolumn{1}{c}{$\log N_{\rm tot}$} \\
\multicolumn{1}{c}{(1)} & & \multicolumn{1}{c}{(2)} &
\multicolumn{1}{c}{(3)} & & \multicolumn{1}{c}{(4)} &
\multicolumn{1}{c}{(5)} & & \multicolumn{1}{c}{(6)} &
\multicolumn{1}{c}{(7)} & & \multicolumn{1}{c}{(8)} &
\multicolumn{1}{c}{(9)} & & \multicolumn{1}{c}{(10)}} 
\startdata H\,I 1215$^a$ & & $>14.8$ & ... & & $> 15.0$ & ... & & $>15.0$ & ... & & $14.3\pm 0.1$ & $12\pm 2$ & & $>15.5$ \nl 
Si\,III 1206$^b$ & & $12.5\pm 0.2$ & $2.4$ & & $12.4\pm 0.2$ & $2.4$ & & $12.5\pm 0.1$ & $2.4$ & & $<11.9$ & ... & & $12.9\pm 0.2$ \nl 
C\,III 977 & & $13.6\pm 0.2$ & $17\pm 7$ & & $13.7\pm 0.1$ & $25\pm 5$ & & $13.3\pm 0.1$ & $26\pm10$ & & $13.4\pm 0.1$ & $36\pm 12$ & & $14.1\pm 0.2$ \nl
O\,VI 1031,1037$^a$ & & $14.5\pm 0.2$ & $13\pm 4$ & & $14.0\pm 0.1$ & $24\pm 9$ & & $13.8\pm 0.2$ & $29\pm 17$ & & $14.1\pm 0.1$ & $42\pm 13$ & & $14.8\pm
0.2$ 
\enddata 

\tablenotetext{a}{See also Tripp \etal\ (2008).}
\tablenotetext{b}{The lines appear to be saturated.  The reported column
densities are estimated for the adopted Doppler parameter, $b=2.4$ \kms, that
corresponds to gas of $\approx 10^4$ K.}
\end{deluxetable*}

\section{ANALYSIS}

  We have obtained a spectroscopic sample of galaxies in fields around
three QSOs, for which ultraviolet echelle spectra from FUSE and
HST/STIS are available for identifying intervening hydrogen \lya\ and
\ovi\ absorbers.  The galaxies span a broad range in the rest-frame
$R$-band magnitude from $M_R-5\log\,h>-16$ to $M_R-5\log\,h<-22$ and a
broad range in the projected physical distance from $\rho<30\ h^{-1}$
kpc to $\rho>4\ h^{-1}$ Mpc.  We have shown in \S\ 4.2 that the
completeness of the spectroscopic survey is well understood and
characterized by the angular selection function presented in Figure 6.
Together with a complete sample of intervening \lya\ and \ovi\
absorbers found at $z=0.01-0.5$ in these fields (\S\ 5 and Figure 7),
the galaxy sample offers a unique opportunity for studying the origin
of \lya\ and \ovi\ absorbers based on their cross-correlation
amplitude with known galaxies, and for investigating the gas content
in halos around galaxies.

\subsection{The Galaxy--Absorber Cross-Correlation Functions}

  To quantify the origin of \lya\ and \ovi\ absorbers, we first
measure the projected two-point galaxy auto-correlation function using
a flux-limited sample that has been assembled from our spectroscopic
survey.  The flux-limited galaxy sample contains 670 spectroscopically
identified intervening galaxies of $R\le 22$ within
$\Delta\,\theta=11'$ of the background QSOs, 222 of which
show absorption-line dominated spectral features.  The median
rest-frame $R$-band absolute magnitude of the 448 emission-line
dominated galaxies is $\langle M_R\rangle-5\log\,h=-19.2$
(corresponding to $\approx 0.33\,L_*$ for $M_R*-5\log\,h=-20.44$ from
Blanton \etal\ 2003), while the 222 absorption-line dominated galaxies
have $\langle M_R\rangle-5\log\,h=-20.43$ (corresponding to $\approx
L_*$).  We calculate the projected two-point correlation function
$\omega_{gg}(r_p)$ versus co-moving projected distance $r_p$, using
the Landy \& Szalay (1993) estimator
\begin{equation}
\omega_{gg}(r_p)=\frac{D_g\,D_g-2\,D_g\,R_g+R_g\,R_g}{R_g\,R_g},
\end{equation}
where $D_g$ represents the input galaxy sample and $R_g$ represents a
random galaxy sample.  The random galaxy sample is generated following
the completeness function presented in Figure 6 for each field.  To
minimize possible counting noise introduced by the random galaxy
sample, we have produced a random galaxy sample per field that is a
factor of ten larger than the true galaxy sample.  The two-point
function is then calculated by counting the appropriate $D_g\,D_g$,
$D_g\,R_g$, and $R_g\,R_g$ pairs within different $r_p$ intervals,
from $r_p<250\ h^{-1}$ kpc, to $r_p=250-800\ h^{-1}$ kpc, to
$r_p=800-1500\ h^{-1}$ kpc, and to $r_p=1.5-3\ h^{-1}$ Mpc.

  We present in panel (a) of Figure 13 the auto-correlation functions
intergrated over $40\,h^{-1}$ co-moving Mpc in redshift space for all
galaxies (open circles), absorption-line dominated galaxies (open
triangles), and emission-line dominated galaxies (open squares) in the
flux-limited sample.  Errorbars represent poisson counting
uncertainties\footnote{We note that on small scales ($r_p\apll 1\
h^{-1}$ Mpc) the correlation amplitude is dominated by satellite
galaxies (e.g.\ Zheng \etal\ 2007).  Errors in the measured clustering
signals on these scales are expected to be roughly poisson counting
errors.  On larger scales, however, field-to-field variations are
expected to dominate the errors in the observed clustering signals and
the reported poisson errors represent a lower limit to the true
uncertainties.  Given that our survey covers only three QSO fields, we
are unable to evaluate the uncertainties due to field-to-field
variations using a jackknife method.  }.  These data points for each
of these subsamples are also repeated in panels (b), (c), and (d),
respectively, to be compared with their corresponding galaxy--absorber
cross-correlation functions.  Panel (a) of Figure 13 confirms that
absorption-line dominated galaxies indeed cluster more strongly than
emission-line dominated galaxies.  The observed clustering amplitude
of absorption-line dominated galaxies in our sample is comparable to
what is found by Zehavi \etal\ (2005) for SDSS galaxies of
$M_R-5\log\,h<-19$ at $z\sim 0.1$.  The stronger clustering amplitude
indicates that on average these absorption-line galaxies originate in
higher overdensity regions and presumably more massive dark matter
halos.

\begin{figure*}
\begin{center}
\includegraphics[scale=0.5]{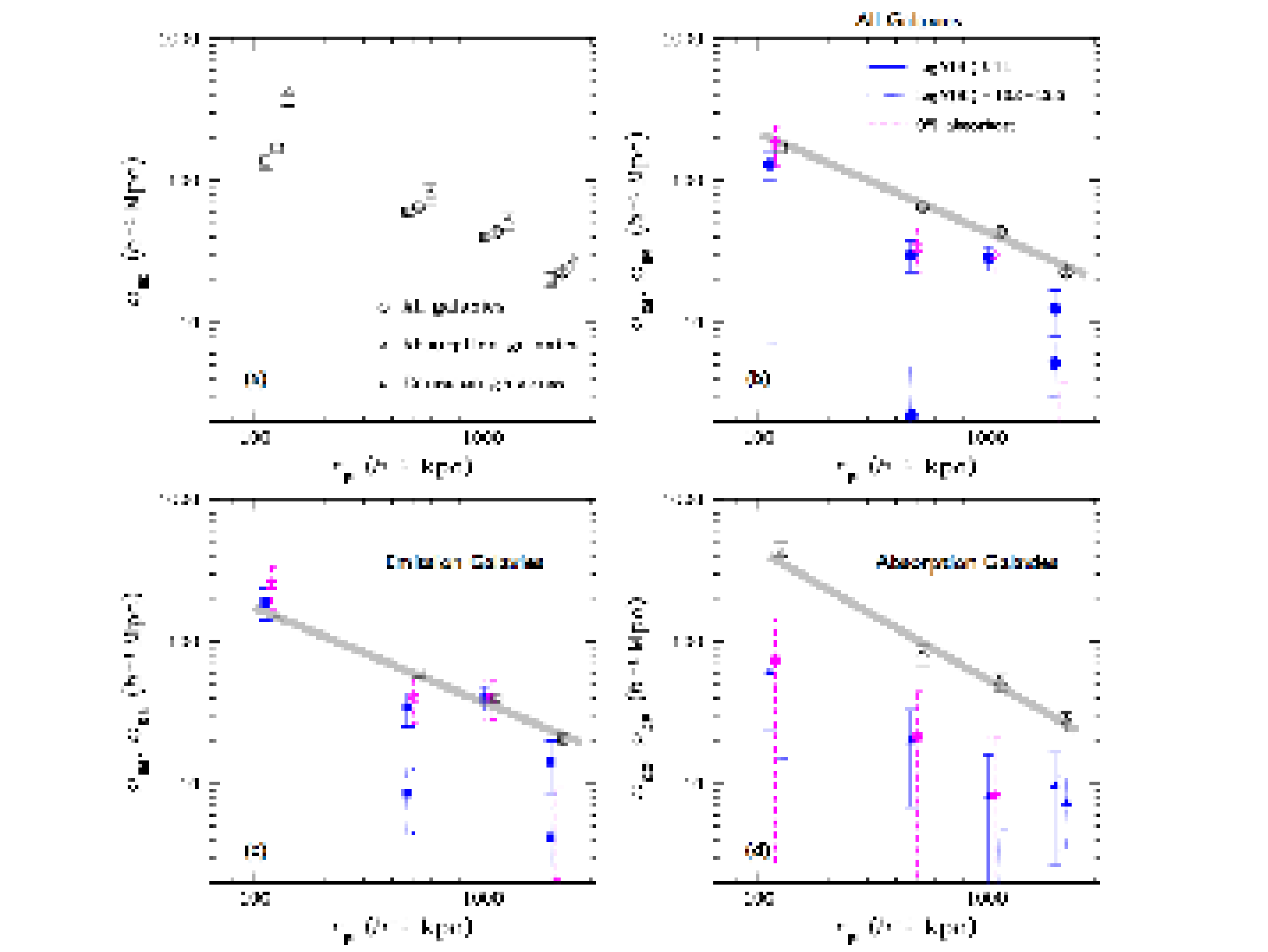}
\caption{({\it a}) Projected auto-correlation functions of galaxies in
our flux-limited sample.  Open circles represent the auto-correlation
function including all galaxies; open triangles represent the
auto-correlation function of absorption-line dominated galaxies only;
and open squares represent the auto-correlation function of
emission-line dominated galaxies only.  We have applied a small offset
in projected co-moving distance $r_p$ to measurements for different
subsamples for clarity.  Each of the auto-correlation functions is
repeated in panels ({\it b}), ({\it c}), and ({\it d}), respectively,
for the corresponding type for direct comparison with the
galaxy--absorber cross-correlation functions.  ({\it b}) Projected
galaxy--absorber cross-correlation functions, including all galaxies
in the flux-limited sample.  The solid points represent galaxy and
\lya\ absorber cross-correlation functions; points with solid
errorbars are for strong \lya\ absorbers of $\log\,N(\hI)\ge 14$;
points with dotted errorbars are for weak \lya\ absorbers of
$\log\,N(\hI)=12.5-13.5$.  Stellar symbols with dashed errorbars
represent the galaxy and \ovi\ absorber cross-correlation
function. The shaded line highlights the galaxy auto-correlation
function to guide the comparisons with galaxy--absorber
cross-correlation functions. ({\it c}) Similar to panel (b), but for
emission-line dominated galaxies only.  (d) Similar to panels (b) and
(c), but for absorption-line dominated galaxies only.}
\end{center} 
\end{figure*}

  Next, we measure the projected two-point galaxy--absorber
cross-correlation function using the flux-limited galaxy sample
discussed above and the absorber catalogs discussed in \S\ 5.  We
exclude absorbers that are within $\Delta\,v=3000$ \kms\ of the
background QSO to reduce contaminations due to QSO associated
absorbers.  There are 195 \lya\ absorbers and 15 \ovi\ absorbers
included in our analysis.  We calculate the projected two-point
cross-correlation function $\omega_{ga}(r_p)$ versus co-moving
projected distance $r_p$ between galaxies and absorbers, using the
Landy \& Szalay (1993) estimator
\begin{equation}
\omega_{ga}(r_p)=\frac{D_g\,D_a-D_g\,R_a-R_g\,D_a+R_g\,R_a}{R_g\,R_a},
\end{equation}
where $D_g$ and $D_a$ represent the input galaxy and absorber samples,
and $R_g$ and $R_a$ represent random galaxy and absorber samples.  We
have produced random \lya\ and \ovi\ absorber catalogs for each field,
assuming a flat absorber selection function in the redshift interval
between $z=0.01$ and $z_{\rm QSO}$.  The random absorber catalogs are
also ten times larger than the true absorber catalogs.  The two-point
cross-correlation function is then calculated by counting the
appropriate $D_g\,D_a$, $D_g\,R_a$, $R_g\,D_a$, and $R_g\,R_a$ pairs
within the same $r_p$ intervals of the auto-correlation function
calculation.

  Panels (b), (c), and (d) of Figure 13 show the galaxy--\lya\
absorber and galaxy--\ovi\ absorber cross-correlation functions
intergrated over $40\,h^{-1}$ co-moving Mpc in redshift space.
Comparisons between galaxy--absorber cross-correlation and galaxy
auto-correlation functions of different galaxy type show four
interesting features.  

  First, while both strong and weak \lya\ absorbers exhibit on average
weaker clustering amplitudes than the galaxies as a whole (solid
points in panel b), strong \lya\ absorbers of $\log\,N(\hI)\ge 14$
appear to exhibit a comparable clustering amplitude on large scales of
$r_p=0.25-3\ h^{-1}$ Mpc and a higher clustering amplitude on small
scales of $r_p\le 250\ h^{-1}$ kpc with the emission-line dominated
galaxies (solid points with solid errorbars panel c).  The comparable
clustering signal on large scales indicates that strong \lya\
absorbers and emission-line galaxies share common halos.  The higher
galaxy--absorber cross-correlation signal relative galaxy
auto-correlation signal on small scales is understood if the gas
covering fraction is high in halos around these galaxies.  Taking into
account previous findings of Chen \etal\ (1998; 2001a), we conclude
that emission-line dominated galaxies are surrounded by extended gas
of nearly unity covering fraction and that strong \lya\ absorbers
primarily probe extended gaseous halos around young star-forming
galaxies.

  Second, these strong \lya\ absorbers exhibit on average $\approx 6$
times weaker amplitude with the absorption-line dominated galaxies
(solid points with solid errorbars panel d) at all separations of
$r_p<3\ h^{-1}$ Mpc.  The weaker clustering amplitude between strong
\lya\ absorbers and absorption-line dominated galaxies is
qualitatively consistent with the clustering amplitude of Mg\,II
absorbers and luminous red galaxies (LRGs) found by Gauthier \etal\
(2009) using a flux-limited LRG sample.  It suggests that the
incidence of strong \lya\ absorbers is very low around these
absorption-line dominated galaxies.

  Third, weak \lya\ absorbers of $\log\,N(\hI)<13.5$ exhibit only a
weak clustering signal to galaxies of all type at separations $r_p<3\
h^{-1}$ Mpc (solid points with dotted errorbars in Figure 13).  The
weak clustering amplitude indicates that the majority of these weak
absorbers do not share the same dark matter halos as the galaxies.
This result extends the finding of Grogin \& Geller (1998) to $\apll
0.3\,L_*$ galaxies that weak \lya\ absorbers of $\log\,N(\hI)<13.5$
occur far more frequently in underdense regions in comparison to these
dwarf galaxies.  Clustering amplitudes measured at large separations
of $r_p \approx 10\ h^{-1}$ Mpc are necessary to constrain the mean
overdensities where these weak \lya\ absorbers reside.

  Finally, \ovi\ absorbers exhibit similar clustering amplitudes as
strong \lya\ absorbers.  Specifically, they show a comparable
clustering amplitude (crosses in panel c of Figure 13) with
emission-line galaxies on scales of $r_p=0.25- 1.5\ h^{-1}$ Mpc but a
factor of $\approx 6$ times lower (petagon points in panel d of Figure
13) than absorption-line dominated galaxies.  Despite large
uncertainties, the differential clustering amplitudes suggest a direct
association between \ovi\ absorbers and star-forming galaxies and a
lack of physical connections between the majority of \ovi\ absorbers
and the gaseous halos around these massive, early-type galaxies.

  We have also attempted to investigate possible dependence of the
cross-correlation amplitude on the strength of O\,VI absorbers using a
mark two-point correlation statistic (see e.g.\ Sheth et al.\ 2005).
Applying the absorber column density as a mark, we could not
distinguish whether such dependence is present.  It appears that the
sample size is still too small for such study to yield statistically
significant results.

  Nevertherless, the comparable clustering amplitude among emission-line
galaxies, strong \lya\ absorbers, and \ovi\ absorbers is, however,
difficult to interpret, because the number density of \ovi\ absorbers
is found to be $n_{\displaystyle\ovi}\approx 10-17$ per unit redshift
interval per line of sight (e.g.\ Tripp \etal\ 2008; Thom \& Chen
2008a) and the number density of strong \lya\ absorbers is found to be
$n_{\displaystyle\lya}\approx 25-30$ (Weymann \etal\ 1998; Dobrzycki
\etal\ 2002).  It shows that on average roughly half of the strong
\lya\ absorbers have associated \ovi.  We defer a more detailed
discussion of this discrepancy to \S\ 8.2.

\subsection{Incidence and Covering Fraction of O\,VI Absorbing Gas Around Galaxies}

  The galaxy sample also allows us to examine the incidence and
covering fraction of \ovi\ absorbing gas around galaxies.  We have
identified a total of 20 foreground galaxies of $R\le 22$ in our
spectroscopic sample that occur at co-moving projected distances $r_p
\le 250\ h^{-1}$ kpc from the lines of sight toward HE\,0226$-$4110,
PKS\,0405$-$123, and PG\,1216$+$069.  Six of the galaxies are
absorption-line dominated galaxies and eight are ``isolated'' galaxies
without additional neighboring galaxies found at $r_p\le 250\ h^{-1}$
kpc from the QSO sightline and within velocity offsets of
$\Delta\,v\le 300$ \kms.  The adopted search volume is typical of the
halo size around $L_*$ galaxies.  Galaxies and absorbers that occur in
this small volume are likely to share a common halo.

  For each galaxy in this sample, we search for corresponding \ovi\
absorption features in the available echelle spectra of the QSOs and
measure the rest-frame absorption equivalent width $W(1031)$.  In the
absence of an absorption feature, we determine a 2-$\sigma$ upper
limit to $W(1031)$ for the galaxy.  Figure 14 shows $W(1031)$ versus
$r_p$ for these galaxies of $R\le 22$ in our sample.  Emission-line
dominated galaxies are marked by squares and absorption-line dominated
galaxies are marked by triangles.  Arrows indicate the 2-$\sigma$
upper limit in $W(1031)$ for galaxies that do not have an associated
\ovi\ absorber.  For those galaxies with neighbors, we include only
the galaxy at smallest $r_p$.  These galaxies are marked by an
additional open circle (indicating at least one neighboring galaxy is
present at $r_p\le 250\ h^{-1}$ kpc from the absorber) in Figure 14.

\begin{figure}
\begin{center}
\includegraphics[scale=0.4]{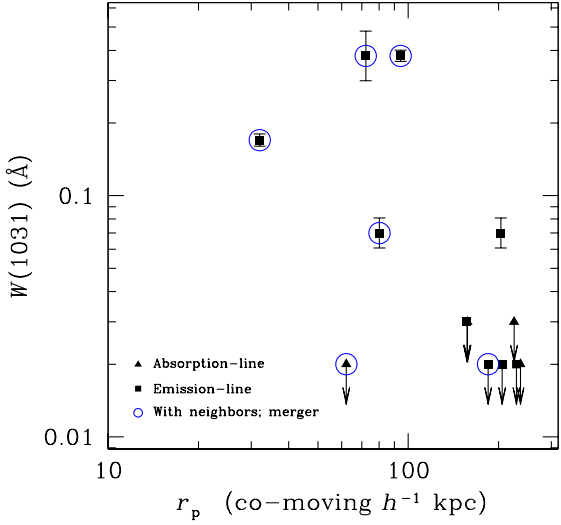}
\caption{Observed \ovi\ absorption strength for all galaxies of $R\le
22$ found at $z=0.01-0.5$ and co-moving projected distances $r_p\le
250\ h^{-1}$ kpc in the fields around HE\,0226$-$4110,
PKS\,0405$-$123, and PG\,1216$+$069.  Squares represent emission-line
dominated galaxies, and triangles represent absorption-line dominated
galaxies.  Arrows indicate the 2-$\sigma$ upper limit to the
rest-frame 1031 \AA\ absorption equivalent width for galaxies that do
not have an associated \ovi\ absorber.  Note that there are two
non-absorbing galaxies occur at $r_p\approx 157\ h^{-1}$ kpc.  Several
galaxies are found to have {\it at least} one neighboring galaxy
located within $r_p=250\ h^{-1}$ kpc and $|\Delta\,v|\le 300$ \kms\
from the absorber, in which cases we include only the closest galaxy
in the plot.  Points with open circles represent galaxies that either
have additional neighboring galaxies or exhibit disturbed morphologies
indicative of a merger event.}
\end{center} 
\end{figure}

  It is clear that none of the absorption-line dominated galaxies in
Figure 14 has a corresponding \ovi\ absorber to a sensitve upper
limit, consistent with the lower clustering amplitude seen in panel
(d) of Figure 13.  Considering only emission-line dominated galaxies
in our sample, we find that nine of the 14 galaxies at $r_p\le 250\
h^{-1}$ kpc have an associated \ovi\ absorber.  This translates to a
covering fraction of $\kappa \approx 64$\% for \ovi\ absorbing gas
within $250\ h^{-1}$ co-moving projected distance of star-forming
galaxies.  The covering fraction appears to be $\kappa\approx 100$\%
at $r_p\le 100\ h^{-1}$ kpc, as all four emission-line dominated
galaxies have an associated \ovi\ absorber.  The observed
$\kappa\approx 64$\% \ovi\ covering fraction at $r_p\le 250\ h^{-1}$
kpc from emission-line galaxies is consistent with the known number
density statistics of strong \lya\ and \ovi\ absorbers, but it further
underscores the difficulty in interpreting the comparable clustering
measurements presented in panel (c) of Figure 13.

  We note that the \ovi\ absorbing galaxy at $r_p=72\ h^{-1}$ in
Figure 14 is the complex absorber at $z=0.1242$ toward PG\,1216$+$069.
As shown in \S\ 6.3 and Figure 12, the galaxy exhibits a disturbed
morphology that suggests an on-going merger.  It is therefore likely
that this galaxy is located in a multiple galaxy environment.  In
addition, the only galaxy at $r_p<100\ h^{-1}$ that does not have a
corresponding \ovi\ absorber to a sensitive upper limit is an
absorption-line dominated galaxy at $z=0.2678$.  At this redshift, an
emission-line dominated galaxy of $R=21.6$ at $r_p \approx 140\
h^{-1}$ and an absorption-line dominated galaxy of $R=18.6$ at
$r_p=251\ h^{-1}$ kpc are also found from the QSO sightline.  Although
our sample is still small and the results have large uncertainties
because of systematic bias from possible field to field variations, it
is interesting to find that star-forming galaxy ``groups'' at
$r_p\apll 100\ h^{-1}$ kpc appear to show a higher incidence of \ovi\
absorbers.

\section{DISCUSSION}

  Using a flux-limited ($R\le 22$) sample of 670 intervening galaxies
spectroscopically identified at $z<0.5$ in fields around three QSOs,
HE\,0226$-$4110, PKS\,0405$-$123, and PG\,1216$+$069, we have
calculated the projected two-point correlation functions of galaxies
and QSO absorption-line systems.  Our analysis confirms early results
of Chen \etal\ (2005; see also Figure 4 of Wilman \etal\ 2007) that
strong \lya\ absorbers of $\log\,N(\hI)> 14$ share comparable
clustering amplitude with emission-line dominated galaxies and that
weak \lya\ absorbers of $\log\,N(\hI)\le 13.5$ cluster only very
weakly around galaxies.  Combining the observed $\approx 100$\%
covering fraction around galaxies (Chen \etal\ 1998, 2001a) and the
comparable clustering amplitude, we conclude that strong \lya\
absorbers at $z<0.5$ primarily probe extended gaseous halos around
star-forming galaxies and that a large fraction of weak \lya\
absorbers of $\log\,N(\hI)\le 13.5$ originate in low overdensity
regions that occur more frequently than $\approx 0.3\,L_*$ galaxies.
The differential clustering amplitude of strong \lya\ absorbers with
different types of galaxies at projected co-moving distances $r_p\le
250\ h^{-1}$ kpc further indicates that the incidence of strong \lya\
absorbers is very low around absorption-line dominated galaxies.

  We have also studied the correlation between galaxies and \ovi\
absorbers.  While our spectroscopic survey has uncovered multiple
large-scale galaxy overdensities in the three fields, Figures 6\&7
show that only two \ovi\ absorbers (at $z=0.0966$ toward
PKS\,0405$-$123 and at $z=0.2823$ toward PG\,1216$+$069) occur at the
redshifts of these overdensities.  An interesting result from the
two-point correlation analysis is that while the majority of \ovi\
absorbers do not probe the gaseous halos around massive, early-type
galaxies, they appear to share common halos with emission-line
dominated star-forming galaxies (and therefore with strong \lya\
absorbers as well).  The interpretation of the clustering analysis is,
however, complicated by additional observations that the covering
fraction of \ovi\ absorbing gas is $\kappa\approx 64$\% around
emission-line dominated galaxies.  While our sample is still small,
this low gas covering fraction is qualitatively consistent with the
observed difference in the number densities of strong \lya\ and \ovi\
absorbers.  It indicates that only a subset of star-forming galaxies
are associated with the observed \ovi\ absorption features.  The
discrepancy between the observed clustering amplitude and covering
fraction of \ovi\ absorbers implies that additional variables need to
be accounted for.
   
  In this section, we review the known properties of individual \ovi\
absorbing galaxies, and discuss the implications of our results on the
origin of \ovi\ absorbers and on the extended gaseous envelopes around
galaxies.

\subsection{The Properties of \ovi\ Absorbing Galaxies}

  There are 13 \ovi\ absorption systems (15 well-separated components)
identified at $z=0.017-0.495$ along the sightlines toward
HE\,0226$-$4110, PKS\,0405$-$123, and PG\,1216$+$069 (Tripp \etal\
2008; Thom \& Chen 2008a,b), including one at $z=0.207$ with
associated Ne\,VIII features (Savage \etal\ 2005).  Our spectroscopic
survey, together with previous searches in these fields, has uncovered
11 galaxies at projected co-moving distances of $r_p<250\ h^{-1}$ kpc
and velocity offsets of $\Delta\,v\le 300$ \kms\ from six of the 13
\ovi\ absorbers\footnote{In the absence of peculiar velocity field,
$\Delta\,v = \pm 300$ \kms\ would correspond to a co-moving pathlength
of $\pm 3.3\ h^{-1}$ Mpc.  Adopting the luminosity function of Blanton
et al.\ (2003) and the two-point galaxy auto-correlation function of
Zehavi et al.\ (2005), we estimate that the probability of finding a
random galaxy of $L > 0.1\,L_*$ at $r_p < 100\ h^{-1}$ co-moving kpc
due to large-scale overdensity is $\approx 3$\%.  The probability
increases to 20\% for $r_p < 250\ h^{-1}$ kpc.}.  Although half of the
\ovi\ absorbers do not yet have an associated galaxy identified, this
small sample of 11 galaxies allows us to examine the common properties
of \ovi\ absorbing galaxies, including the luminosity, galaxy
environment, morphology, and spectral features.

  First, the rest-frame $R$-band absolute magnitude of the galaxies
associated with six \ovi\ absorbers span a broad range from
$M_R-5\log\,h=-15.4$ to $M_R-5\log\,h=-20.9$.  Adopting the magnitude
limit at which our spectroscopic survey is 100\% complete, we further
place a conservative upper limit on $M_R$ for the underlying absorbing
galaxies of six remaining \ovi\ absorbers\footnote{We have excluded
the \ovi\ absorber at $z=0.01746$ toward HE\,0226$-$4110, because at
this low redshift our spectroscopic survey area only covers $\approx
170\ h^{-1}$ kpc co-moving radius of the absorber.}.  A summary of
galaxies found in the vicinity of \ovi\ absorbers is presented in
Table 8, which lists from colums (2) through (11) the absorber
redshift $z_{\rm abs}$, the velocity centroid relative to $z_{\rm
abs}$ of the O\,VI feature $\Delta\,v_{\rm OVI}$, the O\,VI absorbing
gas column density $N({\rm OVI})$, the rest-frame absorption
equivalent width $W_{\rm rest}(1031)$, the velocity centroid relative
to $z_{\rm abs}$ of the H\,I feature $\Delta\,v_{\rm HI}$, the H\,I
absorbing gas column density $N({\rm HI})$, references of the absorber
measurements, the projected co-moving distance $r_p$ and absolute
magnitude $M_R$ of the galaxies, and references of the galaxy
measurements.  We find that while the known \ovi\ absorbing galaxies
exhibit a broad range of intrinsic luminosity, from $<0.01\,L_*$ to
$\approx L_*$, a clear correlation between $N(\ovi)$ and $M_R$ is
absent.
 
  At the same time, four of the six \ovi\ absorbers that have known
associated galaxies are surrounded by either a merging galaxy pair or
by more than one galaxy in the small volume of $r_p<250\ h^{-1}$ kpc
and $\Delta\,v\le 300$ \kms.  In contrast, deep spectroscopic surveys
to search for the galaxies producing absorption features of
low-ionization species such as Mg\,II\,$\lambda\lambda\,2796, 2803$
have yielded $\apll 16$\% of Mg\,II absorbers originating in multiple
galaxy environment (e.g.\ Steidel \etal\ 1997).  On the other hand,
not all galaxy ``groups'' within a similar volume produce an
associated \ovi\ absorber in the spectrum of the background QSO.  We
have shown in \S\ 7.2 that a group of three galaxies at $z\approx
0.268$ and a group of two galaxies at $z\approx 0.199$ within $r_p\le
250\ h^{-1}$ kpc from the sightline toward HE\,0226$-$4110 do not have
an associated \ovi\ absorber.  The most luminous members of both
groups exhibit absorption-line dominated spectral features (see e.g.\
the bottom panel of Figure 15).  It appears that \ovi\ originates
preferentially in groups of gas-rich galaxies.  A detailed
investigation using a statistically representative sample of galaxy
groups close to QSO sightlines is necessary to confirm this result
(see Thom \etal\ 2009, in preparation).

  Using available high-resolution images in the HST archive, we are
also able to examine the detailed optical morphology of five galaxies
in the sample of 11 \ovi\ absorbing galaxies.  Individual images
presented in Figures 10 \& 12 show that four of the galaxies
(responsible for three \ovi\ absorbers) exhibit disk-like morphology
with mildly disturbed features on the edge.  This asymmetric disk
morphology suggests that tidal disruption may be in effect.  The \ovi\
absorbing galaxy at $z=0.4942$ (right panel of Figure 10) appears to
be a regular disk, although the available image depth is insufficient
to reveal more details.

  Finally, all but one of these 11 \ovi\ absorbing galaxies show
prominent emission-line features that suggest a range of ISM
metallicity and star formation history.  The exception is a galaxy at
$z=0.2077$ and projected co-moving distance $r_p=238\ h^{-1}$ kpc from
the sightline toward HE\,0226$-$ 4110, which exhibits primarily
absorption features in the spectrum (see Mulchaey \& Chen 2009).  The
spectra of three \ovi\ absorbing galaxies, one at $z=0.0965$ and two
at $z=0.1670$ toward PKS\,0405$-$123, have been published by previous
authors (see Spinrad \etal\ 1993; Prochaska \etal\ 2006).  The spectra
of three galaxies identified for the \ovi\ $+$ Ne\,VIII absorber at
$z=0.207$ toward HE\,0226$-$4110 are presented in Mulchaey \& Chen
(2009).  Two of these galaxies show emission-line dominated spectral
features.  The spectrum of the \ovi\ absorbing galaxy at $z=0.1239$
toward PG\,1216$+$069 is presented in Figure 12.  It exhibits strong
emission features due to H$\alpha$, [N\,II], and [S\,II] that imply
roughly solar metallicity in the ISM.  In the top four panels of
Figure 15, we present the spectra of four additional \ovi\ absorbing
galaxies identified in our survey, three of which are for the \ovi\
absorber at $z=0.0918$ toward PKS\,0405$-$123 and one for the \ovi\
absorber at $z=0.4951$ along the same sightline.  Emission line
features, such as H$\beta$ and [O\,III], are present in all four
galaxies.  While the \ovi\ absorbing galaxy at $z=0.4942$ exhibits
Balmer absorption series that implies a post-starburst nature, the
lack of [N\,II] emission features together with the presence of strong
H$\alpha$ emission in the three \ovi\ absorbing galaxies at $z=0.091$
places 2-$\sigma$ upper limits to the ISM metallicity at $\approx
10$\% to 30\% of solar values.  In contrast to the spectral features
of \ovi\ absorbing galaxies, we include in the bottom panel of Figure
15 the spectrum of an absorption-line dominated galaxy at $z=0.2678$
and $r_p=62\ h^{-1}$ co-moving kpc from the sightline toward
HE\,0226$-$4110.  The galaxy has $M_R-5\log\,h=-19.9$ and does not
have a corresponding \ovi\ absorber to a 2-$\sigma$ upper limit of
$W(1031)<0.02$ \AA.

\begin{figure}
\begin{center}
\includegraphics[scale=0.4]{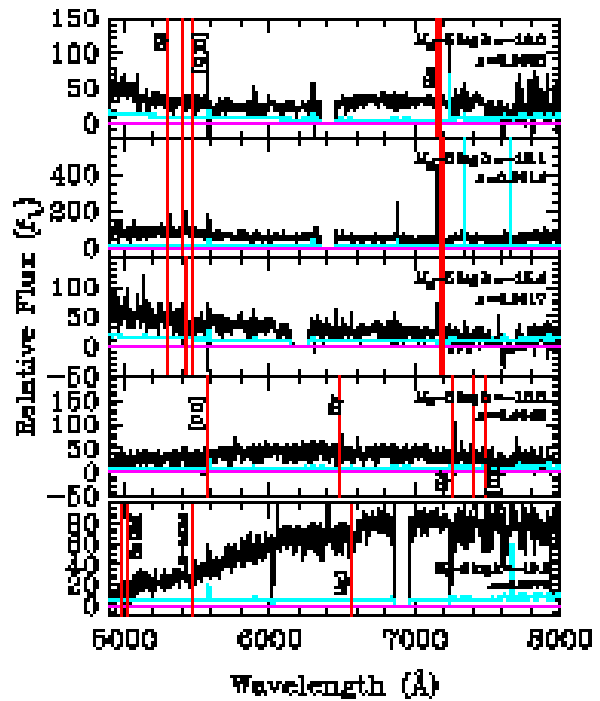}
\caption{Spectra and the associated error arrays of four new \ovi\
absorbing galaxies identified in our survey (top four panels), three
of which are for the \ovi\ absorber at $z=0.0918$ toward
PKS\,0405$-$123 and one for the \ovi\ absorber at $z=0.4951$ along the
same sightline.  The rest-frame absolute magnitude $M_R$ and the
best-fit redshift of each galaxy are given in the upper-right corner
of each panel.  Dotted lines indicate the emission features found at
the best-fit galaxy redshift.  For comparison, we have included in the
bottom panel the spectrum of an absorption-line dominated galaxy at
$z=0.2678$ and $r_p=62\ h^{-1}$ co-moving kpc from the sightline
toward HE\,0226$-$4110.  The galaxy has $M_R-5\log\,h=-19.9$ and does
not have a corresponding \ovi\ absorber to a 2-$\sigma$ upper limit of
$W(1031)<0.02$ \AA.}
\end{center} 
\end{figure}

\begin{deluxetable*}{lrrrrrrrcrrr}
\tabletypesize{\tiny}
\tablecaption{Summary of Galaxies in the Vicinity of \ovi\ Absorbers at $z<0.5$\tablenotemark{a}}
\tablewidth{0pt}
\tablehead{\colhead{} & \multicolumn{7}{c}{Absorbers} & & \multicolumn{3}{c}{Galaxies} \\
\cline{2-8}
\cline{10-12} \\
\colhead{} & \colhead{} & \colhead{$\Delta\,v_{\displaystyle\ovi}$} & \colhead{} & \colhead{$W_{\rm rest}(1031)$} & 
\colhead{$\Delta\,v_{\displaystyle\hI}$} & \colhead{} & \colhead{} & \colhead{} & 
\colhead{$r_p$} & \colhead{$M_R$\tablenotemark{b}} &\colhead{} \\
\colhead{Sightline} & \colhead{$z_{\rm abs}$} & \colhead{(\kms)} & \colhead{$\log\,N(\ovi)$} &
\colhead{(m\AA)} & \colhead{(\kms)} & \colhead{$\log\,N(\hI)$} & \colhead{Reference\tablenotemark{c}} & \colhead{} &
\colhead{($h^{-1}$ kpc)} & \colhead{$-5\,\log\,h$} & \colhead{Reference\tablenotemark{c}} \\
\colhead{(1)} & \colhead{(2)} & \colhead{(3)} & \colhead{(4)} & \colhead{(5)}
& \colhead{(6)} &\colhead{(7)} & \colhead{(8)} & & \colhead{(9)} & \colhead{(10)} & \colhead{(11)}}
\startdata
HE\,0226$-$4110 & 0.01746 &     0 &   $13.6\pm 0.1$ &  $40\pm 10$ &     0 & $13.28\pm 0.06$ &         (1) & &  ...  &    ...   &     \nl
                & 0.20701 &     0 & $14.37\pm 0.03$ & $169\pm 15$ & $-24$ & $15.06\pm 0.04$ &         (2) & &  32.0 &  $-17.2$ & (3) \nl
                &         &       &                 &             &  $+5$ & $14.89\pm 0.05$ &         (2) & &  92.3 &  $-18.9$ & (3) \nl
                &         &       &                 &             &       &                 &             & & 238.1 &  $-18.9$ & (3) \nl
                & 0.32639 &     0 &   $13.6\pm 0.2$ &   $43\pm 8$ &     0 &        $< 12.5$ &         (4) & &   ... & $>-17.3$ & (3) \nl
                & 0.34034 &     0 &   $13.9\pm 0.1$ &   $62\pm 7$ &     0 &   $13.6\pm 0.1$ &         (4) & &   ... & $>-17.4$ & (3) \nl
                & 0.35529 &     0 &   $13.7\pm 0.1$ &   $46\pm 8$ &     0 &   $13.6\pm 0.2$ &         (4) & &   ... & $>-17.5$ & (3) \nl
PKS\,0405$-$123 & 0.09180 & $+20$ & $13.80\pm 0.04$ &   $73\pm 8$ &     0 & $14.52\pm 0.04$ &         (5) & &  80.2 &  $-16.0$ & (3) \nl
                &         &       &                 &             &       &                 &             & & 101.2 &  $-16.1$ & (3) \nl
                &         &       &                 &             &       &                 &             & & 228.5 &  $-15.4$ & (3) \nl
                & 0.09658 &     0 &   $13.7\pm 0.2$ &   $71\pm 9$ &     0 & $14.65\pm 0.05$ &         (5) & & 203.5 &  $-18.2$ & (6) \nl
                & 0.16710 &     0 & $14.78\pm 0.07$ & $361\pm 41$ &     0 & $16.45\pm 0.05$ &     (5),(7) & &  94.3 &  $-20.9$ & (8) \nl
                &         &       &                 &             &       &                 &             & &  79.1 &  $-18.1$ & (8) \nl
                & 0.18291 & $-87$ &   $13.7\pm 0.2$ &  $42\pm 13$ & $-82$ & $14.90\pm 0.05$ &     (4),(5) & &  ...  & $>-18.9$ & (3) \nl
                &         &     0 &   $14.0\pm 0.2$ &  $65\pm 15$ &     0 &   $14.1\pm 0.1$ &     (4),(5) & &  ...  &    ...   &     \nl
                & 0.36332 &     0 &   $13.5\pm 0.1$ &   $30\pm 5$ &     0 &   $13.6\pm 0.1$ &     (4),(5) & &  ...  & $>-20.6$ & (3) \nl
                & 0.49510 &     0 &   $14.5\pm 0.1$ & $213\pm 16$ &     0 &   $14.3\pm 0.2$ & (4),(5),(9) & & 114.9 &  $-18.8$ & (3) \nl
PG\,1216$+$069  & 0.12420 &$-158$ &   $14.5\pm 0.2$ & $225\pm 25$ &$-158$ &         $>14.8$ &         (3) & &  72.2 &  $-20.0$ & (3) \nl
                &         & $-82$ &   $14.0\pm 0.1$ &             & $-82$ &         $>15.0$ &         (3) & &  ...  &    ...   &     \nl
                &         &$+120$ &   $13.8\pm 0.2$ & $194\pm 28$ &$+120$ &         $>15.0$ &         (3) & &  ...  &    ...   &     \nl
                &         &$+188$ &   $14.1\pm 0.1$ &             &$+188$ &   $14.3\pm 0.1$ &         (3) & &  ...  &    ...   &     \nl
                & 0.28232 &     0 &   $13.4\pm 0.2$ &   $26\pm 6$ &     0 & $16.70\pm 0.04$ &    (4),(10) & &  ...  & $>-19.9$ & (3) \\
\enddata
\tablenotetext{a}{Galaxies are found at $r_p\le 250\ h^{-1}$ projected co-moving kpc and $\Delta\,v\le 300$ \kms.}
\tablenotetext{b}{In cases where no absorbing galaxies have been found, we place a conservative upper limit for $M_R$ based on the $R$-band threshold that corresponds to a 100\% completeness out to $\Delta\,\theta<2'$ in our spectroscopic survey.  The magnitude thresholds are $R=23$ for the field around HE\,0226$-$4110, and $R=20$ for the other two fields. }
\tablenotetext{c}{(1) Lehner \etal\ (2006); (2) Savage \etal\ (2005); (3) This work; (4) Thom \& Chen (2008b); (5) Prochaska \etal\ (2004); (6) Prochaska \etal\ (2006); (7) Chen \& Prochaska (2000); (8) Spinrad \etal\ (1993); (9) Howk \etal\ (2009); (10) Tripp \etal\ (2008).}
\end{deluxetable*}

  Excluding the \ovi\ absorber at $z=0.01746$ toward HE\,0226$-$4110
for which our galaxy survey does not cover beyond $160\ h^{-1}$ kpc,
we note that four of the five strong \ovi\ absorbers with
$\log\,N({\rm O\,VI})\ge 14$ have been identified with galaxies at
$r_p<250\ h^{-1}$ kpc and $\Delta\,v\le 300$ \kms, while only two of
the seven weaker \ovi\ absorbers with $\log\,N({\rm O\,VI})< 14$ have
been identified with galaxies at $r_p<250\ h^{-1}$ kpc and
$\Delta\,v\le 300$ \kms.  Given the small sample size, we cannot rule
out possible correlations between absorber strengths and galaxy
properties or the possibility that some of the weaker O\,VI absorbers
may originate in underdense IGM.  For \ovi\ absorbers that do not have
galaxies found at $r_p<250\ h^{-1}$ kpc, we estimate the magnitude
limit of the absorbing galaxies based on our survey depth and present
these numbers in Column (9) of Table 8.

\subsection{The Origin of \ovi\ Absorption Systems}

  The galaxy--\ovi\ absorber cross-correlation analysis presented in
\S\ 7.1 has clearly identified emission-line dominated galaxies to be
principally responsible for the observed \ovi\ absorbers, but the
physical mechanism that distributes the metal-enriched gas to large
galactic distances is unclear.  Two competing scenarios to explain the
origin of \ovi\ absorbers in regions around emission-line galaxies are
(1) starburst driven outflows (e.g.\ Heckman \etal\ 2001; Kawata \&
Rauch 2007; Oppenheimer \& Dav\'e 2009) and (2) satellite accretion
(e.g.\ Wang 1993; Bournaud \etal\ 2004; Elmegreen \etal\ 2007).  Given
that these emission-line galaxies are presumably low-mass (given the
sub-$L_*$ nature) and star-forming (given the presence of strong
emission lines), the strong correlation seems qualitatively consistent
with the expectations of \ovi\ originating in starburst outflows
(c.f.\ Kawata \& Rauch 2007; Oppenheimer \& Dav\'e 2009).  On the
other hand, satellite accretions or galaxy interactions are also
expected to induce star formation (see e.g.\ Li \etal\ 2008).  Here we
examine whether additional insights for the origin of \ovi\ absorbers
can be gleaned based on known properties of the absorbing galaxies.

  We have noted an apparent discrepancy between the observed partial
covering fraction of \ovi\ absorbing gas, $\kappa\approx 64$\%, at
projected co-moving distances $r_p\apll 250\ h^{-1}$ kpc and the
comparable clustering amplitudes of emission-line galaxies and \ovi\
absorbers on scales $r_p\le 1.5\ h^{-1}$ Mpc.  While the partial
covering fraction indicates that not all emission-line galaxies
contribute to the observed \ovi\ absorber statistics, the comparable
clustering amplitude suggests otherwise.  Namely, every emission-line
galaxy has a corresponding \ovi\ absorber.  Likewise, the comparable
clustering amplitudes of \ovi\ and strong \lya\ absorbers also imply
that the two absorber populations share common halos, but
absorption-line surveys show that not all strong \lya\ absorbers are
accompanied by an \ovi\ absorber (see e.g.\ Tables 3\&8).  Although
both starburst outflows and tidal debris due to galaxy interactions
may explain the observed partial covering fraction, it is not clear
that they can both explain the comparable clustering amplitude at
$r_p\apll 3\ h^{-1}$ Mpc.

  To understand these apparent discrepancies, we first note that the
correlation amplitude on scales $r_p < 1\ h^{-1}$ Mpc is dominated by
galaxies that share common halos (e.g.\ Zheng \etal\ 2007; Tinker
\etal\ 2007).  The comparable clustering amplitudes of \ovi\ and
emission-line galaxies (and strong \lya\ absorbers) found in our small
sample from three QSO fields may be understood, if \ovi\ absorbers
arise preferentially in groups of emission-line galaxies.  This
hypothesis is supported by two features found in Figure 14.  First,
four of the five galaxy--\ovi\ absorber pairs are indeed found in an
environment of multiple emission-line galaxies.  Second, only one of
the four ``isolated'' emission-line galaxies has a correponding \ovi\
absorber.  Available observations are indeed consistent with the
expectations that \ovi\ absorbers arise primarily in gas-rich galaxy
groups.  Taking into account the mildly disturbed disk-like
morphologies of \ovi\ absorbing galaxies, we further argue that the
\ovi\ absorbers originate in tidal debris produced by galaxy
interactions in a group or pair environment and that tidal
interactions may be principally responsible for distributing
chemically enriched gas to large galactic distances.

  A larger sample is necessary to improve the uncertainties in the
measurements of $\kappa$ for both group/pair and ``isolated'' galaxies,
and to address possible bias due to field to field variation.
Extending the cross-correlation analysis to scales of $r_p\approx 10\
h^{-1}$ Mpc is also necessary for constraining the mean halo mass of
the absorbers based on their large-scale clustering amplitudes.

\subsection{Extended Gas Around Galaxies}

  Previous studies have shown that luminous galaxies are surrounded by
extended gaseous envelopes that may be probed by C\,IV and Mg\,II
absorption transitions out to projected physical distances of
$\rho\approx 100\ h^{-1}$ kpc (e.g.\ Steidel \etal\ 1994; Chen \etal\
2001b; Tinker \& Chen 2008; Chen \& Tinker 2008) and by strong \lya\
absorption out to $\rho\approx 200\ h^{-1}$ kpc (Lanzetta \etal\ 1995;
Chen \etal\ 1998,2001a).  The covering fraction of extended gas probed
by the presence of these transitions is found to be $>80$\%.  Our
analysis extends these studies to \ovi\ absorbing gas.  We find
extended \ovi\ gas is indeed present around emission-line galaxies out
to projected co-moving distance $r_p\approx 250\ h^{-1}$ kpc
(corresponding to a projected physical distance $\rho\approx 200\
h^{-1}$ kpc at $z=0.3$) with a mean covering fraction of
$\kappa\approx 64$\%.  While \lya\ and Mg\,II absorption features are
found to be progressively weaker in galaxies at larger impact
parameters (Chen \etal\ 1998, 2001a; Chen \& Tinker 2008), we find a
lack of such correlation in either C\,IV (Chen \etal\ 2001b) or \ovi\
(Figure 14).  Including known \ovi\ absorbing galaxies from Cooksey
\etal\ (2008) and Lehner \etal\ (2009) further increases the scatter
in the $W(1031)$ versus $r_p$ distribution at $r_p<250\ h^{-1}$ kpc.

  In a simple two-phase medium model, C\,IV and Mg\,II absorbers arise
in photo-ionized cold clumps pressure-confined in a hot medium.  The
lack of correlation between \civ\ absorber strength and galaxy impact
parameter can therefore be understood by higher gas pressure (and
therefore reduced abundances of C$^{3+}$) at smaller radii (e.g.\ Mo
\& Miralda-Escud\'e 1996).  The same model cannot explain the presence
of \ovi\ around galaxies of a broad range of luminosity due to the
high ionization potential.  Instead, the mildly disturbed disk
morphologies observed in \ovi\ absorbing galaxies suggest that tidal
debris in small groups or close galaxy pairs may be principally
responsible for the observed \ovi\ absorption features.

  Additional insights may be learned from comparisons with known
properties of the Milky Way halos.  While exquisite details, including
tidal streams, infalling clouds, and outflows, have been recorded for
extended gas around the Milky Way (e.g.\ Sembach \etal\ 2003),
applications of these observations for constraining models of galaxy
growth has been limited due to a lack of knowledge for the distances
to these Halo clouds.  In contrast, observations of distant
galaxy--absorbers pairs offer direct measurements of the distances
between absorbing gas and star-forming regions, but the origin of the
absorbing clouds (either due to accretion, outflow, or tidal
stripping) is often more uncertain.

  A systematic survey of \ovi\ absorbers toward 100 extragalactic
sources has yielded positive detections in the Galactic thick disk and
halo along $60-85$\% of these sightlines (Sembach \etal\ 2003).
Because some of these detections are associated with known high
velocity clouds at $\approx 10$ kpc distances (e.g.\ Thom \etal\
2008), the observed incidence of $60-85$\% represents an upper limit
for the covering fraction of \ovi\ absorbing gas over $\sim 250$ kpc
radius to an external observer.  This partial covering of \ovi\ gas in
the halo of the Milky Way, representing typical $\sim\,L_*$ galaxies
in the nearby universe, is qualitatively consistent with the partial
covering fraction found in our study for halos around distant
sub-$L_*$ galaxies.  The lack of evidence for large-scale outflows in
the Milky Way halo based on detections of \ovi\ absorbers also appears
to be consistent with our interpretation that the majority of \ovi\
absorbers are not produced in starburst outflows around distant
star-forming galaxies.  A larger sample of galaxy--\ovi\ absorber
pairs that covers a broad range of impact parameters will allow a
conclusive characterization of the nature of the \ovi\ gas, which in
turn will shed light for the halo gas content around the Milky Way.

  Finally, our analysis has also shown that the majority of strong
\lya\ and \ovi\ absorbers do not probe the gaseous halos around
early-type galaxies.  To study the warm-hot gas around early-type
galaxies or galaxy groups would require a different probe.

\section{SUMMARY}

  We have carried out a deep, wide-area survey of galaxies in fields
around HE\,0226$-$4110, PKS\,0405$-$123, and PG\,1216$+$069.  The QSO
fields are selected to have ultraviolet echelle spectra available from
FUSE and HST/STIS.  The high-resolution UV spectra have uncovered 195
\lya\ absorbers of $\log\,N(\hI)=12.5-16.3$ and 13 \ovi\ absorbers of
$\log\,N(\ovi)=13.4-14.7$ (15 individual components) along the
sightlines toward the background QSOs.  The spectroscopic program has
yielded redshift measurements for 1104 galaxies in three fields.  The
rest-frame $R$-band magnitudes of these galaxies range from
$M_R-5\log\,h>-16$ to $M_R-5\log\,h\apll -22$.  The projected physical
distances of these galaxies range from $\rho<30\ h^{-1}$ kpc to
$\rho>4\ h^{-1}$ Mpc.  While our spectroscopic survey has uncovered
multiple large-scale galaxy overdensities in the three fields, only
two \ovi\ absorbers (at $z=0.0966$ toward PKS\,0405$-$123 and at
$z=0.2823$ toward PG\,1216$+$069) occur at the redshifts of these
overdensities.  Including previously found absorbing galaxies, we have
collected a sample of ten galaxies that coincide with six \ovi\
absorbers found along the three sightlines.  Adopting our
spectroscopic survey limit, we further place upper limits to the
intrinsic luminosities of six remaining \ovi\ absorbing galaxies that
are still missing (three of these are expected to be fainter than
$M_R-5\log\,h=-17.5$).  Two of the fields, PKS\,0405$-$123 and
PG\,1216$+$069, have high quality HST/WFPC2 images available that
cover the $2'\times 2'$ area roughly centered at the QSO.  These
images also allow us to examine the optical morphologies of individual
absorbing galaxies.

  Combining various absorber and galaxy data, we have performed a
cross-correlation study to understand the physical origin of \lya\ and
\ovi\ absorbers and to constrain the properties of extended gas around
galaxies.  The main results of our study are summarize below.

  1.  Using a flux-limited sample of 670 foreground galaxies within
$\Delta\,\theta=11'$ of the QSO, we find based on a cross-correlation
analysis that both strong \lya\ absorbers of $\log\,N(\hI)\ge 14$ and
\ovi\ absorbers exhibit a comparable clustering amplitude as
emission-line dominated galaxies on scales of $r_p<3\ h^{-1}$
co-moving Mpc.  The clustering amplitudes of these absorbers are found
to be $\approx 6$ times lower than those of absorption-line dominated
galaxies.  At the same time, weak \lya\ absorbers of
$\log\,N(\hI)<13.5$ exhibit only a weak clustering signal to galaxies
of all type.  These findings confirm early results of Chen \etal\
(2005; see also Figure 4 of Wilman \etal\ 2007) that strong \lya\
absorbers of $\log\,N(\hI)\le 14$ probe extended halos of
emission-line dominated galaxies and that a large fraction of weak
\lya\ absorbers of $\log\,N(\hI)\le 13.5$ are likely to originate in
low-overdensity regions that occur more frequently than $\approx
0.3\,L_*$ galaxies.

  2.  \ovi\ absorbers exhibit a similar behavior as strong \lya\
absorbers.  These absorbers also show a comparable clustering
amplitude as emission-line galaxies but a factor of six lower
clustering amplitude relative to absorption-line dominated galaxies on
scales of $r_p\le 3\ h^{-1}$ Mpc.  The results imply that the majority
of \ovi\ absorbers do not probe the gaseous halos around massive,
early-type galaxies, and that they probe primarily halos around
emission-line dominated star-forming galaxies.

  3.  Using a small sample of 20 $R\le 22$ galaxies found at co-moving
projected distances $r_p \le 250\ h^{-1}$ kpc from the lines of sight
toward the three QSOs, we find that none of the absorption-line
dominated galaxies in the sample has a corresponding \ovi\ absorber to
a sensitive upper limit of $W(1031)\apll 0.03$ \AA, and that the
covering fraction of \ovi\ absorbing gas around emission-line
dominated galaxies is $\kappa\approx 64$\%.

  4. Four of the six \ovi\ absorbers that have associated galaxies
identified are surrounded by either a merging galaxy pair or by more
than one galaxy in the small volume of $r_p<250\ h^{-1}$ kpc and
$\Delta\,v\le 300$ \kms.  On the other hand, two galaxy ``groups''
found at $r_p\le 250\ h^{-1}$ kpc from a QSO sightline do not have an
associated \ovi\ absorber.  A more detailed examination of galaxies in
these latter two groups shows that the most luminous members of both
groups exhibit absorption-line dominated spectral features, suggesting
that \ovi\ may originate preferentially in gas-rich galaxy groups.

  5. Available high-resolution HST/WFPC2 images of five \ovi\
absorbing galaxies show that these galaxies exhibit disk-like
morphology with mildly disturbed features on the edge, suggesting that
tidal disruption may be in effect.

\acknowledgments

  JSM would like to acknowledge the visitor's program at the Kavli
Institute for Cosmological Physics, where part of the work presented
here was completed.  We thank helpful discussions with J.\ Tinker and
C.\ Thom.  We thank L.\ Matthews, M.\ Rauch, A.\ Kravtsov for helpful
comments on an early version of the manuscript.  We thank R.\ Simcoe
and A.\ Bolton for assistance on obtaining part of the IMACS images
presented in this paper and R.\ Marzke for helpful discussions on
reducing and assembling geometrically distorted multi-CCD imaging
data.  We are grateful to D.\ Kelson and G.\ Walth for assistance on
the reduction of the IMACS spectra.  We also thank J.\ Scott for
providing the combined FUSE spectrum of PG\,1216$+$069.  This research
has made use of the NASA/IPAC Extragalactic Database (NED) which is
operated by the Jet Propulsion Laboratory, California Institute of
Technology, under contract with the National Aeronautics and Space
Administration.  H.-W.C. acknowledges partial support from NASA Long
Term Space Astrophysics grant NNG06GC36G and an NSF grant AST-0607510.
JSM acknowledges partial support for this work from NASA grant
NNG04GC846.



\end{document}